\documentclass[12pt]{article}
\pdfoutput=1
\usepackage[T1]{fontenc}
\usepackage[utf8]{inputenc}
\usepackage{jheppub}

\usepackage[nottoc]{tocbibind}

\usepackage{bbm}
\usepackage{amsmath, mathrsfs}
\usepackage{booktabs}
\usepackage{tikz}
\tikzset{>=stealth}
\usetikzlibrary{calc,decorations.pathmorphing, decorations.pathreplacing, decorations.shapes}
\usepackage{graphicx,subcaption}
\usepackage{mathtools}
\usepackage{bm}
\usepackage{enumitem}
\usepackage{verbatim}
\usepackage{lipsum}
\usepackage{braket}
\usepackage{slashed}
\usepackage[abs]{overpic}
\usepackage[makeroom]{cancel}

\renewcommand{\Re}{\operatorname{Re}}
\renewcommand{\Im}{\operatorname{Im}}

\DeclareMathOperator{\Tr}{Tr}

\DeclareMathAlphabet{\mathbfsf}{OT1}{cmss}{bx}{n}

\newcommand{\cA}{\mathcal{A}}
\newcommand{\cB}{\mathcal{B}}

\newcommand{\cD}{\mathcal{D}}

\newcommand{\cG}{\mathcal{G}}

\newcommand{\cI}{\mathcal I}

\newcommand{\cL}{\mathcal{L}}

\newcommand{\cN}{\mathcal{N}}
\newcommand{\cO}{\mathcal{O}}

\newcommand{\cQ}{\mathcal{Q}}

\newcommand{\Z}{\mathbb{Z}}
\newcommand{\C}{\mathbb{C}}
\newcommand{\R}{\mathbb{R}}

\newcommand{\mb}[1]{\mathbf{#1}}
\newcommand{\mc}[1]{\mathcal{#1}}

\newcommand\beq{\begin{equation}}
\newcommand\eeq{\end{equation}}

\newcommand{\hook}{\mathbin{\rule[.2ex]{.4em}{.03em}\rule[.2ex]{.03em}{.9ex}}}
\newcommand{\e}{{\rm e}}

\newcommand{\rd}{{\rm d}}

\newcommand{\abs}[1]{\left\lvert #1 \right\rvert}

\newcommand{\vol}{{\rm vol}}
\newcommand{\ph}[1]{\phantom{#1}}

\newcommand{\ii}{{\rm i}}
\newcommand{\tE}{t_E}

\renewcommand{\tilde}{\widetilde}

\tikzset{
    partial ellipse/.style args={#1:#2:#3}{
        insert path={+ (#1:#3) arc (#1:#2:#3)}
    }
}

\begin{document}

\title{Introduction to black hole thermodynamics}

\author[]{Pietro Benetti Genolini}
\emailAdd{pietro.benettigenolini@unige.ch}
\affiliation[]{Départment de Physique Théorique, Université de Genève,\\
24 quai Ernest-Ansermet, 1211 Genève, Suisse}

\abstract{ 
These are the lecture notes for a course at the \textit{Roberto Salmeron School in Mathematical Physics} held at the University of Brasilia in September 2025, to be published in the proceedings book \textit{Modern topics in mathematical physics}.
The course provides a concise and biased introduction to black hole thermodynamics. It covers the laws of black hole mechanics, Hawking radiation, Euclidean quantum gravity methods, and $AdS$ black holes.
\\[20pt]
}


\maketitle


\newpage

\section*{Introduction}
\addcontentsline{toc}{section}{Introduction}

General relativity is a very successful theory, but it is also a fundamentally classical theory, in the sense that no probabilities are involved in discussing observables. A simple dimensional argument tells us that its description of the physics breaks down at the scale of the Planck energy, which is the unique combination of the fundamental constants involved in the game, $E_P \equiv \sqrt{ \hbar c^5/G } \sim 10^{19} $  GeV.
Its extraordinary size is due to the weakness of gravity. For reference, the typical energy scale probed by the LHC is $E_W \sim 10^4$ GeV. Therefore, we will almost certainly not observe quantum gravity effects in everyday life or any experiment currently built on Earth. Why should we care, then?

General relativity itself tells us to do so. Spacetime singularities are necessarily present in classical solutions describing either gravitational collapse or cosmology. At these singularities, the classical theory is incomplete, as it doesn’t predict a way of prescribing boundary conditions. Therefore, a theory of quantum gravity is needed.

\medskip

Black holes provide an arena where quantum gravity effects come sharply into focus. They are a striking prediction of general relativity: regions of spacetime bounded by a horizon out of which nothing can escape, and inside which lies hidden a spacetime singularity. Yet, they are not just exotic curiosities, rather the result of quite generic gravitational collapse, and they are real astrophysical objects observed in our universe.

Classically, a black hole is the perfect absorber, as it can emit nothing. However, its defining features (geometry of the horizon, conserved charges) are related by a set of equations that bear a striking resemblance to the laws of thermodynamics.

The analogy becomes physical once one includes quantum effects. Hawking studied a black hole surrounded by a quantum field, and famously showed that it behaves like a thermal object, whose temperature end entropy are fixed by the geometry of the horizon.

Since then, understanding the thermodynamics of black holes has been one of the key questions guiding research into quantum gravity, and it has led, among others, to profound insights into quantum field theory in curved spacetime, the discovery of the holographic nature of gravity, the development of gravitational path integral techniques, and the statistical interpretation of black hole entropy.

\newpage

These notes provide a very concise and very biased introduction to the subject of black hole thermodynamics. 

We begin in Section \ref{sec:1_BH_Mechanics} with a review of some results in the classical description of black holes, focusing on the properties of Killing horizons. In Section \ref{sec:2_BH_Thermodynamics}, we use techniques from thermal quantum field theory to argue that an observer along the orbit of a Killing vector with a Killing horizon detects a temperature that is directly related to the geometry of the horizon itself. We cover the Unruh effect, concerning accelerated observers in flat space, and summarize the Hawking effect, which is instead a phenomenon related to more general event horizons arising from gravitational collapse. In Section \ref{sec:3_GPI}, we introduce a framework for the quantization of gravity, the gravitational path integral, and we look at the thermodynamics of gravity in anti-de Sitter spacetime. Finally, in Section \ref{sec:4_Selected_Topics}, we conclude with a brief review of additional topics related to the gravitational path integral, and issues swept under the rug in Section \ref{sec:3_GPI}.

\medskip

The literature on the topics of the course is enormous, a large number of lecture notes and books exist, and I have drawn from them. I make no claim of originality for the contents of the notes, besides the fact that the choice of topics and the presentation is biased by my own (perhaps idiosyncratic) tastes. Most of the topics were developed within the span of a decade between mid 1970s and mid 1980s. Many of the original papers are beautifully written, are still relevant today and well worth reading. I try to refer to the most relevant ones as we go along, and some have been reprinted in a single book in \cite{Gibbons:1993cg}. For the global structure of the notes, and the presentation, I was very influenced by the lecture notes by Harvey Reall \cite{Reall} and Simon Ross \cite{Ross:2005sc}.\\
As for additional lecture notes and books, I have found the following resources useful.

\medskip

On black hole mechanics and thermodynamics via quantum field theory on curved spacetime (Sections \ref{sec:1_BH_Mechanics} and \ref{sec:2_BH_Thermodynamics})

\begin{itemize}
    \item Birrell, Davies, \textit{Quantum Fields in Curved Space} \cite{Birrell:1982ix}
    \item Fulling, Ruijsenaars, \textit{Temperature, periodicity and horizons} \cite{Fulling:1981}
    \item Jacobson, \textit{Introduction to quantum fields in curved space-time and the Hawking effect} \cite{Jacobson:2003vx}
    \item Reall, \textit{Part 3 Black Holes} \cite{Reall}
    \item Ross, \textit{Black hole thermodynamics} \cite{Ross:2005sc}
    \item Townsend, \textit{Black holes: Lecture notes} \cite{Townsend:1997ku}
    \item Wald, \textit{Quantum Field Theory in Curved Space-Time and Black Hole Thermodynamics} \cite{Wald:1995yp}
    \item Witten, \textit{Introduction to black hole thermodynamics} \cite{Witten:2024upt}
\end{itemize}

\medskip

On Euclidean quantum gravity approach and applications (Sections \ref{sec:3_GPI} and \ref{sec:4_Selected_Topics})

\begin{itemize}
   \item Cassani, \textit{Black Holes and Semiclassical Quantum Gravity} \cite{Cassani:Lectures}
    \item Hawking, \textit{Euclidean Quantum Gravity} \cite{Hawking:EuclideanQG}
   \item Hawking, \textit{The path integral approach to quantum gravity} \cite{Hawking:PathIntegral}
\end{itemize}

\medskip

\subsection*{Acknowledgments} 

I would like to thank the students and the organizers of the \textit{Roberto Salmeron School in Mathematical Physics}, in particular Carolina Matté Gregory, and the students of the course \textit{Topics in gravity} held at Université de Genève in 2024, in particular Pietro Pelliconi. I would also like to thank Davide Cassani, Jerome Gauntlett, Sameer Murthy, Julian Sonner, and James Sparks for multiple conversations on the topics covered in the notes. My work is supported by the SNSF Ambizione grant PZ00P2$\_$208666, and by CP.

\newpage

\section{The laws of black hole mechanics}
\label{sec:1_BH_Mechanics}

\subsection{Rindler horizon}
\label{subsec:1_Rindler_horizon}

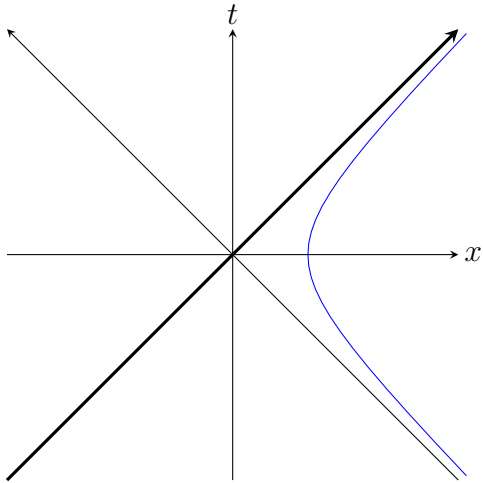
\begin{figure}
    \centering
    \begin{tikzpicture}
    \draw[->] (0,-3) -- (0,3);
    \draw[->] (-3,0) -- (3,0);
    \draw[->] (3,-3) -- (-3,3);
    \draw[very thick,->] (-3,-3) -- (3,3);
    \node at (0,3.2) {$t$};
    \node at (3.2,0) {$x$};
    \draw[blue] plot[domain=-1.8:1.8] ({ cosh(\x)},{sinh(\x)});
    \end{tikzpicture}
    \caption{The worldline of an accelerated observer in flat space. The line $\{ x = t \}$ is the Rindler horizon.}
    \label{fig:1_Accelerated_Observer}
\end{figure}

We begin, counterintuitively for a course on black holes, with flat two-dimensional space
\begin{equation}
\label{eq:1_Minkowski_2d}
	\rd s^2 = - \rd t^2 + \rd x^2 \, ,
\end{equation}
and we look at an observer experiencing a constant acceleration $\alpha>0$ (in her frame). Her worldline parametrized by proper time $s$ is (with an appropriate choice of origin)
\begin{equation}
\label{eq:1_Accelerated_Worldline}
	t = \frac{1}{\alpha} \sinh \alpha s \, , \qquad x = \frac{1}{\alpha} \cosh\alpha s \, , \qquad - t^2 + x^2 = \alpha^{-2} \, ,
\end{equation}
so it's the hyperbola with asymptotes $\{ t = x \, , \ t = -x \}$ represented in Figure \ref{fig:1_Accelerated_Observer}.
The tangent to the worldline is
\begin{equation}
\label{eq:1_Boost_Generator}
	b = \alpha \left( x \frac{\partial \ }{\partial t} + t \frac{\partial \ }{\partial x} \right) \, ,
\end{equation}
with length $b^2\rvert_{\rm obs}=-1$, when evaluated on \eqref{eq:1_Accelerated_Worldline}, and the magnitude of the proper acceleration $A_a = b^c\nabla_c b_a$ is indeed $\alpha$, again when evaluated on the worldline \eqref{eq:1_Accelerated_Worldline}. From our frame, we see her approach (and never get to) the speed of light, and in her frame, eventually she's going to be able to receive information from the region $\{ x > t \}$, but from nowhere left of the line $\{ x = t\}$, which is why we refer to this line as the \textit{Rindler horizon}. This shows that the physics measured by the accelerated observer is quite different from that seen by an inertial one, e.g. with worldline $\{ t = s\, , \ x=0 \}$, but there is an easy way of getting rid of the Rindler horizon: she could simply stop accelerating (unlike the case of black holes, where no such escape is possible).

\medskip

\begin{figure}
    \centering
    \begin{tikzpicture}
    \draw[thick] (3,-3) -- (-3,3);
    \draw[thick] (-3,-3) -- (3,3);
    \draw[->] (0,-3) -- (0,3);
    \draw[->] (-3,0) -- (3,0);
    \draw[->] (3,-3) -- (-3,3);
    \draw[->] (-3,-3) -- (3,3);
    \node at (0,3.2) {$t$};
    \node at (3.2,0) {$x$};
    \node at (3.2,3.2) {$V$};
    \node at (-3.2,3.2) {$U$};
    \node at (3.9,0.7) {\small\color{purple}$\eta=$ const};
    \node at (4.3,2.7) {\small\color{blue}$\xi=$ const};
    \node at (2.8,1) {R};
    \node at (-2.5,1) {L}; 
    \draw[blue] plot[domain=-1.8:1.8] ({ cosh(\x)},{sinh(\x)});
    \draw[blue] plot[domain=-0.95:0.95] ({2.3*cosh(\x)},{2.3*sinh(\x)});
    \draw[purple] plot[domain=0:3] ({\x},{0.2*\x)});
    \draw[purple] plot[domain=0:3] ({\x},{0.6*\x)});
    \end{tikzpicture}
    \caption{Minkowski spacetime divided in the wedges determined by the norm of the boost Killing vector $b^a$. In blue are two loci of constant $\xi$, or equivalently, $\rho$, as defined in \eqref{eq:1_Rindler_Coordinates} and by $\rho = \e^{\alpha\xi}/\alpha$, which are trajectories of the Rindler observers following orbits of $b^a$. In purple are two loci of constant $\eta$.}
    \label{fig:1_RindlerSpace}
\end{figure}
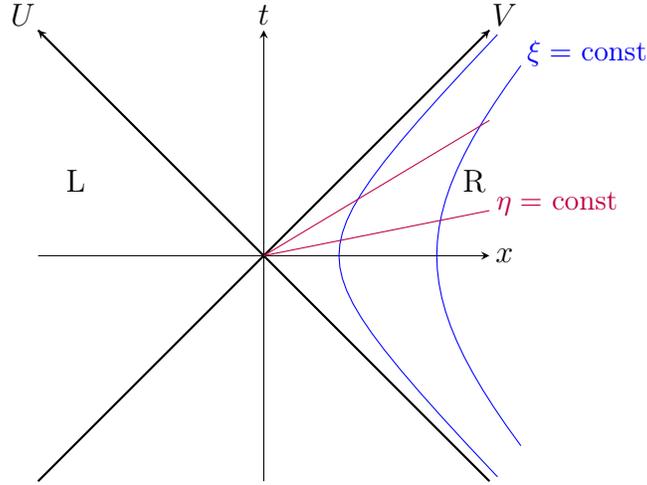

Now we generalize slightly, and consider now the family of observers corresponding to all the orbits of $b^a$ in \eqref{eq:1_Boost_Generator}. These are the hyperbolas $\{-t^2 + x^2 = {\rm constant}\}$, including the degenerate case of the straight asymptotes $\{ t = x\}$, $\{ t = -x \}$.
We remark two important properties of $b^a$: first, since we are in flat space, it is easy to compute that
\[
	\nabla_\mu b_\nu = \alpha \begin{pmatrix}
		0 & 1 \\ -1 & 0
	\end{pmatrix} \, ,
\]
so $b^a$ is a Killing vector (in fact, it's the generator of boosts). Moreover, its length is not constant: $b^2 = - \alpha^2 (x^2 - t^2)$. This signals that the observers are not ``free.'' It's a general property of motion along a Killing vector field $\xi^a$ that it describes an affinely parameterized geodesic if and only if the norm is constant, since
\[
    \xi^b \nabla_b \xi_a = - \xi^b \nabla_a \xi_b = - \frac{1}{2} \nabla_a \xi^2 \, .
\]
In the case of the generator of the boosts, it's timelike only on the two wedges R and L in Figure \ref{fig:1_RindlerSpace}, null on the two lines $\cN \equiv \{ t = x\} \cup \{ t = -x\}$, with the origin being a special point where $b^a$ vanishes, and becomes spacelike in the top and bottom wedges (this is also clear from the orbits).

Focus on the wedges R and L, where $b^a$ is timelike, and consider an observer along an orbit of $b^a$, with normalized velocity $u^a = \frac{1}{\sqrt{-b^2}}b^a$. Her proper acceleration is
\begin{equation}
\label{eq:1_Acceleration_KV_Orbit}
\begin{split}
    A_a &= u^c\nabla_c u_a = \frac{1}{\sqrt{-b^2}} b^c \nabla_c \frac{b_a}{\sqrt{-b^2}} = -\frac{1}{-b^2}b^c \nabla_a b_c = \nabla_a \log \sqrt{-b^2} \, ,
\end{split}
\end{equation}
with magnitude
\begin{equation}
\label{eq:1_Acceleration_Rindler_Observer}
    A = \frac{1}{\sqrt{x^2-t^2}} \, . 
\end{equation}
So, the magnitude of the proper acceleration measured by each observer is constant, though it changes from orbit to orbit. In particular, it's $\alpha$ on the orbit we started with, where $b^2=-1$, vanishes at spatial infinity ($x\to \infty$), and diverges on the asymptotes $\cN$. This is the acceleration measured by an accelerometer carried by each observer, but it's not the acceleration measured by an observer ``at infinity.'' Now, we introduce this quantity in a more general way that will be useful later. 

If $k^a$ is a timelike Killing vector field (in an asymptotically flat spacetime), and there's a particle with velocity $u$, we define the ``energy per unit mass measured at infinity'' to be $E_\infty = - u \cdot k$. It is a constant if the particle moves along a geodesic, but we're interested in the motion along an accelerated trajectory. Interestingly, we are in fact considering motion along an orbit of $k^a$ itself (i.e., a stationary observer), so $E_\infty = \sqrt{-k^2}$, in which case
\[
    \nabla_a E_\infty = \nabla_a \sqrt{-k^2} = \sqrt{-k^2} A_a \, , 
\]
where $A_a$ is the proper (local) acceleration measured by the stationary observer, as showed in \eqref{eq:1_Acceleration_KV_Orbit}. By construction, this is the force per unit mass measured at infinity, so we find the relation
\begin{equation}
\label{eq:1_Acceleration_Infinity}
    A_\infty = \sqrt{-k^2} A \, .
\end{equation}
We can interpret this relation physically: the acceleration of the stationary observer requires a force, which we can imagine being provided by some observer at infinity. Equation \eqref{eq:1_Acceleration_Infinity} is saying that the force measured locally by the accelerating observer will be different from that of the observer at infinity, because of a ``redshift'' factor due to the gravitational pull. In the case of the Rindler observer, we find that
\[
    A_\infty = \alpha \, .
\]
Importantly, this is constant even as $x\to \abs{t}$, we get closer to the horizon, and the local acceleration diverges.\footnote{In flat spacetime, this definition is formal, but it anticipates the genuinely physical notion of force measured by asymptotic observers in curved spacetimes.}

\medskip

Another important property of $b^a$ is that it is normal to $\cN$. This can be more easily seen introducing the light-cone coordinates
\begin{equation}
\label{eq:1_Minkowski_LightConeCoordinates}
    U = t - x \, , \quad V = t + x \, , \quad \Rightarrow \quad \rd s^2 = - \rd U \rd V \, , \quad b = \alpha \left( V \frac{\partial \ }{\partial V} - U \frac{\partial \ }{\partial U} \right) \, .
\end{equation}
Indeed
\begin{equation}
\label{eq:1_Minkowski_NormalKV}
	b_a = \frac{\alpha}{2} \left( - V \, \rd U + U \, \rd V \right)_a = \begin{cases} - \frac{\alpha}{2} V \, (\rd U)_a \quad U=0 \\ \frac{\alpha}{2} U \, (\rd V)_a \quad V= 0 \end{cases} \, , 
\end{equation}
which shows that it's normal to $\cN$.\footnote{If $f$ is a function, the normal to the set $\{ f = \text{constant}\}$ is proportional to $\rd f$.} \\
Since $b^2\rvert_{\cN}=0$, its gradient there is normal to $\cN$ and hence proportional to $b_a$: indeed we have
\begin{equation}
\label{eq:1_Rindler_SurfaceGravity}
    b^2 = \alpha^2 VU \quad \Rightarrow \quad \nabla_a b^2 = \begin{cases}
        - 2\alpha \, b_a \quad & U=0 \\
        2\alpha \, b_a \quad & V=0 
    \end{cases} \, .
\end{equation}
Finally, before leaving our accelerating observer, we define a set of adapted coordinates, that is, coordinates $(\eta,\xi)$ such that $b = \partial_\eta$: one such choice that covers the wedge R is
\begin{equation}
\label{eq:1_Rindler_Coordinates}
\begin{aligned}
	t &= \frac{\e^{\alpha\xi}}{\alpha} \sinh\alpha\eta \, , &\qquad x &= \frac{\e^{\alpha \xi}}{\alpha} \cosh\alpha\eta \, , \\[5pt] 
    \eta &= \frac{1}{2\alpha} \log \frac{x+t}{x-t} \, , &\qquad \xi &= \frac{1}{2\alpha} \log \alpha^2 (x^2-t^2) \, .
\end{aligned}
\end{equation}
Note that $\eta$ and $\xi$ range from $-\infty$ to $+\infty$, though they only cover the wedge R: going back to Figure \ref{fig:1_RindlerSpace}, the hyperbolas in blue are the orbits of $b = \partial_\eta$ corresponding to constant $\xi$, and the straight purple lines correspond to constant $\eta$. In these coordinates, the Minkowski metric \eqref{eq:1_Minkowski_2d} has the form
\begin{equation}
\label{eq:1_Minkowski_2d_Rindler}
	\rd s^2 = \e^{2\alpha \xi } \left( - \rd\eta^2 + \rd \xi^2 \right) \, .
\end{equation}
Looking at this form, we confirm that these coordinates, though defined along the worldline of the observer, are not inertial, as we would expect, since the observer is accelerated. However, as it's always the case, they are locally inertial: here, in a neighbourhood of $\xi=0$, corresponding to the original worldline \eqref{eq:1_Accelerated_Worldline}, $x^2 - t^2 = \alpha^{-2}$.

Rindler-adapted coordinates are often presented in a slightly different way, trading $\xi$ for $\rho = \e^{\alpha \xi}/\alpha > 0$. Then, the metric has the form
\begin{equation}
\label{eq:1_RindlerSpacetime_Geodesic}
    \rd s^2 = - \alpha^2 \rho^2 \, \rd \eta^2 + \rd \rho^2 \, .
\end{equation}
In these coordinates, which again cover the right wedge, the locus $\{x = \abs{t}\}$ is at $\rho = 0$, and the proper acceleration of the observer along $b^a$ (that is, at constant $\rho$) from \eqref{eq:1_Acceleration_Rindler_Observer} is $A = \rho^{-1}$. We shall see \eqref{eq:1_RindlerSpacetime_Geodesic} appear when looking close to the event horizon of a black hole in Section \ref{subsec:2_Near_Horizon}.

\subsection{Schwarzschild horizon}
\label{subsec:1_Schwarzschild_horizon}

We will see that the geometry of the Rindler horizon is also characteristic of regions of black hole spacetimes. As a first example, look at the Schwarzschild metric\footnote{The Schwarzschild metric is the most general spherically symmetric solution to the Einstein equations (a statement going under the name of Birkhoff theorem), so it doesn't just describe a spherically symmetric static black hole, but also the outside of any spherically symmetric configuration, including a spherical gravitational collapse. This is not true in general: the spacetime outside a rotating star is not described by the metric describing the final state of the gravitational collapse, the Kerr metric (see Section \ref{subsec:4_Rotation} for the asymptotically anti-de Sitter case).\label{footnote:1_Schwarzschild}}
\begin{equation}
\label{eq:1_Schwarzschild_Metric}
\begin{split}
    \rd s^2 &= - f(r) \, \rd t^2 + \frac{\rd r^2}{f(r)} + r^2 \, \rd \Omega_2^2 \, , \\ 
    f(r) &\equiv 1 - \frac{2M}{r} \, , \qquad \rd \Omega_2^2 = \rd\theta^2 + \sin^2\theta \, \rd\phi^2 \, ,
\end{split}
\end{equation}
where the coordinates have ranges $t\in \R$, $r>2M$, $\theta \in [0,\pi)$, $\phi \in [0,2\pi)$ (the latter two covering a 2-sphere).
There is a timelike Killing vector field $k = \partial_t$, with non-constant norm $k^2 = - f(r)$. This means that the motion of the stationary observer along $k^a$ is not ``free,'' as it is not along geodesics. 

\begin{figure}
    \centering
    \begin{tikzpicture}
    \draw[thick,->] (3,-3) -- (-3,3);
    \draw[thick,->] (-3,-3) -- (3,3);
    \node at (3.2,3.2) {$V$};
    \node at (-3.2,3.2) {$U$};
    \node at (3.9,1.3) {\small\color{purple}$t=$ const};
    \node at (3.9,2.5) {\small\color{blue}$r=$ const};
    \node at (3.85,-3) {$r=2M$}; 
    \node at (0,1.4) {\color{red}$r=0$}; 
    \draw[blue] plot[domain=-1.35:1.35] ({ 1.5*cosh(\x)},{1.5*sinh(\x)});
    \draw[blue] plot[domain=-1.35:1.35] ({ -1.5*cosh(\x)},{1.5*sinh(\x)});
    \draw[thick,red,dashed] plot[domain=-1.8:1.8] ({ -sinh(\x)},{cosh(\x)});
    \draw[thick,red,dashed] plot[domain=-1.8:1.8] ({ -sinh(\x)},{-cosh(\x)});
    \draw[purple] plot[domain=-3:3] ({\x},{0.4*\x)});
    \end{tikzpicture}
    \caption{Maximal analytic extension of the Schwarzschild black hole, covered by the coordinates $(U,V)$ (each point represents a two-sphere). In blue is a locus of constant $r>2M$, corresponding to an orbit of $k=\partial_t$. In purple is a locus of constant $t$.
    The event horizon $\{ r = 2M\}$ is the union of the two axes $\{U=0\} \cup \{ V=0 \}$, and the singularity $r=0$ is in red.}
    \label{fig:1_KruskalSpace}
\end{figure}
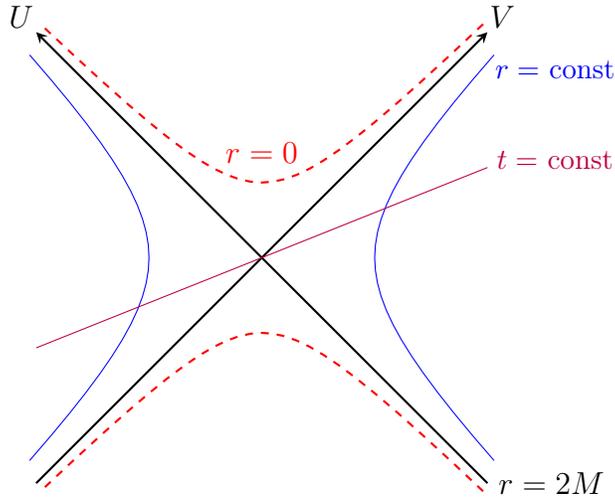

Famously, the apparent singularity at $r=2M$ is only a coordinate singularity, and one can construct a maximal extension covered by the Kruskal--Szekeres coordinates $(U,V,\theta,\phi)$ with\footnote{Coordinates covering the maximal analytical extension had also been independently introduced by Synge (1950) and Fronsdal (1959), before Kruskal (1960) and Szekeres (1960) \cite{Misner:1973prb}.}
\[
    \rd s^2 = - \frac{32 M^3 \e^{- \frac{r(U,V)}{2M}}}{r(U,V)} \rd U \rd V + r(U,V)^2 \, \rd\Omega_2^2 \, ,
\]
where $U, V \in \R$ and $r(U,V)$ is the unique solution to
\begin{equation}
\label{eq:1_Kruskal_Coords}
    UV = - \e^{\frac{r}{2M}} \left( \frac{r}{2M} - 1 \right) \, .
\end{equation}
A diagram of the $U-V$ plane is in Figure \ref{fig:1_KruskalSpace}, where each point represents a two-sphere in the four-dimensional geometry, and the subset covered by the original Schwarzschild coordinates in \eqref{eq:1_Schwarzschild_Metric} is the right wedge $\{U<0,V>0\}$. We have also drawn an orbit of $k^a$ at constant $r>2M$ (in blue) and a straight line corresponding to constant $t$. Looking at \eqref{eq:1_Kruskal_Coords}, we see that the red hyperbola $UV=1$ corresponds to $r=0$, whereas the axes $UV=0$ correspond to $\{ r =2M\}$, the coordinate singularity in \eqref{eq:1_Schwarzschild_Metric}. Note the similarity between the worldlines of the stationary observers along $k^a$ in Figure \ref{fig:1_KruskalSpace} and those along $b^a$ in Figure \ref{fig:1_RindlerSpace}. Again, there is a part of spacetime (namely $U>0$) from which no information can reach an observer along an orbit of $k^a$ with $U<0$, so $\{U=0\}$ acts as a horizon, but now this property is global and cannot be removed by changing the observer’s motion (see later for a formalization of this notion).

It's also straightforward to compute the proper acceleration and the acceleration measured at infinity using \eqref{eq:1_Acceleration_KV_Orbit} and \eqref{eq:1_Acceleration_Infinity}
\[
    A = \frac{1}{\sqrt{f(r)}} \frac{M}{r^2} \, , \qquad A_\infty = \frac{M}{r^2} \, .
\]
As is the case for the Rindler horizon, $A$ diverges near the Schwarzschild horizon, meaning that the local acceleration measured by an observer near the Schwarzschild horizon would diverge. However, if this observer was held by an observer ``at infinity,'' the latter one would only measure a finite value $A_\infty = \frac{1}{4M}$ (where we have measured at the horizon $r=2M$).

To conclude the analogies between the two spacetime, we remark that in the Kruskal coordinates, $k^a$ has the form
\[
    k = \frac{1}{4M} \left( V \frac{\partial \ }{\partial V} - U \frac{\partial \ }{\partial U} \right) \, ,
\]
which is entirely analogous to \eqref{eq:1_Minkowski_LightConeCoordinates}, including the fact that the constant prefactor is $A_\infty\rvert_{\rm horizon}$. We see that $k^a$ is a well-defined Killing vector field on the entire Kruskal spacetime, its norm is $k^2= - f(r)$ even on the extension, and so it's timelike for $r>2M$, spacelike if $0<r<2M$, and null on $\cN = \{r=2M\} = \{UV=0\}$, with a special locus at $\{ U = V = 0\}$ where the Killing vector vanishes \textit{tout court}. It is then straightforward to find the analog of \eqref{eq:1_Minkowski_NormalKV}, showing that $k^a$ is again normal to $\cN$, and to show that 
\begin{equation}
\label{eq:1_Schwarzschild_SurfaceGravity}
     \nabla_a k^2 = \begin{cases}
        - \frac{1}{2M} \, k_a \quad & U=0 \\
        \frac{1}{2M} \, k_a \quad & V=0 
    \end{cases} \, .
\end{equation}

\medskip

Therefore, we have found an analogous structure in the two cases, which is worth defining more generally. We define a \textit{Killing horizon} to be a null hypersurface $\cN$ such that there is a Killing vector $\xi^a$ normal to it. In fact, we are mostly interested in a special class of Killing horizons: a \textit{bifurcate Killing horizon} is the union of two null hypersurfaces that are both Killing horizons intersecting at a codimension-2 spacelike surface, the \textit{bifurcation surface}, where the Killing vector $\xi$ vanishes.

Since $\xi^2=0$ on a Killing horizon, its gradient must be proportional to $\xi_a$ itself (which is normal to $\cN$): we refer to the proportionality constant $\kappa$ as \textit{surface gravity}
\begin{equation}
\label{eq:1_SurfaceGravity_Definition}
	\nabla_a \xi^2 \rvert_{\cN} = - 2 \kappa \, \xi_a \rvert_{\cN} \qquad \Rightarrow \qquad \xi^b \nabla_b \xi^a \rvert_{\cN} = \kappa \, \xi^a \rvert_{\cN} \, .
\end{equation}
The name is justified by the fact that $\kappa$ is the limit at the horizon of the force per unit mass measured at infinity \eqref{eq:1_Acceleration_Infinity}.

Notice that the surface gravity is not a property of the Killing horizon alone, as it depends on the normalization of the Killing vector: $\cN$ is a Killing horizon also for $c\xi^a$ for any real $c$, and the surface gravity would be $c\kappa$. Therefore, in order to assign a physical meaning to the surface gravity, we should also specify a normalization for $\xi^a$. In the two cases considered earlier, we find from \eqref{eq:1_Rindler_SurfaceGravity} and \eqref{eq:1_Schwarzschild_SurfaceGravity} that\footnote{We can avoid the different sign on $\{U=0\}$ and $\{V=0\}$ if we define the surface gravity on $\{V=0\}$ with the opposite sign in \eqref{eq:1_SurfaceGravity_Definition}. This is justified by the fact that on the portion of $\{V=0\}$ and $\{U=0\}$ to the future of the bifurcation surface, $\xi^a$ necessarily has opposite directions.\label{footnote:1_SurfaceGravity}}
\begin{equation}
\label{eq:1_Horizons_SurfaceGravities}
\begin{aligned}
    &\text{Rindler horizon} &\qquad b^2 &= - \alpha^2 (x^2-t^2) \, ,  &\qquad \kappa &= \pm \alpha \, , \\
    &\text{Schwarzschild horizon} &\qquad k^2 &= - 1 + \frac{2M}{r} \, , &\qquad \kappa &= \pm \frac{1}{4M} \, .
\end{aligned}
\end{equation}
Note that in the Schwarzschild case, we can argue that there is a ``canonical'' normalization of the Killing vector defining the horizon: $k^a$ as above generates (future-directed) time translations, normalized to have length $-1$ ``at infinity.'' In contrast, such a ``canonical'' normalization is absent for the Rindler horizon.

\medskip

So, we see that though both cases have the structure of a bifurcate Killing horizon, they are different. This is good: one spacetime is flat and the other isn't, so they shouldn't describe the same physics. How can we formalize the this intuition?

First, consider a family of observers, i.e. a family of (inextendible) timelike curves $\{ \gamma_\alpha\}$: a non-empty boundary of the chronological pasts of their union $\partial I^-\left(\bigcup_\alpha\gamma_\alpha\right)$ is called a \textit{future event horizon}. The future event horizon of the family of observers along orbits of $b^a$ in the right wedge of Minkowski space in Figure \ref{fig:1_RindlerSpace} is $\{U=0, V>0\}$. The same equation also describes the future event horizon of the family of observers along orbits of $k^a$ as in Figure \ref{fig:1_KruskalSpace}. However, physically, we know that there is a big difference between the two cases: in Minkowski space, if the observer stops accelerating and moves on a geodesic, e.g. an orbit of $\partial_t$, then her chronological past includes the entire spacetime. In contrast, in the Kruskal spacetime there is no observer escaping at arbitrarily large distances at arbitrarily late times that is able to receive information from the region $\{0<r<2M\}$. Slightly more formally, we introduce a notion of asymptotic (null) infinity $\mathscr{I}^+$, where null geodesics end, and define a black hole as the region of spacetime that doesn't belong to the chronological past of $\mathscr{I}^+$, that is
\[
    \mathscr{B} \equiv M \setminus I^-(\mathscr{I}^+) \, .
\]
The future event horizon is the boundary $\partial \mathscr{B}$. In the Kruskal spacetime, there is a black hole region corresponding to $\{0<r<2M\}$, which is not there in the (flat) Rindler spacetime.

\subsection{Killing horizons}
\label{subsec:1_Killing_horizons}

We have showed by analysis of the analytic metric that the event horizon of the Schwarzschild black hole is a bifurcate Killing horizon. However, we stress that while the notion of Killing horizon is local and geometric, the notion of event horizon necessarily involves the global structure of spacetime. Nonetheless, the relation between the two notions runs deeper: it is possible to show that the future event horizon of an asymptotically flat stationary black hole with $\kappa\neq 0$ is always a portion of a bifurcate Killing horizon, in presence of ``physically interesting'' matter.\footnote{The proper way to say this is to assume that the stress-energy tensor of the matter $T_{ab}$ satisfies the \textit{dominant energy condition}, that is, $-T^a_{\ph{a}b}V^b$ is a future-directed causal or zero vector for any future-directed timelike vector $V^a$. This guarantees that an observer along $V^a$ would not measure a spacelike energy-momentum current $-T^a_{\ph{a}b}V^b$.} Indeed, this holds for all the known black hole solutions with $\kappa \neq 0$, and in particular for the Kerr--Newman black hole, which is the unique (analytic) black hole solution in four-dimensional Einstein--Maxwell theory with a Killing vector that is timelike in a neighbourhood of asymptotic infinity (that is, it's \textit{stationary}). The other (very interesting!) possibility is a \textit{degenerate Killing horizon}, defined as one with $\kappa = 0$.

This quite general characterization of black hole event horizons allows us to identify properties of these null hypersurfaces that don't rely on knowing analytically the solutions, and in fact hold more broadly than just within general relativity. They are often referred to as \textit{laws of black hole mechanics}, from a 1973 paper by Bardeen, Carter and Hawking \cite{Bardeen:1973gs}. 
The precise statements of the laws vary somewhat depending on assumptions and setups, but the overarching message is that even at the level of classical physics, black holes exhibit a striking analogy with an ordinary thermal system. Such analogy is non-sensical at the level of classical physics, since by definition a black hole absorbs radiation but never emits it: <<\textit{the effective temperature of a black hole is absolute zero. [...] a black hole can be said to transcend the second law of thermodynamics}>> \cite{Bardeen:1973gs}. As we will see in Section \ref{sec:2_BH_Thermodynamics}, things are very different once we include quantum effects.

\paragraph{Zeroth law}

\textit{The surface gravity is constant on a bifurcate Killing horizon.}

Notice that this statement requires the change in sign mentioned in footnote \ref{footnote:1_SurfaceGravity}. To prove this, one first uses Frobenius theorem to find an expression for the square of the surface gravity of a Killing horizon $\cN$ generated by $\xi^a$:
\[
    \kappa^2 = - \frac{1}{2} \nabla^a \xi^b \nabla_a \xi_b \rvert_{\cN} \, , 
\]
from which we can show that the variation of $\kappa$ along a vector field $t^a$ that is tangent to $\cN$ is
\[
\begin{split}
    \kappa t^c\nabla_c \kappa &= - \frac{1}{2} \nabla^a \xi^b t^c \nabla_c \nabla_a \xi_b \rvert_{\cN} \\
    &= - \frac{1}{2} \nabla^a \xi^b t^c R_{bacd}\xi^d \rvert_{\cN} \, .
\end{split}
\]
In the last equality, we used the so-called Killing vector lemma. Choosing $t^a = \xi^a$ shows that $\kappa$ is constant along an orbit of $\xi^a$ in $\cN$, so it's constant along each generator of $\cN$. Moreover, $\kappa$ is also constant on the bifurcation surface: if we restrict to the bifurcation surface, and choose $t^a$ to be a tangent vector, the derivative still vanishes because by definition $\xi^a$ vanishes on the bifurcation surface. So, $\kappa$ is constant everywhere on $\cN$ \cite{Kay:1988mu}.

Note that the constancy of $\kappa$ follows directly from the geometry of bifurcate Killing horizons, so it will apply in any diffeomorphism-invariant theory admitting a horizon that can be extended to a portion bifurcate Killing horizon. However, in the context of general relativity, it is also possible to prove directly that $\kappa$ is constant on the (connected) future event horizon (not necessarily bifurcate) of a stationary black hole obeying the dominant energy condition \cite{Bardeen:1973gs}.

\paragraph{First law}

We can formulate a statement in any theory expressed by a Lagrangian $\cL$ that is only a functional of $g_{ab}$ and other fields $\phi$, the Riemann tensor $R_{abcd}$, and symmetrized covariant derivatives of $R_{abcd}$ and of the fields $\phi$. This is the form of the Lagrangian (if one exists) of any diffeomorphisms-invariant theory.

\hypertarget{1_First_Law}{\textit{We begin with an asymptotically flat stationary black hole solution with a bifurcate Killing horizon, and consider a (not necessarily stationary) asymptotically flat solution of the linearized equations of motion around said solution. Then the following relation between quantities in the perturbation holds
\begin{equation}
\label{eq:1_First_Law}
    \frac{\kappa}{2\pi} \delta S = \delta M - \Omega^i \delta J_i - \Phi^\alpha \delta Q_\alpha \, .
\end{equation}}}
Here
\begin{itemize}
    \item $\kappa$ is the surface gravity of the bifurcate Killing horizon of the black hole
    \item $S$ is defined as follows. Let $\Sigma$ be the codimension-2 bifurcation surface, with binormal $n_{ab}$ (that is, the volume element on the normal space to $\Sigma$ in spacetime). Then
    \begin{equation}
    \label{eq:1_WaldEntropy}
        S \equiv - 2\pi \int_{\Sigma} \frac{\delta \cL}{\delta R_{abcd}} n_{cd} \epsilon_{abc_3\cdots c_n} \, , 
    \end{equation}
    where we have taken the functional derivative of $\cL$ with respect to $R_{abcd}$ holding fixed all the other fields (including the metric) \cite{Wald:1993nt}.
    \item $M$ and $J_i$ are Noether charges associated to the isometries of the black hole (stationary and potentially axisymmetric) obtained integrating Noether currents at asymptotic infinity; $Q_\alpha$ are the electric charges; $\Omega^i$ and $\Phi^\alpha$ are the conjugate angular velocities and electrostatic potentials of the horizon. 
\end{itemize}
We will not prove this form of the first law, which is due to Iyer and Wald \cite{Iyer:1994ys}, but limit ourselves to a few comments.

First, note that we did not need to know anything about the analytic form of the starting solutions, apart from the fact that they solve the equations of motion, and their geometry (that is, we assume that we are discussing black holes with a bifurcate Killing horizon). Second, the first law in \eqref{eq:1_First_Law} is a highly non-trivial relation between quantities measured at the horizon (namely, $\kappa$ and $S$) and quantities measured at asymptotic infinity. These two observations go hand-in-hand. The way to prove the theorem is realizing that for all isometries of solutions in a diffeomorphism-invariant theory there is an exact $(n-1)$-form that can be integrated over a hypersurface extending from the bifurcation surface to asymptotically flat infinity (its existence is guaranteed by the assumptions on the geometry). We then use Stokes' theorem to compute the vanishing integral, thus relating the contribution from the internal boundary (the bifurcation surface) and that from the asymptotic $(n-2)$-sphere, which is then expressed in terms of asymptotic conserved charges.\footnote{An analogous argument in the context of Euclidean gravity is provided in Section \ref{subsec:4_Entropy_Topology}.}

In the context of general relativity, $M$ and $J_i$ on the RHS of \eqref{eq:1_First_Law} reduce to the ADM mass and angular momenta, and it is straightforward to compute \eqref{eq:1_WaldEntropy} from its definition: the relevant part of the Lagrangian is the Einstein--Hilbert action
\[
    \cL = \frac{1}{16\pi} R \, \vol = \frac{1}{16 \pi} g^{ac}g^{bd} R_{abcd} \, \vol \, ,
\]
from which
\begin{equation}
\label{eq:1_Wald_Reduced_BH}
    S = - \frac{1}{8} \int_{\Sigma} g^{ac}g^{bd} n_{cd} \epsilon_{abc_3\cdots c_n} = \frac{1}{4} \int_{\Sigma} \vol_\Sigma = \frac{1}{4} A_{\rm h} \,, 
\end{equation}
which, as we will review in Section \ref{subsec:2_FourLaws}, is the famous expression for the entropy of black holes due to Bekenstein--Hawking. This suggests that \eqref{eq:1_WaldEntropy} could measure the entropy in more general theories of gravity, and thus justifies calling it \textit{Wald entropy}.

\paragraph{Second law} \hypertarget{1SecondLaw}{In contrast to} the First law, the formulation of the Second law is much more restricted. In the context of four-dimensional general relativity, it also goes under the name of Hawking's area law \cite{Hawking:1973uf}. 

\textit{In a strongly asymptotically predictable spacetime} (i.e., provided we have sufficient control over the time evolution) \textit{satisfying the Einstein equations with matter satisfying the null energy condition} (i.e. with ``reasonable'' matter), \textit{the area of the horizon does not decrease in time.}

In order to generalize this result to theories that include derivatives or powers of the curvature, we would need to find a local functional of the geometry of the horizon that is non-decreasing in time evolution and for stationary spacetime reduces to the Wald entropy \eqref{eq:1_WaldEntropy}. As it turns out, this is a tall order: for instance, the definition \eqref{eq:1_WaldEntropy} could suffer from ambiguities \cite{Jacobson:1993vj}, and even resolving them results in a non-decreasing functional only for linear perturbations \cite{Wall:2015raa} or in a restricted sense \cite{Davies:2023qaa}.

\paragraph{Third law}

In the original paper \cite{Bardeen:1973gs}, there was also a conjectural third law of black hole mechanics, which, in a later formulation by Israel, stated

\hypertarget{1_Third_Law}{\textit{A subextremal black hole cannot become extremal
in finite time by any continuous process, no matter how idealized, in which the spacetime and matter fields
remain regular and obey the weak energy condition.}}

Quite recently, this conjecture was shown to be false by Kehle and Unger, who constructed spherically symmetric solutions to Einstein--Maxwell theory with a massless charged scalar that are Schwarzschild near the horizon for a period of advanced time, and then evolve to be exactly an extremal Reissner--Nordstr\"om black hole in a finite amount of time \cite{Kehle:2022uvc}.\footnote{Israel's proof implicitly assumed that the outermost apparent horizon should be connected at all times, whereas Kehle and Unger show that, even under Israel's regularity and energy assumptions, the horizon can become disconnected, allowing the formation of an extremal black hole in a finite time. In fact, they argue that one should read Israel's proof as stating that violations of the third law necessarily have a disconnected apparent horizon.} It is possible to rule out these counterexamples, and thus maintain a third law in the formulation above, if one assumes that the matter stress-energy tensor is constrained by a stronger condition than the dominant energy condition, which takes the form of a bound on the charge to mass ratio \cite{Reall:2024njy}.

There is another statement of the third law of thermodynamics, which states that as the temperature goes to zero, the entropy goes to a universal constant determined by the degeneracy of the ground state of the system. Understanding how to make sense of this statement in the context of black holes requires taking into account quantum effects, and therefore it is beyond the scope of this section.

\newpage

\section{Hawking radiation and black hole thermodynamics}
\label{sec:2_BH_Thermodynamics}

\subsection{Laws of black hole thermodynamics}
\label{subsec:2_FourLaws}

Classical results on the mathematical theory of black holes in four dimensions imply that the gravitational collapse of an isolated body, though temporarily a messy and complicated process where this body rotates, pulsates and throws matter in the universe, settles into a stationary black hole configuration described by the Kerr--Newman solution. Therefore, independently of the (potentially quite complicated) initial state, the final state of the system on and outside the event horizon is completely described by three physical quantities $(M,J,Q)$: mass, angular momentum and electric charge.

This is quite puzzling, as it seems to contradict thermodynamic intuition. Take an object with entropy (say a container filled with gas) and let it fall into the black hole. Once the object has crossed the event horizon, no signal from it can ever reach you, the observer far from the black hole, and thus the entropy of the universe has effectively been lowered.

This would contradict the Second Law of Thermodynamics, which is not good. Bekenstein was the first to articulate this concern, and he also came up with a remarkable solution to the problem: black holes themselves have an entropy, and the total entropy of the black hole and the universe outside the event horizon does not decrease \cite{Bekenstein:1972tm, Bekenstein:1973ur}.

What should the entropy of the black hole be? Bekenstein's idea was to look at information theory. Loosely speaking, in information theory, the entropy of a system is a measure of lack of information about its internal configuration. Analogously, the entropy of a black hole is a measure of the inaccessibility of information to an outside observer about the internal configurations of the black hole. Thus, given a choice of mass, angular momentum and electric charge, the entropy of the black hole is a measure of the size of its equivalence class, which is trivial in the classical viewpoint. In hindsight, it measures the number of \textit{quantum} mechanical microstates, but this was not at all clear at the time. 

Moreover, Bekenstein observed that Hawking had already proved that black holes are characterized by a quantity that never decreases: the area of the horizon (cf. the \hyperlink{1SecondLaw}{second law of black hole mechanics}). Guided once again by the analogy with the Second Law of Thermodynamics, he suggested that the entropy of a black hole should be proportional to its area. He further pointed out that using only general relativity (that is, using only Newton's constant) we cannot construct a combination with dimensions of an area that would allow us to obtain the \textit{dimensionless} entropy, and we have to resort to quantum mechanics. So, he obtained the remarkable prediction that a black hole with horizon of area $A_{\rm h}$ should carry an entropy\footnote{In this expression, we reinstate $c$, $G$ and $\hbar$, keeping $k_B=1$, so $S$ is dimensionless, since $[c]=LT^{-1}$, $[G]=L^3M^{-1}T^{-2}$, $[\hbar]=ML^2T^{-1}$.}
\begin{equation}
\label{eq:2_BekensteinEntropy}
    S_{\rm bh} = \eta A_{\rm h} \frac{c^3}{G \hbar} \, ,
\end{equation}
where $\eta$ is a dimensionless number, unspecified at this stage, and the combination $\ell_P = \sqrt{G\hbar/c^3}$ is the Planck length. 

Contemporaries immediately realized the importance of Bekenstein's observation, but were very skeptical of the physical interpretation.\footnote{A whirlwind account of the events of the time is at the beginning of \cite{Page:2004xp}.} In fact, Bardeen, Carter and Hawking pushed even further the analogy between black hole mechanics and thermodynamics, writing down the first version of the laws introduced in Section \ref{subsec:1_Killing_horizons} \cite{Bardeen:1973gs}. In particular, the first law of black hole mechanics \eqref{eq:1_First_Law} for a $4d$ Kerr--Newman black hole, with the Wald entropy evaluated for the Einstein--Hilbert action as in \eqref{eq:1_Wald_Reduced_BH}, reads
\[
    \delta M = \frac{\kappa}{8\pi} \, \delta A_{\rm h} + \Omega \, \delta J + \Phi \, \delta Q \, ,
\]
where $A_{\rm h}$ is the area of the horizon, $\Omega$ is the angular velocity of the horizon and $\Phi$ is the difference of electrostatic potential measured between asymptotic infinity and the horizon.

Compare with the First Law of Thermodynamics: in any process involving a closed system with energy $E$, entropy $S$ and charges $Q_i$, their variations are related by
\[
    \rd E = T \, \rd S + \sum_i \mu_i \, \rd Q_i \, ,
\]
where $T$ is the temperature and $\mu_i$ is the chemical potential associated to $Q_i$. In order to push further the analogy between black hole mechanics and thermodynamics suggested by the Bekestein's formula \eqref{eq:2_BekensteinEntropy}, we identify mass and energy of the black hole, we recall that indeed $(J,Q)$ are conserved charges, and $(\Omega, \Phi)$ are the conjugate variables, and this leads us necessarily to associate to the black hole a temperature\footnote{Here again we reinstate $c$, $G$, $\hbar$, knowing that $\kappa$ is an acceleration, e.g. for Schwarzschild $\kappa = c^4/(4MG)$, and since $k_B=1$, $[T_{\rm bh}]=ML^2T^{-2}$.}
\begin{equation}
\label{eq:2_Tbh_v1}
    T_{\rm bh} =  \frac{\kappa}{8\pi \eta} \frac{\hbar}{c} \, .
\end{equation}
The relation between temperature and surface gravity is also consistent with the correspondence between the other laws of black hole mechanics and the Laws of Thermodynamics: surface gravity is constant on a bifurcate Killing horizon, just as temperature is constant on a system in thermal equilibrium; and the third law \hyperlink{1_Third_Law}{in the formulation of the original paper} mirrors one of the formulations of the third law of thermodynamics.

However, as remarked by \cite{Bardeen:1973gs}, <<\textit{a black hole cannot be in equilibrium with black body radiation at any non-zero temperature, because no radiation could be emitted from the black hole whereas some radiation would always cross the horizon into the black hole.}>> In fact, temperature is (always) a quantum effect, as confirmed in \eqref{eq:2_Tbh_v1} by the presence of $\hbar$. Therefore, in this chapter we will look at the quantization of fields on a curved background without the Poincaré isometries. As it turns out, this is quite subtle, and will eventually lead us to the proof that black holes \textit{are} indeed thermodynamical objects emitting with a blackbody spectrum with temperature \cite{Hawking:1974rv}
\[
    T_H = \frac{\kappa}{2\pi} \frac{\hbar}{c} \, , 
\]
This fixes the constant $\eta = 1/4$ in \eqref{eq:2_BekensteinEntropy}, as anticipated in \eqref{eq:1_Wald_Reduced_BH}. Knowing this, we can interpret the four laws of black hole mechanics reviewed in Section \ref{subsec:1_Killing_horizons} as truly laws of black hole thermodynamics.

\subsection{Quantum field theory on curved spaces}

In order to show that black holes have a temperature, we need to investigate the behaviour of quantum field theory near the horizon. This requires studying quantum field theory on a curved background.

When one first studies the quantization of a scalar field, one does so by considering a basis of plane wave solutions of the Klein--Gordon equation on flat space that have \textit{positive frequency}, that is they are functions $u_{\mb{p}}$ such that
\begin{equation}
\label{eq:2_PositiveFrequency}
    \ii \cL_{\partial_t} u_{\mb{p}} = \omega \, u_{\mb{p}} \, , \qquad \omega > 0 \, .
\end{equation}
On the space of positive-frequency solutions, one can introduce an inner product that is non-degenerate, Hermitian and positive-definite. One then expands a scalar field on orthonormal elements $u_{\mb{p}}$ and $\overline{u_{\mb{p}}}$, and promotes the coefficients of the expansion to operators, $a_{\mb{p}}$ and $a_{\mb{p}}^\dagger$, which satisfy the algebra of the creation and annihilation operators. Finally, the Hilbert space of the theory is defined to be the symmetric Fock space: the vacuum $\ket{0}$ is defined to be the state annihilated by all $a_{\mb{p}}$, and the other states are constructed by successive applications of $a_{\mb{p}}^\dagger$.

An immediate problem with the generalization of this approach to a curved background is that, even assuming a good causal structure,\footnote{More formally, we consider a \textit{globally hyperbolic spacetime}. This spacetime has a Cauchy surface $\Sigma$ such that the domain of dependence (which, informally, is the union of the points that are reached from $\Sigma$ via a causal curve) is the entire manifold. Simple examples of globally hyperbolic Lorentzian manifolds are flat space and the Kruskal extension of Schwarzschild, but many more are known. Global hyperbolicity is our notion of ``good behaviour'' for a spacetime, since given data on $\Sigma$ we can compute the solution to hyperbolic equations everywhere on $M$. Moreover, it excludes obviously problematic situations, such as closed timelike curves. A globally hyperbolic spacetime is ``good'' for various reasons, among which is the existence of a global time function such that surfaces of constant $t$ are Cauchy surfaces with topology $\Sigma$ and the topology of spacetime is $\R_t \times \Sigma$. Even though the naming may be misleading, on a generic globally hyperbolic spacetime $(\partial_t)^a$ is not a Killing vector.\label{footnote:2_GloballyHyperbolic}} there is no guarantee that there is a globally timelike Killing vector that could play the role of $(\partial_t)^a$ in \eqref{eq:2_PositiveFrequency} and define the positive-frequency subspace. But this is needed to define the annihilation operators, so there is no well-defined notion of ``vacuum.'' Therefore, there is no sensible interpretation of the states of the Hilbert space as ``containing a fixed number of particles.'' 
This looks like a radical departure from the case of flat space, but recall that we are doing quantum field theory. This is of course related to the concept of ``particle,'' but as we see not equivalent. In fact, even on flat space it would be wrong to interpret the ``vacuum'' as the state characterized by the absence of fluctuations, and ``particle'' may not be a useful concept. Talking about the presence of a particle also requires talking about the state of motion of the detector, because the mode decomposition (and consequently the notion of vacuum and particles) is global in nature: the canonical plane waves are agreed upon by all inertial detectors, who will thus also agree on a definition of the vacuum, but we will see in Section \ref{subsec:2_UnruhEffect} that this is not true for accelerating observers.


In a sense, this shows that we’re using the wrong approach, because we are trying to select a preferred set of observers that agree on the splitting of the space of solutions to the Klein--Gordon equation, in analogy with inertial observers in flat spacetime. However, just as there is no preferred coordinate system in general relativity, where coordinate systems are irrelevant, we should imagine that there is no preferred Hilbert space of states when discussing quantum fields on spacetime. The corresponding mathematical statement is the Stone–von Neumann theorem: for systems with finite number of degrees of freedom there is a unique way (modulo unitary equivalence) of representing the canonical commutation relations on a Hilbert space, but this is not true for systems with an infinite number of degrees of freedom. In flat spacetime, this issue is avoided because Lorentz symmetry selects a preferred representation.

There is a way of introducing and discussing general relativity in a manifestly diffeomorphism-invariant way, which is the way we all learnt it. Is there a way of presenting quantum field theory without referring to the construction of a Hilbert space? Yes, it’s the algebraic approach, in which the key role is played by the algebra of observables, and states are defined by the value of each observables (rather than the number of particles). In a more mathematical rephrasing, a quantum field theory is an assignment of an algebra of observables to each subset of spacetime, and a state is a linear map from the algebra of observables to $\C$ that is normalized and positive. Notice that indeed the Hilbert space realization of the algebra of observables is not relevant to the discussion, though we can construct one (following an approach first introduced by Gelfand, Naimark and Segal).

We will not delve into algebraic quantum field theory, because it's too sophisticated for the questions we want to address. Instead, in the next sections we will focus on the properties that define a thermal state from the expectation value of the two-point function. We note that algebraic quantum field theory has been successful in providing rigorous proofs of the Unruh and Hawking effect for free theories, and establishes a mathematically sound framework to discuss questions such as the definition of the entropy of quantum field theories. To get a feeling for this approach, in addition to the book \cite{Wald:1995yp}, one can read \cite{Hollands:2014eia, Witten:2021jzq}.

\subsubsection{The KMS condition}
\label{subsubsec:2_Thermal_QFT}

In a first course in quantum field theory, one usually describes correlators computed in the vacuum state, which is a \textit{pure} state, that is, a unique ray in the projective Hilbert space. A system with finite temperature, instead, is a statistical ensemble of pure states: it's a \textit{mixed} state described (only) by a density matrix.

As a concrete example, we look at a system with Hamiltonian $H$, and assume that the energy is the only quantum number labelling pure states and allowed to vary, so we are describing a system with temperature $T$ in the canonical ensemble. The probability that the system at temperature $T$ is in a pure state with energy $E$ is given by
\[
    p_{E} = \frac{\e^{ - \beta E }}{Z} \, , 
\]
where $\beta = 1/T$, and $Z$ is the canonical partition function of the system obtained tracing over a complete set of states
\[
    Z(\beta) = \sum_{E} \e^{ - \beta E } = \Tr \e^{ - \beta H } \, .
\]
The expectation value of any observable $\cO$ at a temperature $T$ is given by the Gibbs formula, which expresses it as a weighted average over the value of $\cO$ on all the pure states:
\begin{equation}
\label{eq:2_GibbsFormula}
    \braket{\cO}_\beta = \sum_{E} p_{E} \braket{E|\cO | E} = \frac{\Tr \left( \cO \, \e^{ - \beta H } \right) }{\Tr \left( \e^{ - \beta H } \right)} \, .
\end{equation}
Notice that in order for this to make sense, both numerator and denominator must be defined separately, which imposes restrictions on the spectrum of the Hamiltonian.\footnote{For completeness, we write the result also using the density matrix. This is an operator defined by
\[
    \rho = \frac{1}{Z} \e^{ - \beta H } \, , 
\]
and thus \eqref{eq:2_GibbsFormula} can be also written as $\braket{\cO}_\beta = \Tr \rho \cO$.
\label{footnote:2_DensityMatrix}}

\medskip

Now, consider two observables $\cA$ and $\cB$ that evolve according to the Heisenberg picture
\[
    \cA_t = \e^{\ii H t} \cA \e^{- \ii H t} \, , \qquad \cB_t = \e^{\ii H t} \cB \e^{- \ii H t} \, .
\]
We define their correlators at time $t$ as
\begin{equation}
\label{eq:2_Correlators_v1}
\begin{split}
    G_+^\beta(t, \cA, \cB) &\equiv \braket{\cA_t \cB}_\beta = Z^{-1} \Tr \left( \cA_t \cB \, \e^{-\beta H} \right) \, , \\ 
    G_-^\beta(t, \cA, \cB) &\equiv \braket{\cB \cA_t }_\beta = Z^{-1} \Tr \left( \cB \cA_t \, \e^{-\beta H} \right) \, .
\end{split}
\end{equation}
This definition, and the following use of the properties of the trace, requires the trace to be convergent, even if the operator is unbounded, and this is guaranteed, for instance, by choosing a compact spatial manifold. These correlators are defined by the physics for $t\in\R$, but we can analytically extend them introducing a complex variable $z = t + \ii \tE$ where $t$ and $\tE$ are both real, and using the Heisenberg picture
\[
\begin{split}
    G_+^\beta(z, \cA, \cB) &= Z^{-1} \Tr \left[ \e^{\ii (z + \ii\beta) H} \cA \, \e^{-\ii z H} \cB  \right] \, , \\
    G_-^\beta(z, \cA, \cB) &= Z^{-1} \Tr \left[ \cB \, \e^{\ii z H} \cA \, \e^{ - \ii (z - \ii\beta) H} \right] \, .
\end{split}
\]
Requiring that the exponents have negative real parts restricts the domain of holomorphicity of these functions, namely they are holomorphic in
\[
\begin{aligned}
    & G_+^\beta(z, \cA, \cB) &\quad -\beta &< \Im z < 0 \, , \\
    & G_-^\beta(z, \cA, \cB) &\quad 0 &< \Im z < \beta \, , \\    
\end{aligned}
\]
and $G_\pm^\beta(t, \cA, \cB)$ is their limiting value when $\Im z \to 0^\mp$.
In fact, more is true: provided $-\beta \leq \Im z \leq 0$, we can use the cyclity of the trace to show the functional relation
\begin{equation}
\label{eq:2_KMS_v1}
    G_+^\beta( z, \cA, \cB) = G_-^\beta (z + \ii \beta, \cA, \cB) \, .
\end{equation}
This relation goes under the name of \textit{KMS condition} (after the original papers by Kubo and Martin--Schwinger). It is sometimes written on the real axis as
\begin{equation}
\label{eq:2_KMS_v2}
    \braket{\cA_t \cB}_\beta = \braket{\cB \cA_{t+\ii\beta} } \, .
\end{equation}

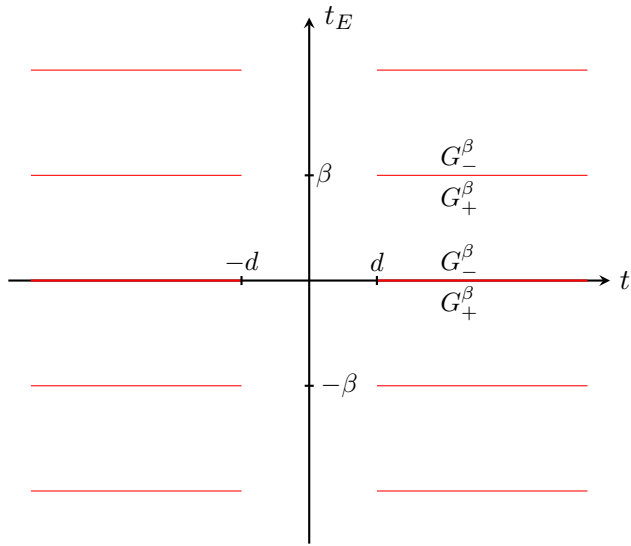
\begin{figure}
    \centering
    \begin{tikzpicture}
    \draw[thick,->] (0,-3.5) -- (0,3.5);
    \draw[thick,->] (-4,0) -- (4,0);
    \node at (0.4,3.5) {\small$\tE$};
    \node at (4.2,0) {\small $t$};
    \draw [red] (-3.7,-2.8) -- (-0.9,-2.8);
    \draw [red] (0.9,-2.8) -- (3.7,-2.8);
    \draw [red] (-3.7,-1.4) -- (-0.9,-1.4);
    \draw [red] (0.9,-1.4) -- (3.7,-1.4);
    \draw [red,thick] (-3.7,0) -- (-0.9,0);
    \draw [red,thick] (0.9,0) -- (3.7,0);
    \draw [red] (-3.7,1.4) -- (-0.9,1.4);
    \draw [red] (0.9,1.4) -- (3.7,1.4);
    \draw [red] (-3.7,2.8) -- (-0.9,2.8);
    \draw [red] (0.9,2.8) -- (3.7,2.8);
    \draw [thick] (-0.06,1.4) -- (0.06,1.4);
    \draw [thick] (-0.06,-1.4) -- (0.06,-1.4);
    \node at (0.4,-1.4) {\footnotesize $-\beta$};
    \node at (0.2,1.4) {\footnotesize $\beta$};
    \draw [thick] (0.9,-0.06) -- (0.9,0.06);
    \draw [thick] (-0.9,-0.06) -- (-0.9,0.06);
    \node at (-0.9,0.25) {\footnotesize $-d$};
    \node at (0.9,0.25) {\footnotesize $d$};
    \node at (2,0.25) {\footnotesize $G^\beta_-$};
    \node at (2,-0.3) {\footnotesize $G^\beta_+$};   
    \node at (2,1.65) {\footnotesize $G^\beta_-$};
    \node at (2,1.1) {\footnotesize $G^\beta_+$};  
    \end{tikzpicture}
    \caption{Thanks to the KMS condition \eqref{eq:2_KMS_v1} and the causality requirements of relativistic quantum field theory, we can extend the correlators to a unique analytic function on the complex plane for $z=t+\ii \tE$, except for the horizontal branch cuts from $t=\pm d$ to $t\to \pm \infty$ at $\tE = \ell \beta$ ($\ell \in \Z$).}
    \label{fig:2_Complex_z_Plane}
\end{figure}

We can use the KMS relation \eqref{eq:2_KMS_v1} to construct a periodic function throughout the $z$-plane apart from the lines $\Im z = \ell \beta$ with $\ell \in \Z$, defined as
\[
\begin{aligned}
    \cG^\beta(z,\cA, \cB) &= G_+^\beta(z, \cA, \cB) &\qquad  -\beta &< \Im z < 0 \, , \\
    \cG^\beta(z,\cA, \cB) &= G_-^\beta(z, \cA, \cB) &\qquad  0 &< \Im z < \beta \, ,
\end{aligned}
\]
and
\[
    \cG^\beta(z, \cA, \cB) = G_+^\beta( z - \ii \ell \beta, \cA, \cB) = G_-^\beta( z - \ii (\ell-1)\beta), \cA, \cB) \, ,
\]
for an integer $\ell$ chosen based on the strip where $z$ is. At this stage, though, we still don't know about analyticity on the real axis, so we cannot \textit{yet} define a unique holomorphic function. On the other hand, if we preserve the causality requirements of relativistic quantum field theories, then $\cA_t\cB = \cB \cA_t$ for $t$ in some open interval $(-d,d)$ on the real axis, so $G^\beta_\pm(t,\cA, \cB)$ are equal on that interval, and we can use the edge-of-the-wedge theorem\footnote{This is a generalization of Morera's theorem applied to contours passing through the ``window'' $\{ \Im z = 0, \ \abs{\Re z}< d \}$, and then breaking them along the real axis.}
to prove that $G^\beta_\pm(z,\cA, \cB)$ are analytic continuations of each other and thus define a single holomorphic function on a connected region of the complex plane (excluding parts of the lines $\Im z = \ell \beta$ with $\ell \neq 0$), see also Figure \ref{fig:2_Complex_z_Plane}. Thus, in thermal relativistic quantum field theories (even interacting ones), using the Gibbs formalism we can construct a single holomorphic function that is periodic in imaginary time, as guaranteed by the KMS condition.

In fact, in the algebraic approach to quantum field theory this reasoning is turned on its head, and the KMS condition is taken to be the defining property of the state of thermal equilibrium at a temperature $1/\beta$. 

\subsubsection{Unruh effect}
\label{subsec:2_UnruhEffect}

As a first application of the formalism just introduced, we consider a scalar quantum field theory on flat space at zero temperature, eventually taking the viewpoint of the accelerated observer introduced in Section \ref{subsec:1_Rindler_horizon}.

We want to construct an analytic extension of the Wightman 2-point functions\footnote{There has been a straightforward change of notation for the observables compared to \eqref{eq:2_Correlators_v1}, and the superscript $\infty$ refers to the fact that they are computed at zero temperature, that is, at infinite $\beta = 1/T$.}
\begin{equation}
\label{eq:2_WightmanFunctions_Scalars}
\begin{split}
    G_+^\infty(t, \mb{x}, \mb{y}) &\equiv \braket{0|\phi(t,\mb{x}) \phi(0,\mb{y})|0} \, , \\
    G_-^\infty(t, \mb{x}, \mb{y}) &\equiv \braket{0 | \phi(0,\mb{y}) \phi(t,\mb{x}) | 0} \, .
\end{split}
\end{equation}
We assume that the theory satisfies the causal requirements of relativistic quantum field theory, so the commutators of the fields at spacelike separation vanishes, and we have
\[
    G_+^\infty(t, \mb{x}, \mb{y}) = G_-^\infty(t, \mb{x}, \mb{y}) \qquad \abs{t} < \abs{\mb{x} - \mb{y}} \, ,
\]
therefore, we can repeat the discussion in the previous section, and construct a holomorphic function $\cG^\infty(z,\mb{x},\mb{y})$ such that
\begin{equation}
\label{eq:2_mathcalG_Infinity}
   \cG^\infty(z,\mb{x},\mb{y}) = \begin{cases}
        G_+^\infty(z,\mb{x}, \mb{y}) &\quad \Im z < 0 \\
        G_-^\infty(z,\mb{x}, \mb{y}) = G_+^\infty(z,\mb{x}, \mb{y}) &\quad \Im z = 0, \ \abs{\Re z}< \abs{\mb{x} - \mb{y}} \\
        G_-^\infty(z,\mb{x}, \mb{y}) &\quad \Im z > 0 \\
   \end{cases}
\end{equation}
though there may be branch cuts along the real axis for $\abs{\Re z} > \abs{\mb{x} - \mb{y}}$. In fact, this is a function only of a continuation of the Lorentzian distance between the two points, that is $F(- z^2 + \abs{\mb{x} - \mb{y}}^2)$, where $F(w)$ is a holomorphic function of $w$ except at the branch point $w = 0$, that is, at $z^2 = \abs{\mb{x} - \mb{y}}^2$.

Now, consider the correlator as measured by the accelerated observer introduced in Section \ref{subsec:1_Rindler_horizon}: she follows an orbit of the boost generator $b^a$. If we are in dimension greater than 2, we align the plane $(t,x^1)$ to the boost, and we introduce Rindler coordinates on the right wedge of the plane, as in \eqref{eq:1_RindlerSpacetime_Geodesic}, via
\begin{equation}
\label{eq:2_RindlerCoordinates}
    t = \rho \sinh \alpha \eta \, , \quad x^1 = \rho \cosh\alpha\eta \, .
\end{equation}
Then, the boost generator has the form $b=\partial_\eta$, and we leave untouched the coordinates that may be transverse the boost plane, that is, $\mb{x_\perp}$. In these coordinates the flat space Wightman function becomes
\begin{equation}
\label{eq:2_cG_RindlerCoords}
\begin{split}
    \cG^\infty(t,\mb{x}, \mb{y})\rvert_{\text{Rindler coords}} &= F( \rho_x^2 + \rho_y^2 - 2 \rho_x \rho_y \cosh\alpha\eta + \abs{\mb{x_\perp} - \mb{y_\perp}}^2) \, .
\end{split}
\end{equation}
We want to show that the resulting function can be extended to a holomorphic function periodic with period $\beta$ that is the analytic continuation of the propagators of a scalar field on Rindler space, i.e., we want to identify the right-hand side of the equation above with $\cG_{\text{Rin}}^\beta(\eta, (\rho_x, \mb{x_\perp}), (\rho_y, \mb{y_\perp}))$ and determine $\beta$. To do so, we need to perform an analytic continuation to the complex plane, show that the resulting function has the appropriate analytic structure (and in particular periodicity along imaginary translations) and that its values along the imaginary axis are the Green's function for the Wick rotation of the Klein--Gordon propagator. This latter property would define it uniquely, because the operator $\left(\partial_{\tE}^2 + \sum_i \partial^2_{x^i} - m^2 \right)$ is elliptic, and its Green's function is unique once boundary conditions are fixed. This is in constrast to the Lorentzian Klein--Gordon operator, which is hyperbolic, and even with fixed boundary conditions, there are multiple Green's functions corresponding to different causal supports (e.g. Feynman's, ``retarded'', ``advanced'').

We first extend \eqref{eq:2_cG_RindlerCoords} to the complex plane introducing $\zeta = \eta + \ii \eta_E$, and we observe that this extension has poles when $\zeta$ satisfies
\[
    \cosh \alpha\zeta = \frac{\rho_x^2 + \rho_y^2 + \abs{ \mb{x_\perp} - \mb{y_\perp} }^2 }{ 2 \rho_x \rho_y } \, , 
\]
and is periodic when $\zeta$ is shifted by $\frac{2\pi}{\alpha} \ii$. Finally, we should show that the would-be Wightman function $\cG_{\text{Rin}, E}^\infty(\eta_E, (\rho_x, \mb{x_\perp}), (\rho_y, \mb{y_\perp}))$ is the Green's function for the Wick-rotated Klein--Gordon operator (that is, after $t=-\ii \tE$). To do so, we notice that after Wick rotation, the Rindler coordinates \eqref{eq:2_RindlerCoordinates} are just polar coordinates in a Euclidean two-dimensional space
\[
    \tE = \rho \sin \alpha\eta_E \, , \qquad x^1 = \rho \cos\alpha\eta_E \, ,
\]
so the Wick-rotated differential operator is just a rewriting in polar coordinates of the flat Klein--Gordon operator, for which $\cG_E^\infty(t,\mb{x}, \mb{y})\rvert_{\text{polar coords}}$ is the Green's function (reading \eqref{eq:2_cG_RindlerCoords} right to left).

Therefore, we confirm that for the Rindler observer, the Minkowski vacuum state behaves like a thermal state with temperature
\begin{equation}
\label{eq:2_Unruh_Temperature_T0}
	T_0 = \frac{\alpha}{2\pi} \, .
\end{equation}
It's important to remark that this result is not restricted to the free scalar theory we just looked at, but can in fact be generalized using algebraic QFT to an arbitrary interacting quantum field theory on flat space (a result proved by Bisognano and Wichmann).

\medskip

As a matter of fact, this is \textit{not} the actual temperature measured by the accelerated observer. Recall that the worldline of a Rindler observer has tangent vector $b = \partial_\eta$, so the normalized four-velocity according to the metric \eqref{eq:1_RindlerSpacetime_Geodesic} is
\[
    u^a = \frac{1}{\alpha\rho } \left(\frac{\partial \ }{\partial \eta} \right)^a = \frac{1}{\alpha\rho} b^a \, .
\]
On the other hand, the frequency $\omega$ above is measured using directly $b^a$. Therefore, the frequency measured by the observer is $\omega_{\rm obs} = \omega / (\alpha\rho)$, leading to a Planck spectrum
\[
    \frac{1}{ \e^{2\pi \omega_{\rm obs} \rho}  - 1  } \, ,
\]
and thus to the \textit{Unruh temperature}\footnote{The fact that temperature changes with the observer is phenomenon often called Tolman--Ehrenfest law: the local temperature $T_{\rm obs}$ measured by an observer travelling along the orbit of a timelike Killing vector $k^a$ is such that $\sqrt{-k^2} T_{\rm obs} = const$. Here we labelled the constant $T_0$. Note the analogy with \eqref{eq:1_Acceleration_Infinity}.\label{footnote:2_TolmanLaw}}
\begin{equation}
\label{eq:2_UnruhTemperature}
    T_U = \frac{1}{2\pi\rho} = \frac{A}{2\pi} \, ,
\end{equation}
where $A$ is the magnitude of the proper acceleration of the observer \eqref{eq:1_Acceleration_Rindler_Observer}. Notice that an observer at infinity, for which $\rho \to + \infty$, will find $T_U \to 0$, which is sensible, because the spacetime we are working on it's still just Minkowski.

The peculiar phenomenon we just discovered is named \textit{Unruh effect}: in flat space, the vacuum defined by an inertial observer will be perceived by an accelerated observer as a \textit{thermal} state with temperature \eqref{eq:2_UnruhTemperature}. This is a physical effect: an accelerated observer with a particle detector will indeed detect particles, and one can show that the Minkowski vacuum is not a pure state for the Rindler observer. However, there is no paradox or contradiction: the mode decomposition (and consequently the notion of vacuum and ``particles'') is \textit{global} in nature, and it requires knowledge of the observer's history. A better notion would be provided by a \textit{local} quantity, such as the expectation value of the stress-energy tensor: if the inertial observer measures $\braket{0_M|T_{ab}|0_M} = 0$, then the stress-energy tensor measured by the accelerated observer, or any other observer related by diffeomorphism, is related by the induced tensor transformation, so it still vanishes: $\braket{0_M|T_{ab}'|0_M} = 0$. The renormalized stress–energy tensor probes the local geometry and the quantum state, but is insensitive to the observer’s acceleration. Its vanishing reflects the fact that the underlying spacetime–state pair is that of flat space quantum field theory, even though the observer’s particle notion may be thermal.

\subsection{Black holes}

We would like to extend the discussion made in the previous section to curved spacetime, and in particular to quantum field theory on the region outside the event horizon of a black hole.

\subsubsection{Near the horizon of a black hole}
\label{subsec:2_Near_Horizon}

First, we show that using the previous arguments we have secretly connected thermal equilibrium with solving an elliptic differential equation on a \textit{smooth} Riemannian space. This will provide a way of detecting the temperature (which, again, is related to the periodicity of the Green's function) simply looking at the geometry of a background with an analytically continued metric \cite{Hartle:1976tp, Gibbons:1976pt}.

\medskip

Consider again the metric of the Rindler wedge of flat space, in Rindler coordinates \eqref{eq:1_RindlerSpacetime_Geodesic}. After Wick rotation $\eta = - \ii \eta_E$, the metric is
\begin{equation}
\label{eq:2_EuclideanRindler}
    \rd s_E^2 = \alpha^2 \rho^2 \, \rd\eta_E^2 + \rd\rho^2 + \rd \mb{x_\perp}^2 \, .
\end{equation}
Focusing on the $\R^2$ spanned by $\rho$ and $\eta_E$, we see that this is locally isometric to $\R^2$ in polar coordinates when one identifies $\rho>0$ as the radial distance and $\eta_E$ as the angle, so the curvature vanishes. Whether it's actually $\R^2$ depends on the identification of $\eta_E$. In $\R^2$, taken a circle of radius $r$, it's a fact that the ratio of proper length and proper radius is fixed to be
\[
    \frac{\text{circumference}}{\text{radius}} = \frac{2\pi r}{r} = 2\pi \, .
\]
Suppose we identify $\eta_E \sim \eta_E + \beta$, then in the Euclideanization of Rindler space we would have, for a circle of constant $\rho_*$
\[
    \frac{\text{circumference}}{\text{radius}} = \frac{\beta \alpha \rho_*}{\rho_*} = \beta \alpha \, .
\]
Therefore, the geometry has a locally flat behaviour, and so it is smooth, if and only if we take $\beta = 2\pi/\alpha$, which is the same result obtained from the analysis of the Wightman functions! If this identification for the period of $\eta_E$ is not made, the line element \eqref{eq:2_EuclideanRindler} is said to have a \textit{conical singularity}. This is because indeed a cone can be obtained by cutting an angle from a flat plane and gluing together the two sides: the geometry would be smooth everywhere but at the apex of the cone, and parallel transport of vectors around the apex would result in a change of their orientation equal to the angle removed from flat space.

With this observation, together with the discussion in the previous section, we have connected the regularity of the Wick-rotated geometry to the periodicity of the Euclidean Wightman function (unique after imposing boundary conditions, because of the ellipticity of the differential equation it solves) and then, using the KMS condition, with the temperature. The thermal system on the Lorentzian geometry is mapped to the field theory on a smooth Euclidean background with a circle direction. Isn't this amazing?

\medskip

We can apply the same Euclidean method to a much more general class of solutions. The Unruh effect is essentially related to the bifurcate Killing horizon for the generator of boosts $b^a$ in Minkowski flat space. However, one of the points of Section \ref{sec:1_BH_Mechanics} was that bifurcate Killing horizons share many properties, among which the fact that observers along orbits of the generators are accelerated in an analogous way, with acceleration related to the surface gravity: do they all measure a temperature? We now repeat the Unruh argument in a geometrical setting where acceleration is enforced by spacetime curvature rather than by external forces.

Let's focus on a special case: let's look at a static spherically symmetric spacetime with line element
\begin{equation}
\label{eq:2_StaticSphericalBH}
    \rd s^2 = - f(r) \, \rd t^2 + \frac{\rd r^2}{f(r)} + r^2 \, \rd\Omega^2_2 \, , 
\end{equation}
where $\rd\Omega_2^2$ is the line element of the round metric on $S^2$. The Schwarzschild solution \eqref{eq:1_Schwarzschild_Metric} falls in this class, as do many others. There is an obvious Killing vector $k = \partial_t$, with norm $k^2 = - f(r)$. We assume further that there is a largest $r_+$ such that $f(r_+)=0$, and that $f(r)>0$ for $r>r_+$, which is the region we focus on. So, $k^a$ is timelike on our spacetime, and its norm vanishes on the surface $\cN= \{ r = r_+ \}$. To further investigate $\cN$, we need to introduce coordinates such that the metric is well-defined on $\cN$, which are a generalization of the ingoing Eddington--Finkelstein coordinates.

Let $r_*(r)$ be the function such that
\[
    r_*'(r) = \frac{1}{f(r)} \, ,    
\]
and define $v = t + r_*(r)$. In these coordinates, the line element has the form
\[
    \rd s^2 = - f(r) \, \rd v^2 + 2 \, \rd v \rd r + r^2 \, \rd\Omega^2_2 \, ,
\]
which is still Lorentzian and non-degenerate at $r=r_+$. Moreover $k=\partial_v$. We can now compute
\[
    k_a \rvert_{\cN} = (\rd r)_a \, ,
\]
so it's normal to $\cN$, which confirms that $\cN$ is a Killing horizon for $k^a$. Furthermore, we find that
\[
    \nabla_a k^2 \rvert_{\cN} = - f'(r_+) \, k_a \rvert_{\cN} \, ,
\]
and comparing with \eqref{eq:1_SurfaceGravity_Definition}, we identify the surface gravity of the horizon as 
\begin{equation}
\label{eq:2_StaticSphericalBH_SurfaceGravity}
	\kappa = \frac{f'(r_+)}{2}
\end{equation}	
(with a normalization of the Killing vector dependent on $f(r)$). Not only is this a Killing horizon: provided appropriate boundary conditions on $f(r)$ are imposed, this would also be the event horizon of a black hole. In the following, we assume that $f'(r_+) > 0$, so that $\cN$ is a bifurcate horizon.

\medskip

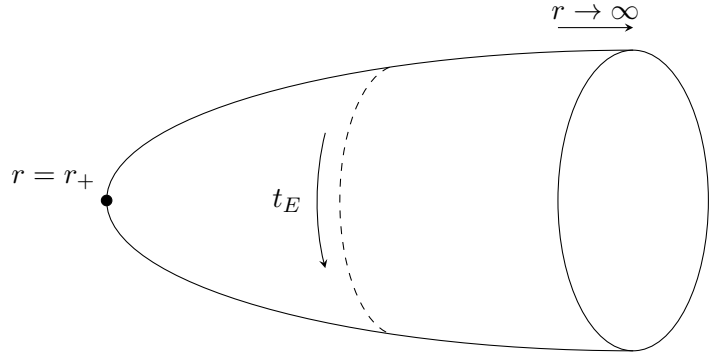
\begin{figure}
    \centering
    \begin{tikzpicture}
        \draw (2,0) [partial ellipse=90:270:7cm and 2cm];
        \draw (2,0) ellipse (1cm and 2cm);
        \filldraw[black] (-5,0) circle (2pt) node[anchor=south east]{\small $r=r_+$};
        \draw [->] (1,2.3) -- (2,2.3);
        \node at (1.5,2.5) {\small $r\to \infty$};
        \draw [dashed] (-1.1,0) [partial ellipse=100:260:0.8cm and 1.8cm];
        \draw [->] (-1.4,0) [partial ellipse=150:210:0.8cm and 1.8cm];
        \node at (-2.6,0) {\small $\tE$};
    \end{tikzpicture}
    \caption{The topology of a static spherically symmetric black hole is $\R^2\times S^2$. The metric on the $\R^2$ factor parametrized by $(\tE, r)$, and the ranges of the coordinates are such that the space caps off smoothly at $r=r_+$ with flat metric in a neighbourhood. The asymptotic behaviour as $r\to \infty$ depends on the cosmological constant: here we represent an asymptotically flat solution, where the $\R^2$ factor, as $r\to \infty$, becomes a cylinder with metric $\rd \tE^2 + \rd r^2$, since the size of the circles at constant $r$ doesn't grow. This is the ``cigar'' geometry.}
    \label{fig:2_CigarGeometry}
\end{figure}

As with Rindler space, we perform a Wick rotation $t=-\ii\tE$ and study the geometry of the resulting Euclidean space
\begin{equation}
\label{eq:2_StaticSphericalBH_Euclidean}
    \rd s^2_E = f(r) \, \rd \tE^2 + \frac{\rd r^2}{f(r)} + r^2 \, \rd\Omega^2_2 \, .
\end{equation}
For all $r>r_+$, the space above is smooth, but that's not necessarily true at $\{r=r_+\}$. Let's look at a neighbourhood of this set, where the line element can be well-approximated by
\[
    \rd s^2 \sim f'(r_+)(r- r_+) \, \rd \tE^2 + \frac{\rd r^2}{f'(r_+)(r-r_+)} + r_+^2 \, \rd\Omega_2^2 \, .
\]
We then introduce a new radial coordinate with origin at the horizon
\[
    \rho^2 = \frac{4}{f'(r_+)} ( r - r_+) \, .
\]
Using this coordinate, the line element in a neighbourhood of the horizon looks like
\begin{equation}
\label{eq:2_RindlerMetric_NearHorizon}
    \rd s^2 = \frac{f'(r_+)^2}{4} \rho^2 \, \rd \tE^2 + \rd\rho^2 + r_+^2 \, \rd \Omega_2^2 \, .
\end{equation}
But this is just the Euclidean Rindler metric \eqref{eq:2_EuclideanRindler} with $\alpha = f'(r_+)/2$! So, we already know how to deal with the smoothness requirement: we should view $\rho$ as a radial coordinate $\rho>0$ and, to avoid conical singularities, we should identify
\begin{equation}
\label{eq:2_Periodicity_EuclideanTime}
	\tE \sim \tE + \beta \, , \qquad \beta = \frac{4\pi}{f'(r_+)}  = \frac{2\pi}{\kappa} \, ,
\end{equation}
which guarantees that the circles of constant $\rho$ shrink smoothly as $\rho \to 0$.
The resulting geometry is a product of a disc $\R^2$ and a 2-sphere, with a smooth metric. The disc is represented in Figure \ref{fig:2_CigarGeometry}, and over each point $(\tE, r)$ there is a sphere with radius $r$. Notice that in requiring that the analytically-continued geometry is smooth, we have excised the part of spacetime beyond the horizon, since effectively $r\geq r_+$.

\medskip

Now, we can appeal again to the chain of reasoning introduced earlier: regularity of the Wick-rotated geometry near the Killing horizon requires a periodicity for the adapted coordinate along the Killing vector, we identify this periodicity with the periodicity of the Euclidean Wightman function using the KMS condition, and then conclude that: an observer following an orbit of $k=\partial_t$ in the (Lorentzian) spacetime \eqref{eq:2_StaticSphericalBH}) is accelerated with acceleration
\[
A = \frac{\kappa}{\sqrt{-k^2}} \, , \, ,
\] 
and she will detect a surrounding thermal bath of particles with temperature \eqref{eq:2_UnruhTemperature}
\[
    T = \frac{A}{2\pi} = \frac{\kappa}{2\pi \sqrt{-k^2}} \, , 
\]
where $\sqrt{-k^2}$ in the denominator represents the redshift factor due to the change to the observer's frame (see \eqref{eq:1_Acceleration_Infinity} and the Tolman--Ehrenfest law mentioned in footnote \ref{footnote:2_TolmanLaw}). Suppose that $f(r)$ is now such that $k^2 \to -1$: then we find an interpretation for the constant in Tolman's law: it's the temperature measured by an observer far from the horizon, and in particular it's
\begin{equation}
\label{eq:2_HawkingTemperature}
    T_H = \frac{\kappa}{2\pi} \, .
\end{equation}
This is referred to as the \textit{Hawking temperature}, which is the temperature measured by observers ``normalized'' at infinity.
Notice again the comments already made after \eqref{eq:1_Horizons_SurfaceGravities}: if the bifurcate Killing horizon is actually the event horizon of a spacetime, the Hawking temperature is non-vanishing, whereas in Rindler space there is no preferred observer along $b^a$, and the would-be ``Hawking temperature'' vanishes, as found also right after \eqref{eq:2_UnruhTemperature}.

\medskip

For future use, we also note that the application of the Green's function (or regularity) method does not pick up the temperature measured by the observer \eqref{eq:2_UnruhTemperature}, but rather the constant $T_0$ \eqref{eq:2_Unruh_Temperature_T0} in the Tolman--Ehrenfest law, which we identified as the Hawking temperature.

We also recall that the Lorentzian Killing horizon $\cN$ is a codimension-1 set, and the Bekenstein--Hawking entropy \eqref{eq:1_Wald_Reduced_BH} is given by the area of the bifurcation surface. After the Wick rotation and the construction of the regular Euclidean manifold, the bifurcation surface, where the Killing vector vanishes, has become the 2-sphere over the origin of the disc, that is, the locus $\{r=r_+\}\times S^2$, and that's what remains of the entire Killing horizon.

Finally, we remind the reader that the Euclidean method just introduced characterizes stationary equilibrium states. Instead, the Hawking radiation in collapse geometries is a dynamical phenomenon whose temperature coincides with the Euclidean result.

\subsubsection{Far from the horizon of a black hole}
\label{subsec:2_Far_From_Horizon}

At this point one may ask whether we have already reached our goal of proving the \textit{Hawking effect}: a black hole with surface gravity $\kappa$ radiates particles with a thermal spectrum at temperature \eqref{eq:2_HawkingTemperature}. We have not. 

\begin{figure}
\begin{subfigure}{.5\textwidth}
    \centering
    \begin{tikzpicture}[scale=0.75]
    \draw (3,-3) -- (-3,3);
    \draw (-3,-3) -- (3,3);
    \draw (3,3) -- (6,0) -- (3,-3);
    \draw (-3,3) -- (-6,0) -- (-3,-3);
    \draw[decorate,decoration={coil,aspect=0}] (-3,3) -- (3,3);
    \draw[decorate,decoration={coil,aspect=0}] (-3,-3) -- (3,-3);
    \node at (4.6,2.2) {$\mathscr{I}^+$};
    \node at (4.4,-2.2) {$\mathscr{I}^-$};
    \node at (1.4,2.2) {$\mc{H}^+$};
    \node at (1.4,-2.2) {$\mc{H}^-$};
    \end{tikzpicture}
  \caption{Extended Schwarzschild}
  \label{fig:2_Schwarzschild}
\end{subfigure}
\begin{subfigure}{.5\textwidth}
  \centering
    \begin{tikzpicture}
    \fill[fill=gray!20!white] (0,3) -- (0,-3) arc (-27.17:8.24:10)--(0,3);
    \draw (0,3) -- (0,-3) -- (4,1) -- (2,3) -- (0,1);
    \draw[decorate,decoration={coil,aspect=0}] (0,3) -- (2,3);
    \node at (3.4,2.2) {$\mathscr{I}^+$};
    \node at (2.4,-1.2) {$\mathscr{I}^-$};
    \node at (1.7,2.2) {$\mc{H}^+$};
    \end{tikzpicture}
  \caption{Stellar collapse}
  \label{fig:2_StellarCollapse}
\end{subfigure}
\caption{Penrose diagrams of (a) the extended Schwarzschild black hole, and (b) a spherically symmetric stellar collapse (in grey the star).}
\label{fig:2_PenroseDiagrams}
\end{figure}
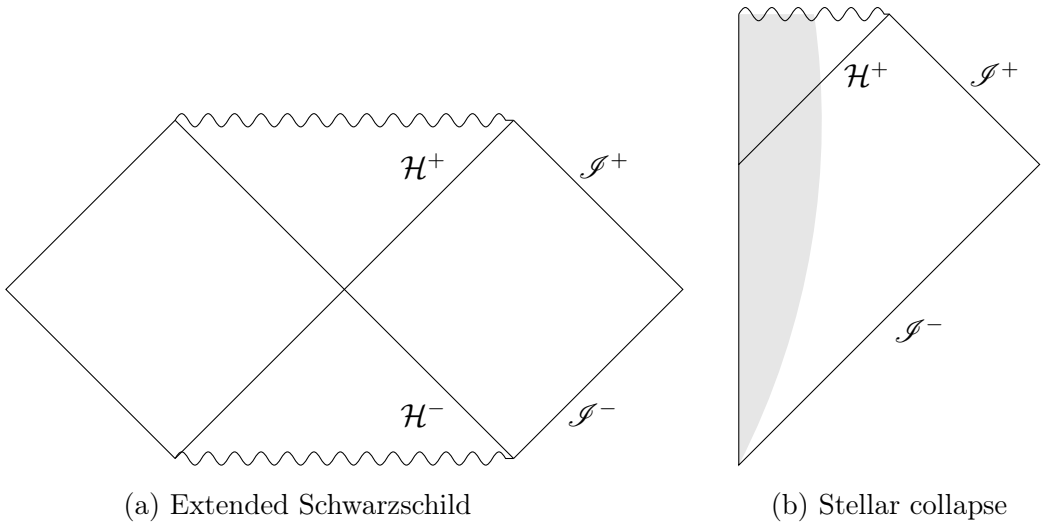

First, we have not because in order to have a bifurcate horizon with the right analiticity properties in, say, the Schwarzschild black hole, we would need the entire conformal diagram \ref{fig:2_Schwarzschild}, but in realistic black holes originated from stellar collapse, part of the diagram is hidden due to the presence of matter, as in \ref{fig:2_StellarCollapse}. Therefore, we need to study a different situation. \\
Second, in applying the analysis of the Unruh effect to a curved spacetime with a Killing horizon we have secretly made an assumption about the existence of a vacuum state. In fact, a more careful analysis (see \cite[sec. 5.3]{Wald:1995yp}) shows that in order to generalize the Unruh effect to a curved spacetime, one requires the existence of a vacuum state that is invariant under the isometry generating the horizon and non-singular, in the sense that the expectation value of the stress-energy tensor in the vacuum should be non-singular. Such a state does not exist if the Killing vector generating the horizon is not globally timelike: whilst this is not a problem for Schwarzschild, it is for the Kerr black hole. Therefore, there is no globally regular, isometry-invariant thermal state for Kerr. Nevertheless, physically relevant states describing gravitational collapse still exhibit Hawking radiation at temperature $T_H$.

\medskip

We will not review the original computation by Hawking \cite{Hawking:1975vcx}, which is quite involved and which the interested reader can find reviewed in detail in \cite{Reall}. One crucial aspect of the computation involving the spherically symmetric gravitational collapse represented in Figure \ref{fig:2_StellarCollapse} is that, even though the metric outside the matter is the \textit{static} Schwarzschild metric (because of Birkhoff's theorem), the geometry is not overall stationary, because the metric inside the matter is not. Therefore, observers on $\mathscr{I}^+$ and $\mathscr{I}^-$ would not agree on the definition of the ``vacuum.'' In particular, considering the propagation of radiation from $\mathscr{I}^-$ in the collapsing matter and then scattering to $\mathscr{I}^+$ leads to the conclusion that in the state that the observer on $\mathscr{I}^-$ calls ``vacuum,'' the observer on $\mathscr{I}^+$ will measure a spectrum of particles that at late retarded times only depends on the surface gravity of the black hole.

Differently from the Unruh effect, the final result for the number of particles measured by the late-time observer in the early-time vacuum will not be a purely Planckian spectrum at the Hawking temperature, but instead will also include a ``greybody'' factor, which takes into account the fact that the black hole emits \textit{and} absorbs radiation. However, the ratio of the absorption and emission rates is independent of the greybody factor, signalling that the black hole with surface gravity $\kappa$ is in equilibrium with a heat bath at temperature $T_H = \kappa/2\pi$, as given in \eqref{eq:2_HawkingTemperature}.

The computation can be generalized to non-scalar field theories, and to different gravitational backgrounds. For instance, we could have also studied the collapse of matter with angular momentum and charge. After a dynamical state, the black hole would settle down to a stationary Kerr--Newman solution, which is described only by its mass, angular momentum and charge. At late time, the black hole is in equilibrium with a heat bath with temperature given again by $T_H$ in \eqref{eq:2_HawkingTemperature} (clearly with the appropriate surface gravity), and the emitted particles preferably have angular momentum and charge with the same sign as the black hole itself (this is the same that one would expect for a rotating charged black body) \cite{Hawking:1975vcx}.

\subsection{Where to now?}
\label{subsec:2_Where_to_now}

We argued that in presence of a gravitational collapse leading to a stationary black hole, at late time an observer would measure an outgoing flux of particles distributed along the spectrum according to Planck's law with a greybody factor and a temperature $T_H$ \eqref{eq:2_HawkingTemperature}. In order to get to this result, we needed to consider the behaviour of quantum fields near the black hole, but in the semiclassical approximation, that is, neglecting the quantization of the gravitational field itself, and the backreaction of the fields on the geometry. Before moving on with the lectures, we stop to make some comments.

\medskip

It is useful to have a heuristic picture of the origin of the black hole radiation. The Hawking effect has an analogue in (flat space) electrodynamics, the Schwinger effect \cite{Schwinger:1951nm}. Recall that the quantum field theory vacuum is not really empty, and pairs of particle-anti particle are continuously created. For instance, let's focus on electron-positron pairs and apply a strong electric field to a region of (supposedly) empty space: as the pair of particles is created, the field pulls the electron in one direction and the positron in the opposite direction. If the field is sufficiently strong, we will be able to observe the creation of \textit{real} electrons and positrons at the opposite ends of the region, and there will be a current flowing.

The picture with gravity is slightly different, as all particles now have the same ``charge.'' In this case, the crucial role is played by the black hole region of spacetime created by the strong gravitational field: one member of the electron-positron pair could fall inside the horizon, and the other could fly off to infinity, having now become a \textit{real} particle measured by the far observer.

\medskip

As already anticipated at the beginning of Section \ref{subsec:2_FourLaws}, we showed that black holes are indeed thermodynamical objects. Classical thermodynamics can be derived, using statistical physics, from quantum microstates. For a black hole, in order to reproduce the Bekenstein--Hawking entropy, we would need to describe $\sim \exp(A_{\rm h}/4)$ microstates. This requires a quantum theory of gravity. One of the most impressive successes of string theory as a theory of quantum gravity has been the counting of the degeneracy of states corresponding to a microscopic description of a supersymmetric black hole, reproducing the Bekenstein--Hawking entropy, and even providing corrections to the result \cite{Strominger:1996sh}.

\medskip

The idea that black holes must be assigned an entropy was based on the Second Law of Thermodynamics, and the suggestion was that in presence of a black hole, this would be modified to a \textit{Generalized Second Law}: in any physical process
\[
    \Delta \left( S_{\rm matter} + S_{\rm bh} \right) \geq 0 \, .
\]
Here, recall that $S_{\rm bh} = A_{\rm h}/4$ is entropy of the black hole, which is just a function of $(M,J,Q)$, so it's oblivious to the entropy of the matter/radiation falling in the black hole and it is not at all obvious that it would increase sufficiently to compensate the decrease in $S_{\rm matter}$. In particular, requiring that the Generalized Second Law must hold would also imply a bound on $S_{\rm matter}$ as a function of its extensive parameters. Such a bound is intrinsically related to gravity, as it has at its core the tenet that one cannot fill a region of space with an arbitrarily high number of degrees of states without encountering a gravitational instability and forming a black hole. In particular, this leads to the idea of a holographic principle: \textit{the physics in a region with boundary area $A$ is fully described by no more than $A/4$ degrees of freedom}. This suggestion is in stark contrast with the predictions of local field theory, and it should be a property of a theory of quantum gravity. The most concrete instance of the holographic principle is the realization of the $AdS$/CFT correspondence in string theory (see \cite{Bousso:2002ju} for a review).

\medskip

In the derivation above, we ignored the backreaction of the radiation on the geometry. A proper study of the backreaction requires a quantum theory of gravity, but  from the presence of radiation itself, one can immediately draw some puzzling conclusions about unitary evolution. Consider matter starting its collapse in a pure (definite) quantum state. It will form a black hole, which will then radiate and eventually evaporate completely, leaving only its thermal radiation in the universe. The latter state can now only be described using a density matrix, it's a mixed state. Thus, we seem to have described the evolution from a pure to a mixed quantum state, which would not be consistent with unitary evolution in quantum mechanics. This is roughly the content of the \textit{information paradox}: does the black hole evolve unitarily in time?

It is fair (and fun) to say that the subject of the evolution of a thermal black hole is still a heated subject, and we will not cover it. For some reviews see, for instance, \cite{Mathur:2009hf, Almheiri:2020cfm}. We limit ourselves to notice that the previous comments on string theory realizations imply that black holes do indeed evolve unitarily.

\medskip

Finally, to conclude this section, let's view some numbers. For a static black hole with mass $M$
\[
\begin{split}
    T_H &= \frac{\kappa}{2\pi} \frac{\hbar}{k_B c} = \frac{1}{8\pi M} \frac{\hbar c^3}{ G k_B} \sim 6.17 \cdot 10^{-8} \cdot \frac{M_\odot}{M} \, {\rm K}  \, , \\
    S_{\rm bh} &= \frac{A}{4} \frac{k_B c^3}{G\hbar} = 4\pi M^2 \frac{G k_B}{c\hbar} \sim 1.05 \cdot 10^{77} \cdot \frac{M^2}{M_\odot^2} k_B \, .
\end{split}
\]
Therefore, modelling with crude approximation an astrophysical black hole ($M\sim 10^6 M_\odot$) with a Schwarzschild solution, we find that its Hawking temperature is minuscule, much lower than the temperature of the CMB radiation, whereas its entropy is enormous, much higher than the entropy of matter in the same volume.



\newpage

\section{The gravitational path integral}
\label{sec:3_GPI}

\subsection{From Lorentzian horizons to Euclidean saddles}

The results of the previous two chapters suggest that black hole thermodynamics is well-represented by the geometry of the solution. In Lorentzian signature, the temperature of a stationary black hole is fixed by the existence of a Killing horizon and by the associated surface gravity, while the entropy is encoded in a Noether-like charge \eqref{eq:1_WaldEntropy} evaluated on the bifurcation surface. Both quantities are defined using purely geometric data and rely only on stationarity and the presence of a bifurcate Killing horizon.

At the same time, thermal equilibrium in quantum field theory is most naturally characterized not through local observables, but through the global analytic structure of correlation functions. In particular, equilibrium states are distinguished by the KMS condition \eqref{eq:2_KMS_v1}, which expresses periodicity in imaginary time with period set by the inverse temperature. As discussed in Chapter \ref{sec:2_BH_Thermodynamics}, this periodicity emerges naturally when quantum fields are studied on stationary spacetimes with horizons.

\medskip

Euclidean methods reorganize these observations into a geometric framework. For a stationary spacetime, the flow generated by the Killing vector defining the horizon can be analytically continued to imaginary time. Under this continuation, the horizon (where the Killing vector vanishes) becomes a fixed point of the ``Euclidean time'' flow. As discussed in Section \ref{subsec:2_Near_Horizon}, requiring the Euclidean metric to be regular at this fixed point forces the imaginary time coordinate to be periodic, with a period uniquely determined by the surface gravity. This geometric regularity condition is the Euclidean counterpart of the KMS condition and reproduces the Hawking temperature without reference to particle notions.

From this viewpoint, thermality is not a dynamical phenomenon tied to particle production, but a consequence of symmetry and regularity. The near-horizon geometry already contains the information required to characterize thermal equilibrium. The analytic continuation makes this structure manifest by trading Lorentzian time evolution for geometric periodicity.

Entropy admits a similarly geometric interpretation. In Lorentzian signature, Wald’s formula \eqref{eq:1_WaldEntropy} associates entropy with a Noether charge evaluated on the bifurcation surface. In Euclidean signature, the same surface appears naturally as the fixed-point set of the Euclidean time symmetry. We will see in Section \ref{subsec:4_Entropy_Topology} that contributions to the on-shell action localize there, and this underlies Euclidean derivations of black hole entropy.

\medskip

It is important to stress the scope of this framework. Euclidean methods are best suited to describing stationary equilibrium configurations and their semiclassical thermodynamics. They do not capture real-time processes such as gravitational collapse or evaporation far from equilibrium. Rather, they provide a geometrically transparent way to extract thermodynamic information (and more!) from gravitational systems that admit a semiclassical description.

Motivated by these considerations, we now turn to a direct implementation of Euclidean ideas at the level of quantum gravity itself. In this section we look at a na\"{i}ve but insightful approach to the study of quantum aspects of gravity: the gravitational path integral.

\subsection{Definition}
\label{subsec:3_PathIntegralReview}

Let's begin by recalling the definition of the path integral in quantum mechanics. This is applied to the computation of the probability amplitude that a particle at position $x_1$ at time $t_1$ is found at position $x_2$ at a later time $t_2$. Feynman's idea is to compute this via the evaluation of the integral over all trajectories between $x_1$ and $x_2$, weighted by an oscillatory contribution due to the classical action:\footnote{In this section we reintroduce $\hbar$.}
\[
    \braket{x_2| \e^{-\ii t \frac{H}{\hbar}} |x_1} = \int_{x=x_1}^{x=x_2} \cD x \, \e^{\ii \frac{S}{\hbar}} \, ,
\]
where $S=\int_{0}^{t}\left( \frac{m}{2}\dot{x}^2 - V(x) \right) \rd s$. An apparently naive but quite far-reaching observation is that the time propagator $\e^{-\ii t \frac{H}{\hbar}}$ is related to the density matrix operator $\e^{- \beta H}$ by $t = - \ii \beta \hbar$ (where $\beta = 1/T$). As we saw and justified in some detail in the previous section, the relation is deeper and more involved, but one can in fact construct the path integral representation of the density matrix operator (see footnote \ref{footnote:2_DensityMatrix}), and view it as a path integral representing evolution in ``imaginary time.'' In fact, the construction of the path integral representation of $\e^{- \beta H}$ can be done in a much more rigorous way than that of $\e^{- \ii t H/\hbar}$, even to the level of satisfaction of a mathematician (and indeed it is named Feynman--Kac formula after the mathematician Mark Kac). The result is formally represented by the expression
\begin{equation}
\label{eq:3_PathIntegralDensityMatrix}
    \braket{x_2 | \e^{- \beta H} | x_1} = \int_{x=x_1}^{x=x_2} \cD x \, \e^{ - \frac{S_E}{\hbar}} \, ,
\end{equation}
where now $S_E$ is the \textit{Euclidean} action
\begin{equation}
\label{eq:3_EuclideanAction_QM}
    S_E = \int_0^{\tE} \left( \frac{m}{2} \dot{x}^2 + V(x) \right) \rd s_E \, ,
\end{equation}
and $\tE = \beta \hbar$. Notice that in this variables, the relation between the time propagator and the density matrix has become the canonical Wick rotation $t = - \ii \tE$, and it can be used to relate $S$ and $S_E$
\[
\begin{split}
   \ii S &= \ii \int_0^{t} \left( \frac{m}{2} \left( \frac{\rd x}{\rd s} \right)^2 - V(x) \right) \rd s = \ii \int_0^{-\ii \tE} \left( \frac{m}{2} \left( \frac{\rd x}{\rd s} \right)^2 - V(x) \right) \rd s \\
   &= \ii \int_0^{\tE} \left( - \frac{m}{2} \left( \frac{\rd x}{\rd s_E} \right)^2 - V(x) \right) (-\ii \, \rd s_E) \\
   &= - S_E
\end{split}
\]
where in the second line we introduced $s = - \ii s_E$. In higher dimensions, this procedure justifies the adjective ``Euclidean'', but for now notice that the quantum mechanical $S_E$ has the same form as the total energy $T+V$. Finally, we can construct the canonical partition function $Z(\beta)$, which is obtained from \eqref{eq:3_PathIntegralDensityMatrix} by identifying $x_1$ and $x_2$ and summing over them, thus obtaining a trace
\[
    Z(\beta) = \int \cD x \, \e^{ - \frac{S_E}{\hbar}} \, ,
\]
where the sum is over all path that close after ``imaginary time'' $\beta$. We now see that the path integral picture of the trace naturally enforces the fact that the thermal partition function should be computed over Euclidean backgrounds where ``time'' is identified in a circle of radius $\beta$, which we discussed from the periodicity of the Green's function in Section \ref{subsubsec:2_Thermal_QFT}. We have come full circle (pun intended)!

\begin{figure}
    \centering
\begin{tikzpicture}[scale=1.2]

    \draw[dashed] (2,-2) arc (0:180:2cm and 0.7cm);
	\shade[shading=radial, inner color=gray!20!white, outer color=gray!60!white, opacity=0.70] (2,2) -- (2,-2) arc (360:180:2cm and 0.7cm) -- (-2,2) arc (180:360:2cm and 0.7cm) ;
	\draw[] (2,2) -- (2,-2) arc (360:180:2cm and 0.7cm) -- (-2,2) arc (180:360:2cm and 0.7cm) ;
  \shade[shading=radial, inner color=gray!20!white, outer color=gray!60!white, opacity=0.70] (2,2) arc (0:360:2cm and 0.7cm); 
  \draw (2,2) arc (0:360:2cm and 0.7cm);
    \node [opacity=0.7] at (0,0) {\small $M$};
	\node at (2.3,2.2) {\small $\Sigma_2$};
	\node at (2.3,-2.2) {\small $\Sigma_1$};
    \node at (2.3,0) {\small $C$};
\end{tikzpicture}
    \caption{A schematic representation of the application of the gravitational path integral. To compute the amplitude \eqref{eq:3_ProbabilityAmplitude} one should integrate over all configurations filling $\partial M$ with the appropriate boundary condtions on $\partial M$.}
    \label{fig:3_SpacetimeRegion}
\end{figure}
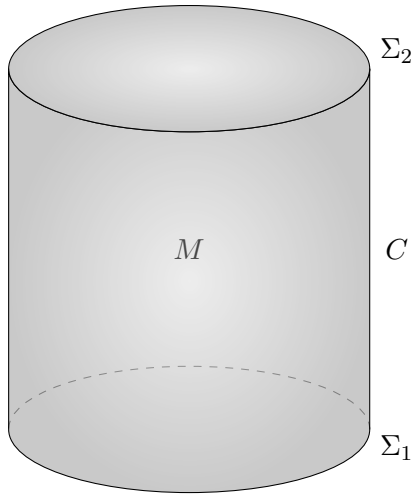

\medskip

Guided by this, and the success of the path integral in quantum field theory, we want to extend it to gravity. This approach is referred to as the \textit{gravitational path integral}, and was first advocated by Gibbons and Hawking in \cite{Gibbons:1976ue}. We want to compute the probability amplitude of evolving from a state described by a metric $g_1$ on a spacelike surface $\Sigma_1$ with matter fields $\Phi_1$ to a state described by metric $g_2$ on a spacelike surface $\Sigma_2$ with matter fields $\Phi_2$, which can be formally written as
\begin{equation}
\label{eq:3_ProbabilityAmplitude}
    \braket{g_2, \Phi_2, \Sigma_2 | g_1, \Phi_1, \Sigma_1} = \int \cD g \cD\Phi \, \e^{\ii \frac{S}{\hbar}} \, .
\end{equation}
Here we introduced the (undefined) path integral measures $\cD g$ and $\cD\Phi$, and the integral is taken over all the field configurations satisfying the appropriate boundary conditions. More precisely, to avoid further divergences we have to assume that either $\Sigma_{1,2}$ are compact or that they are joined by a timelike tube, to make the overall region of spacetime we are concerned with compact. Thus we are working with a compact region of spacetime $M$ with boundary $\partial M = \Sigma_1 \cup \Sigma_2 \cup C$, as in Figure \ref{fig:3_SpacetimeRegion}.

\subsection{Gravity action}

The action for the metric in general relativity is the so-called Einstein--Hilbert action
\begin{equation}
\label{eq:3_SEH}
    S_{\rm EH} = \frac{1}{16\pi} \int \left( R - 2\Lambda\right) \sqrt{-g} \, \rd^4 x \, ,
\end{equation}
to which we should add the Lagrangian describing the matter fields $S_m$ coupled to the metric. If we tried to use $S=S_{\rm EH} + S_m$, we would immediately find an issue: the Einstein--Hilbert action does not generically yield a well-posed variational problem on a space with a boundary! That is, varying the Einstein--Hilbert action does not only give the equations of motion, but also a boundary term. If the space is asymptotically flat and the metric and its derivatives decay sufficiently fast at infinity, this boundary term can be ignored, as it is usually done. However, in general, we cannot do that.

More precisely, varying $S_{\rm EH}$ leads to
\begin{align}
\label{eq:3_VariationSEH}
    \delta S_{\rm EH} &= \frac{1}{16\pi} \int_{M} \left[ \left( R_{ab} - \frac{1}{2} R g_{ab} + \Lambda g_{ab} \right) \delta g^{ab} + \nabla_a X^a \right] \sqrt{\abs{g}} \, \rd^4 x \\
    &= \frac{1}{16\pi} \int_{M} \left( R_{ab} - \frac{1}{2} R g_{ab} + \Lambda g_{ab} \right) \delta g^{ab} \, \sqrt{\abs{g}} \, \rd^4 x + \frac{1}{16\pi} \int_{\partial M} n_a X^a \, \sqrt{\abs{h}} \, \rd^3 x \, , \nonumber
\end{align}
where we have used the divergence theorem in the second step, denoting by $h_{ab}$ the induced metric on $\partial M$. The bulk term in \eqref{eq:3_VariationSEH} consists of the Einstein equations in presence of a cosmological constant, whereas the boundary term is given by
\[
\begin{split}
    X^a &= g^{bc} \delta \Gamma^a_{bc} - g^{ab}\delta \Gamma^c_{bc} \, , \\
    \delta \Gamma^a_{bc} &= \frac{1}{2} g^{ad} \left( \delta g_{db;c} + \delta g_{cd;b} - \delta g_{bc;d} \right) \, ,
\end{split}
\]
and at the end of the day its presence is due to the fact that the Riemann tensor is second order in the derivatives of the metric. In order to define a well-posed variational problem, we need the boundary term to vanish on its own (which would require both the metric and its derivatives to vanish on the boundary) or to add a further boundary term that cancels the variation of the Einstein--Hilbert action. Such a term would have to be a special combination of the first derivatives of the induced metric. Luckily, such a term does exist: it's the \textit{extrinsic curvature} of an embedded surface.

\subsubsection{Gibbons--Hawking--York term}

Consider a timelike or spacelike surface $\Sigma$ with unit normal $n_a$ (that is $n_an^a = \pm 1$ depending on whether it's timelike or spacelike), and induced metric $h_{ab} = g_{ab} \mp n_a n_b$. The latter is sometimes denoted first fundamental form and has the property that $h^a_{\ph{a}b}$ is a projector on the tangent space to $\Sigma$ (indeed $h^a_{\ph{a}b}n^b=0$). The normal is defined on $\Sigma$, but we can extend it to the ambient space $M$ as we want (the final result does not depend on the extension). We can then use the Lie derivative to measure how the induced metric on $\Sigma$ varies as we move along an integral curve of $n^a$: the result is the \textit{extrinsic curvature}
\begin{equation}
\label{eq:3_ExtrinsicCurvature_Def}
    K_{ab} = \frac{1}{2} \cL_{n^\sharp} h_{ab} \, ,
\end{equation}
which is a rank-2 symmetric tensor, sometimes denoted second fundamental form.\footnote{Let $(M,g)$ be a Riemannian manifold and $p$ be a point in $M$. Given an element $\alpha\in T_p^*M$, the notation $\alpha^\sharp$ indicates the vector in $T_pM$ such that $g_p(\alpha^\sharp, v) = \alpha(v)$ for all $v\in T_pM$. In abstract index notation, it corresponds to ``raising the index'': $(\alpha^\sharp)^a \equiv \alpha^a = g^{ab}\alpha_b$.\label{footnote:3_MusicalIsomorphism}} We can also write it in a (non-trivially) equivalent way as
\begin{equation}
\label{eq:3_ExtrinsicCurvature_Def_2}
    K_{ab} = h^c_{\ph{c} a} h^d_{\ph{d}b} \nabla_c n_{d} \, .
\end{equation}

By its definition it is clear that $K_{ab}$ depends also on the way the surface sits in the ambient space. In contrast, the \textit{intrinsic curvature} of $\Sigma$ is measured by the Riemann tensor $R[h]^a_{\ph{a}bcd}$ associated to $h_{ab}$ (and to its Levi-Civita connection): it is related to the extrinsic curvature and the curvature of the ambient space $R[g]^a_{\ph{a}bcd}$ by the \textit{Gauss equation}
\[
    R[h]^a_{\ph{a}bcd} = h^a_{\ph{a}e} h^f_{\ph{f}b} h^g_{\ph{g}c} h^h_{\ph{h}d} R[g]^e_{\ph{e}fgh} \pm 2 K_{[c}^{\ph{[c}a} K_{d]b} \, .
\]
Finally, we also define the trace of the extrinsic curvature by $K \equiv h^{ab}K_{ab}$.

\medskip

To get a feeling for the meaning of the extrinsic curvature, consider a surface $\Sigma$ in $M$ defined by the level set $f=0$ with $f:M\to \R$ and $\rd f \neq 0$ on $\Sigma$. The unit normal is $n_a= \cN (\rd f)_a$ where $\cN$ is a normalization function on $M$ such that $n_a$ is a unit normal, and 
\[
    \nabla_a n_b = \cN \, \partial^2_{ab}f + \partial_a \log \cN \, n_b \, .
\]
Using \eqref{eq:3_ExtrinsicCurvature_Def_2}, the extrinsic curvature is
\[
    K_{ab} = h^c_{\ph{c}_a} h^d_{\ph{d}b} \cN \partial^2_{cd}f \, .
\]
Therefore, if $M=\R^d$, $K_{ab}$ is the projection on the tangent space to the surface of the second derivative of $f$. Since $f(p)=0$ for $p\in\Sigma$, and the projection of the first derivative is vanishing (as the latter is proportional to the normal), $K_{ab}$ is the leading approximation to $f$ in a neighbourhood of $p$, once projected to the tangent plane, or the best approximation of the hypersurface by a paraboloid. It measures how much the hypersurface moves ``away'' from the tangent plane at a point in the direction of the normal to the point, and thus corresponds to our intuitive definition of ``curvature.''\footnote{For instance, you can convince yourself that a cylinder has vanishing intrinsic curvature (you can create it by rolling up a flat piece of paper) but it has non-vanishing extrinsic curvature.}

As an example of computation of extrinsic curvature that will be particularly useful in a little while, consider a timelike hypersurface of constant $r=r_0$ in the static spherically symmetric spacetime \eqref{eq:2_StaticSphericalBH}. The unit normal to such surface if $n= f(r_0)^{-1/2} \rd r$, the dual vector field is $n^\sharp = f(r_0)^{1/2} \partial_r$ and the induced metric is $h = - f(r_0) \, \rd t^2 +  r_0^2 \, \rd\Omega^2_2$. We can then use the definition of the Lie derivative to compute $K_{ab}$ using \eqref{eq:3_ExtrinsicCurvature_Def}, which in a coordinate basis is
\begin{equation}
\label{eq:3_ExtrinsicCurvature_StaticSphericallySymmetric}
\begin{split}
     K_{\mu\nu} &= \frac{1}{2} \left( n^\rho \partial_\rho h_{\mu\nu} + h_{\mu\rho} \partial_\nu n^\rho + h_{\rho\nu} \partial_\mu n^\rho \right) = \frac{1}{2} \sqrt{f(r_0)} \partial_r h_{\mu\nu} \\
     &= \sqrt{f(r_0)} \left( - \frac{f'(r_0)}{2} \rd t^2 + r_0 \, \rd\Omega_2^2 \right)_{\mu\nu} \, ,
\end{split}
\end{equation}
with trace
\begin{equation}
\label{eq:3_TraceExtrinsicCurvature_StaticSphericallySymmetric}
    K = h^{\mu\nu} K_{\mu\nu} = \frac{f'(r_0)}{2\sqrt{f(r_0)}} + \frac{2}{r_0}\sqrt{f(r_0)} \, .
\end{equation}

\medskip

As anticipated, adding a multiple of the extrinsic curvature of the boundary to $S_{\rm EH}$ allows us to construct a well-defined variational problem. More precisely, we add the \textit{Gibbons--Hawking--York term} \cite{York:1972sj, Gibbons:1976ue}
\[
    S_{\rm GHY} = \frac{1}{8\pi G} \int_{\partial M} K \, \sqrt{\abs{h}} \, \rd^3 x \, ,
\]
Together with this term, the variation \eqref{eq:3_VariationSEH} becomes
\begin{equation}
\label{eq:3_VariationSEH_SGHY}
\begin{split}
    \delta ( S_{\rm EH} + S_{\rm GHY} ) &= \frac{1}{16\pi} \int_M \left( R_{ab} - \frac{1}{2}R g_{ab} + \Lambda g_{ab} \right) \delta g^{ab} \, \sqrt{\abs{g}} \, \rd^4 x \\
    & \ \ \ + \frac{1}{16\pi} \int_{\partial M} \Pi_{ab}\delta h^{ab} \, \sqrt{\abs{h}} \, \rd^3 x \, ,
\end{split}
\end{equation}
where
\begin{equation}
\label{eq:3_Piab}
    \Pi_{ab} = K_{ab} - K h_{ab} \, .
\end{equation}
Therefore, the boundary term vanishes if we impose Dirichlet boundary conditions on the metric, that is, $\delta h^{ab}=0$, or if we impose Neumann boundary conditions by fixing the ``derivative'' $\Pi_{ab}=0$, and we are only left with the equations of motion, that is, with a well-posed variational problem. The combination \eqref{eq:3_Piab} is referred to as \textit{Brown--York stress-energy tensor}, for reasons that we'll see later on in Section \ref{subsec:4_Subtleties}.

\subsubsection{Euclidean space}
\label{subsubsec:3_EuclideanSpacetime}

We have established that the action $S$ appearing in \eqref{eq:3_ProbabilityAmplitude} as the weight of each contribution to the gravitational path integral contains $S = S_{\rm EH} + S_{\rm GHY} + S_m$ (and potentially also fixed terms that could be extracted from the definition of the measure, as we shall soon see). An immediate problem, though, is that for Lorentzian metrics and unitary matter fields, $S$ is real, so the integrand oscillates and the path integral will not generically converge. Relatedly, the boundary problem to be solved in order to find the configurations contributing to the path integral involves a hyperbolic equation, which is not a well-posed problem even with boundary conditions, as already mentioned in Section \ref{subsec:2_UnruhEffect}, and as you know from trying to find propagators in Lorentzian quantum field theory: the existence of the solution is not guaranteed and even so it is not unique (e.g. the non-uniqueness of the propagator for a real scalar field).

One solution to improve this state of affairs in stationary spacetime, as discussed in the previous section, is to perform a Wick rotation, in which case we can define a unique vacuum by imposing the asymptotic boundary conditions for the fields. Moreover, the propagator found in this way agrees upon analytic continuation with the Feynman's propagator. As already mentioned, this procedure for quantum mechanics leads to the Feynman--Kac formula, which mathematicians too are happy with.

It is bold but plausible to suggest that the same procedure be applied to the gravitational path integral \eqref{eq:3_ProbabilityAmplitude}. Letting $t$ be the length in time of the timelike tube connecting the surfaces $\Sigma_1$ and $\Sigma_2$, if we define $t = - \ii \tE$, then the induced metric on the tube is now positive-definite and the gravitational path integral involves now the boundary problem of an elliptic differential equation. Namely, the integration is over all the positive-definite metrics $g_E$ on the compact region $M$ that induce the positive-definite metric $h_E$ on $\partial M$. The weight of each configuration is now $\e^{-S_E}$ where
\begin{equation}
\label{eq:3_WickRotation}
\begin{split}
    S_E &= - \ii S \rvert_{\text{Wick rotated}} \\
    &= - \frac{1}{16\pi} \int_M \left( R - 2\Lambda \right) \sqrt{g_E} \, \rd^4 x - \frac{1}{8\pi} \int_{\partial M} K \, \sqrt{h_E} \, \rd^3 x - S_{m,E} \, ,
\end{split}
\end{equation}
and now the Ricci scalar $R$ and the extrinsic curvature $K$ are computed using $g_E$ and $h_E$, respectively. These prescriptions are the essence of \textit{Euclidean quantum gravity}. Even though this procedure defines a semiclassical computational framework rather than a rigorously defined quantum theory of gravity, such framework has been very successful, even (and surprisingly) beyond the explanation of the thermodynamic properties of black holes. There are still some issues to be clarified, some of which we will review in Section \ref{subsec:4_Subtleties}.

For the time being, we ignore all issues of convergence or subtleties in the definition, we put on blinders and compute. What should we begin with? As we already discussed at length, Euclidean field theory shines when it is applied to thermal Lorentzian systems: the description of a Lorentzian system on a background $\R_t\times \Sigma$ at a temperature $T$ is mapped to a Euclidean problem on $S^1\times \Sigma$, where the angular coordinate on $S^1$ has period $\beta = 1/T$. The characteristic thermodynamic state function of the system is the free energy $F$, directly related to the partition function of the system via
\begin{equation}
\label{eq:3_FreeEnergy_Def}
    \beta F(\beta) = - \log Z (\beta) \, .
\end{equation}
As reviewed in Section \ref{subsec:3_PathIntegralReview}, in the path integral approach, the canonical partition function is given by the integral
\[
    Z(\beta) = \int \cD g_E \cD \Phi_E \, \e^{ - \frac{S_E}{\hbar}} \, ,
\]
where the integration is over all metrics and field configurations with the appropriate boundary condition on $\partial M \cong S^1\times \partial \Sigma$.\footnote{Throughout this section we work in the canonical ensemble, fixing the induced metric (and hence the temperature) at the boundary.} Note that this is a tantalizing reminder of the holographic nature of gravity, and will become better defined in $AdS$.

\medskip

How do we approach the Euclidean gravitational path integral? The short answer is that we do not, because we do not know how to make sense of the measure over the space of metrics. Instead, we look at the \textit{semiclassical approximation}: in the limit when $\hbar$ is much smaller than the action of the typical path,\footnote{But we keep $\tE=\beta\hbar$ fixed (see \eqref{eq:3_EuclideanAction_QM}). There are two other limit one can take by instead considering $\beta\hbar\to 0$: the \textit{classical} limit ($\hbar\to 0$, $\beta$ fixed) and the \textit{high temperature limit} ($\beta\to 0$, $\hbar$ fixed). In both these cases, the weight in the path integral reduces to the contribution of the potential evaluated on the paths.} the generalized contour integral would be (hopefully) dominated by the configurations $(\overline{g_E}, \overline{\Phi_E})$ corresponding to stationary points of the action $S_E$, which are the solutions to the classical equations of motion. In this limit, we obtain
\begin{equation}
\label{eq:3_SemiclassicalExpansion}
    S_E(g_E, \Phi_E) = S_E(\overline{g_E}, \overline{\Phi_E}) + S^{(2)}_E(g_E, \Phi_E) + \cdots
\end{equation}
where $S_E(\overline{g_E}, \overline{\Phi_E})$ is the on-shell action evaluated on the solutions of the classical equations of motion and it is commonly denoted $I$, $S^{(2)}_E(g_E, \Phi_E)$ is a functional quadratic in the perturbation around the classical solution, and we ignored higher order terms in the expansion of the action. So, the free energy in \eqref{eq:3_FreeEnergy_Def} is given by
\[
    \beta F = \frac{1}{\hbar}I - \log \int \cD g_E \cD\Phi_E \, \e^{ - \frac{S_E^{(2)}}{\hbar}} + \cdots \, .
\]
The second term in the expansion is referred to as the \textit{one-loop contribution}, and it represents the effects of quantum fluctuations around the classical saddle point: for instance, in absence of other fields it represents the contributions of thermal gravitons on the background. If we ignore this term, we find a very interesting relation \cite{Gibbons:1976ue}:
\begin{equation}
\label{eq:3_QuantumStatisticalRelation}
    I = \hbar \beta F \, .
\end{equation}
This equation goes under the name of \textit{quantum statistical relation} (the ``quantum'' bit refers to Planck constant), and can be used in conjunction with the canonical statistical relation, which relates the free energy of the thermal system with the entropy and energy
\begin{equation}
\label{eq:3_GrandCanonicalStatisticalRelation}
    F = E - T S \, ,
\end{equation}
to obtain an equation relating the Euclidean on-shell action, the entropy and the energy.
The quantum statistical relation is a far-reaching statement that relates thermodynamics and quantum gravity. It is valid, as is the entire semiclassical approximation, if we can ignore the higher order terms in the expansion of the action, which requires that the ``gravitational coupling'' is small, or, that the energies involved are much smaller that the Planck energy. Equivalently, that the length scales of the problem are much larger than the Planck length
\[
    E \ll E_P \qquad \Leftrightarrow \qquad \ell \gg \ell_P = \sqrt{\frac{G\hbar}{c^3}} \, .
\]
Even though it may seem that the path integral reformulation has not provided us with much insight into the quantum structure of gravity, we should not dismiss the importance of the conceptual framework that it provides us, as it will give us a lot of mileage. We now review a classical application of the semiclassical Euclidean framework, comparing competing saddles with identical asymptotic data and discovering a phase transition.

\subsection{Hawking--Page transition}
\label{subsec:3_EuclideanStatic}

We focus on the quantum description of a static thermal system with temperature $T=1/\beta$ in pure gravity. Upon Wick rotation, ``time'' becomes a compact direction $S^1$ with length $\beta$, and in order to follow the path integral prescriptions we should choose a boundary $\partial M$, which we fix to have topology $S^1\times S^2$. Therefore, the problem has been mapped to the question of finding metrics filling $S^1_\beta \times S^2$. In the semiclassical approximation, these metrics should also be solutions of the classical equations of motion coming from the action
\[
    S_E = - \frac{1}{16\pi} \int_M \left( R - 2\Lambda \right) \sqrt{g_E} \, \rd^4 x - \frac{1}{8\pi } \int_{\partial M} K \, \sqrt{h_E} \, \rd^3 x \, ,
\]
namely they are \textit{Einstein manifolds} (i.e. metrics whose Ricci tensor is proportional to the metric) satisfying
\[
    R_{ab} = \Lambda g_{ab} \, .
\]
We are then interested in computing the on-shell action on these solutions.

If we focus on asymptotically flat spacetime ($\Lambda=0$), the Schwarzschild solution will be the dominant contribution (in absence of angular momentum), but, as we mentioned in Section \ref{subsec:2_Where_to_now}, we find an issue with its interpretation as an equilibrium solution, because the black hole evaporates. One can also compute the specific heat of the solution, and confirm that the canonical ensemble is unstable, as the Schwarzschild black hole cannot be in thermal equilibrium with a reservoir. The temperature is given by the inverse of $\beta$ in \eqref{eq:2_Periodicity_EuclideanTime}, from which
\[
	T_H = \frac{1}{8\pi M} \qquad \Rightarrow \qquad C = \frac{\partial E}{\partial T} = - \frac{1}{8\pi T_H^2} \, , 
\]
which is negative.

A much better setup is found by looking at asymptotically \textit{AdS} solutions ($\Lambda < 0$). Black holes in $AdS$ are ``eternal,'' the specific heat can be positive, so they can reach equilibrium with their own thermal Hawking radiation. Intuitively, this is due to the negative cosmological constant, which acts as a ``confining box'' that prevents bulk objects from reaching ``infinity'': reasons for this interpretation are briefly reviewed in Section \ref{subsec:3_Physics_in_AdS}. So, we focus on the study of gravity in presence of a negative cosmological constant, which we normalize to $\Lambda = - 3/\ell^2$ so that the action and equations of motion read
\begin{equation}
\label{eq:3_EH_Action_NegCC}
\begin{split}
	S_E &= - \frac{1}{16\pi} \int_M \left( R + \frac{6}{\ell^2} \right) \sqrt{g_E} \, \rd^4 x - \frac{1}{8\pi } \int_{\partial M} K \, \sqrt{h_E} \, \rd^3 x \, , \\ 
	R_{ab} &= - \frac{3}{\ell^2}g_{ab} \, .
\end{split}
\end{equation}

\subsubsection{Thermal AdS}
\label{subsubsec:3_ThermalAdS}

A first solution to \eqref{eq:3_EH_Action_NegCC} is \textit{(Euclidean) anti-de Sitter spacetime}, aka \textit{hyperbolic space}, which is a space of negative costant curvature, that is, its Riemann tensor satisfies
\[
	R_{abcd} = - \frac{1}{\ell^2} ( g_{ac}g_{bd} - g_{ad}g_{bc} ) \, ,
\]
where the constant $\ell$ is referred to as \textit{radius}, by analogy with the sphere (which satisfies the same condition with a positive sign). It is ``basic,'' in the sense that any other constant curvature space with negative curvature is locally isometric to \textit{EAdS}$_4$. The topology of \textit{EAdS}$_4$ is the trivial one, same as $\R^4$, and one can construct a coordinate system such that the metric is
\begin{equation}
\label{eq:3_AdS_Metric_GlobalCoords_1}
    \rd s^2 = \left( 1 + \frac{r^2}{\ell^2} \right) \rd \tE^2 + \frac{\rd r^2}{ 1+ \frac{r^2}{\ell^2}} + r^2 \, \rd\Omega^2_{2} \, ,
\end{equation}
with $\tE\in\R$ and $r\geq 0$. In these coordinates, the metric on $EAdS_4$ has the same form as \eqref{eq:2_StaticSphericalBH_Euclidean}, with $f(r)$ given by
\[
	f(r) = 1 + \frac{r^2}{\ell^2} 
\]
which never vanishes, so $\partial_{\tE}$ is never zero, and therefore there is no regularity requirement that would impose a specific periodicity for $\tE$. Therefore, in order to match our boundary conditions, we can identify $\tE \sim \tE + \beta$ with any $\beta$. The resulting space has topology $S^1\times \R^3$, and is referred to as \textit{thermal AdS} (of which $EAdS_4$ above is the universal cover).

Following the prescriptions of the gravitational path integral, we consider a region $M$ of spacetime defined by $\{ 0 \leq r \leq r_0 \}$ and bounded by the hypersurface $\partial M = \{ r = r_0 \}$ with induced metric
\begin{align}
\label{eq:3_CutoffBdry_InducedMetric}
    h_E = f(r_0) \, \rd \tE^2 + r_0^2 \, \rd\Omega_2^2 \, .
\end{align}
Because of the periodic identification of $\tE$, this surface has the required topology $S^1\times S^2$: the length of the circle is $\sqrt{f(r_0)}\beta$ and the radius of the 2-sphere is $r_0$. As $r_0/\ell \to \infty$ and we look at ``infinity,'' both the circle and the sphere grow without bound:
\begin{equation}
\label{eq:3_ConformalBdry_S1S2}
	h_E \sim \frac{r_0^2}{\ell^2} \left[ \rd \tE^2 + \ell^2 \rd \Omega_2^2 \right] \, , 
\end{equation}
which is \textit{conformally related} to a metric with finite volume $S^1_\beta \times S^2_\ell$. This is a feature of spaces that asymptotically have the same behaviour as anti-de Sitter (called asymptotically $AdS$ spaces, in a triumph of descriptive naming): we are really interested in their \textit{conformal} boundary. Note that when the cosmological constant vanishes, instead, the radius of the circle decouples from the radius of the sphere, and we have a typical asymptotically flat behaviour.

\medskip

We are interested in the on-shell value of the Einstein--Hilbert action with Gibbons--Hawking--York term \eqref{eq:3_EH_Action_NegCC}
The bulk term on-shell reduces to a multiple of the volume of $M$
\begin{equation}
\label{eq:3_Bulk_ThermalAdS}
    I_{\rm bulk} = \frac{1}{16\pi} \int_M \frac{6}{\ell^2}  \, \sqrt{g_E} \, \rd^4 x = \frac{\beta}{2\ell^2} r_0^3 \, .
\end{equation}
To compute the Gibbons--Hawking--York term we can borrow the result from \eqref{eq:3_TraceExtrinsicCurvature_StaticSphericallySymmetric} 
\begin{equation}
\label{eq:3_GHY_ThermalAdS}
\begin{split}
    I_{\rm GHY} &= - \frac{1}{8\pi} \int_{\partial M} K \, \sqrt{h_E} \, \rd^3 x = - \frac{\beta}{2 } \left( \frac{r_0^2}{2} f'(r_0) + 2r_0 f(r_0) \right) \\
    &= \frac{\beta}{2 \ell^2} \left( - 3 r_0^3 - 2r_0 \ell^2 \right) \, , 
\end{split}
\end{equation}
and the naive result for the on-shell action is
\begin{equation}
\label{eq:3_Divergent_ThermalAdS}
    I = - \frac{\beta}{\ell^2} r_0 \left( r_0^2 + \ell^2 \right) \, .
\end{equation}
This expression is divergent as $r_0 \to \infty$, corresponding to the fact that we are integrating on a non-compact space. 

A surprising feature of asymptotically anti-de Sitter spaces is that the divergences that arise when computing $I_{\rm bulk} + I_{\rm GHY}$ can \textit{always} be cancelled by counterterms computed using only the \textit{intrinsic} geometry of $\partial M$. That is, they are integrals of the induced metric $h_E$, the curvature and its derivative. The resulting expressions are universal, in the sense that they apply to any asymptotically AdS solution in a given dimension.

The computation of these counterterms is a technique called \textit{holographic renormalization}, which is always quite technical and sometimes quite subtle (see \cite{deHaro:2000vlm} for some details).
The result relevant to us is that the on-shell action of \textit{any} asymptotically (locally) \textit{EAdS}$_4$ solution is finite once we add to $I_{\rm bulk} + I_{\rm GHY}$ the counterterm action
\begin{equation}
\label{eq:3_Counterterms}
    I_{\rm ct} = \frac{1}{8\pi } \int_{\partial M} \left ( \frac{2}{\ell} + \frac{\ell}{2} R \right) \, \sqrt{h_E} \, \rd^3 x \, ,
\end{equation}
where $R$ is the Ricci scalar of $h_E$, and then take the limit $r_0 \to \infty$.

\medskip

In order to apply the renormalization to our spacetime, we need to know the Ricci scalar of the induced metric $h_E$, but that's a product metric on round $S^1\times S^2$, so the Ricci scalar is just the sum of the two and one of them vanishes (being a one-dimensional metric):
\[
    R = \frac{2}{r_0^2} \, ,
\]
and
\begin{equation}
\label{eq:3_Counterterm_ThermalAdS}
\begin{split}
    I_{\rm ct} &= \frac{\beta}{2} \left ( \frac{2}{\ell} + \frac{\ell}{r_0^2} \right) r_0^2 \sqrt{f(r_0)} \\
    &= \frac{\beta}{\ell^2} \left[ r_0^3 + r_0 \ell^2 + o (1) \right] \, .
\end{split}
\end{equation}
It's clear that we cancel the divergent terms in \eqref{eq:3_Divergent_ThermalAdS}, and in fact we find that the on-shell action of four-dimensional thermal AdS vanishes:
\[
    I_{\text{Th.AdS}} = 0 \, .
\]
Quite remarkably, this is a property of the particular choice of $\partial M$! The action of $EAdS_4$ in a different ``slicing'' may be non-zero. For instance, the action of \textit{EAdS}$_4$ with $S^3$ boundary is
\[
    I_{\text{EAdS}, S^3} = \frac{\pi\ell^2}{2} \, .
\]

\subsubsection{Static black hole}
\label{subsubec:3_Static_AdS_BH}

There are other static solutions filling the same (conformal) boundary $S^1_\beta \times S^2$. A particularly important one is the Euclidean $AdS$-Schwarzschild solution, which has line element
\begin{equation}
\label{eq:3_AdS_Schwarzschild_Euclidean}
    \rd s^2 = f(r) \, \rd \tE^2 + \frac{\rd r^2}{f(r)} + r^2 \, \rd\Omega^2_2 \,, \qquad f(r) = 1 - \frac{2M}{r} + \frac{r^2}{\ell^2} \, ,
\end{equation}
and so is again a static spherically symmetric metric of the form \eqref{eq:2_StaticSphericalBH_Euclidean}. As before, we consider a region of space with a cutoff at $r=r_0$: the induced metric on the boundary has again the same form as \eqref{eq:3_CutoffBdry_InducedMetric}. Importantly, though $f(r_0)$ is different, it has the same asymptotic behaviour as \eqref{eq:3_ConformalBdry_S1S2} when $r_0/\ell \gg 1$: the space is asymptotically AdS, and near ``infinity'' it has a conformal boundary $S^1_\beta\times S^2_\ell$. This means that we have found two saddle points that \textit{a priori} may contribute to the semiclassical approximation of the same gravitational path integral at fixed $\beta$: in fact, whilst $\beta$ is a free parameter in thermal $AdS$, it is fixed by regularity in the black hole, as we now see.  

\medskip

In order to have a good contribution to the semiclassical approximation of the Euclidean gravity path integral, we should also make sure that it is a smooth spacetime. It is not obvious that the Euclidean metric \eqref{eq:3_AdS_Schwarzschild_Euclidean} is regular everywhere, because $f(r)$ has zeroes. Concretely, we denote by $r_+$ the largest of the real roots of $f(r)$, which satisfies
\begin{equation}
\label{eq:3_AdS_Schwarzschild_Mass}
    M = \frac{r_+}{2} \left( 1 + \frac{r_+^2}{\ell^2} \right) \, ,
\end{equation}
and we restrict to $r\geq r_+$. In contrast to the case of thermal $AdS$, now our regularity discussion really parallels that around \eqref{eq:2_StaticSphericalBH_Euclidean}: the space is smooth if and only if we impose a periodic identification of $\tE$ as in \eqref{eq:2_Periodicity_EuclideanTime}:
\[
	\tE \sim \tE + \beta \, , \qquad \beta = \frac{4\pi r_+}{1 + 3 r_+^2/\ell^2} \, .
\]
The resulting geometry is the product of a disc and a 2-sphere described below \eqref{eq:2_Periodicity_EuclideanTime} and represented in Figure \ref{fig:2_CigarGeometry} (with the provision that the asymptotic behaviour is different: the radius of the circle parametrized by $\tE$ grows rather than remaining constant, so strictly speaking we don't have a ``cigar'').

\medskip

From the Lorentzian viewpoint, the spacetime described by \eqref{eq:3_AdS_Schwarzschild_Euclidean} after the analytic continuation $t = - \ii \tE$ is a static spherically symmetric black hole. There is a singularity at $r=0$, as can be checked computing the Kretschmann scalar
\[
R_{abcd}R^{abcd} = \frac{48 M^2}{r^6} + \frac{24}{\ell^2} \, ,
\]
and, as we showed in Section \ref{subsec:2_Near_Horizon}, there is a Killing horizon $\{ r = r_+\}$ for $k=\partial_t$, with temperature
\begin{equation}
\label{eq:3_AdS_Schwarzschild_Temperature}
    T = \frac{f'(r_+)}{4\pi} = \frac{1}{4\pi r_+} + \frac{3r_+}{4\pi \ell^2} \, .
\end{equation}
The entropy of the horizon is given by the Bekenstein--Hawking formula for the area of the $S^2$ bifurcation surface (in Lorentzian), or, equivalently, the $S^2$ over the origin of the disc (in Euclidean)
\begin{equation}
\label{eq:3_AdS_Schwarzschild_Entropy}
    S_{\rm BH} = \frac{1}{4} A_{S^2_{r_+}} = \pi r_+^2 \, .
\end{equation}

\medskip

The next step is the computation of the on-shell action. At the formal level, we can use many of the expressions already derived for thermal $AdS$, though with different $f(r)$ and different range of $r$.
The bulk on-shell contribution is not \eqref{eq:3_Bulk_ThermalAdS} because $r$ does not extend to $0$
\[
    I_{\rm bulk} = \frac{\beta}{2\ell^2} \left( r_0^3 - r_+^3 \right) \, ,
\]
and the Gibbons--Hawking--York term is formally equal to the first line in \eqref{eq:3_GHY_ThermalAdS}
\[
\begin{split}
    I_{\rm GHY} &= - \frac{\beta}{2} \left( \frac{r_0^2}{2} f'(r_0) + 2r_0 f(r_0) \right) \\
    &= \frac{\beta}{2 \ell^2} \left( - 3 r_0^3 - 2r_0 \ell^2 + 3M \ell^2\right) \, .
\end{split}
\]
Overall, the divergent part is the same as that of thermal $AdS$, with the addition of a finite term 
\[
    I = \frac{\beta}{2 \ell^2} \left( - 2 r_0^3 - 2r_0 \ell^2 + 3M \ell^2 - r_+^3 \right) \, .
\]
To remove the divergences we once again consider the counterterms found via holographic renormalization \eqref{eq:3_Counterterms}, which in the static spherically symmetric background have the structure in the first line of \eqref{eq:3_Counterterm_ThermalAdS}
\[
\begin{split}
    I_{\rm ct} &= \frac{\beta}{2} \left ( \frac{2}{\ell} + \frac{\ell}{r_0^2} \right) r_0^2 \sqrt{f(r_0)} \\
    &= \frac{\beta}{2 \ell^2} \left[ 2 r_0^3 + 2r_0 \ell^2 - 2M \ell^2 + o (1) \right] \, .
\end{split}
\]
Finally, taking the limit $r_0\to \infty$, we find
\begin{equation}
\label{eq:3_AdS_Schwarzschild_OSAction}
    I_{\text{AdS-Schw}} = \frac{\beta}{2} \left( M - \frac{r_+^3}{\ell^2} \right) = \frac{\beta}{4} r_+ \left( 1 - \frac{r_+^2}{\ell^2} \right) \, .
\end{equation}

\medskip

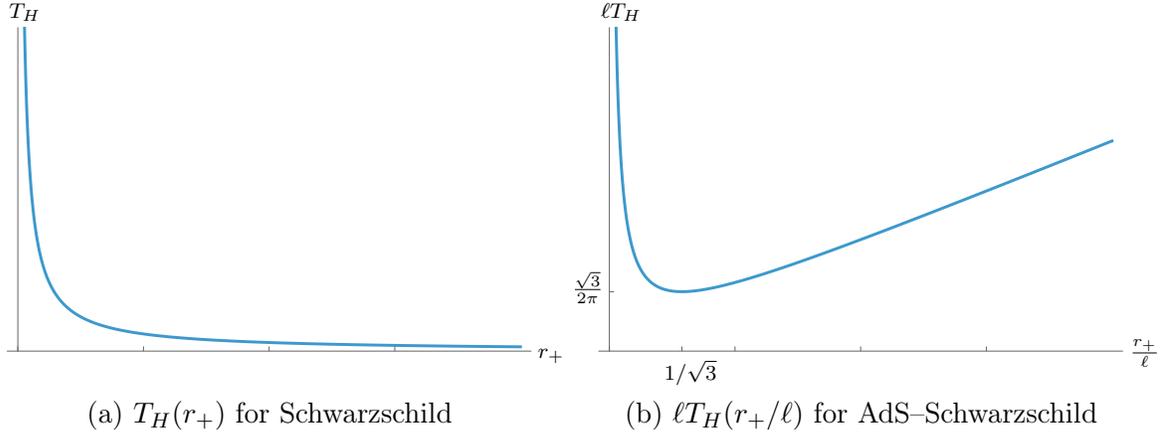
\begin{figure}
\begin{subfigure}{.5\textwidth}
    \centering
    \begin{overpic}[width=0.9\textwidth]{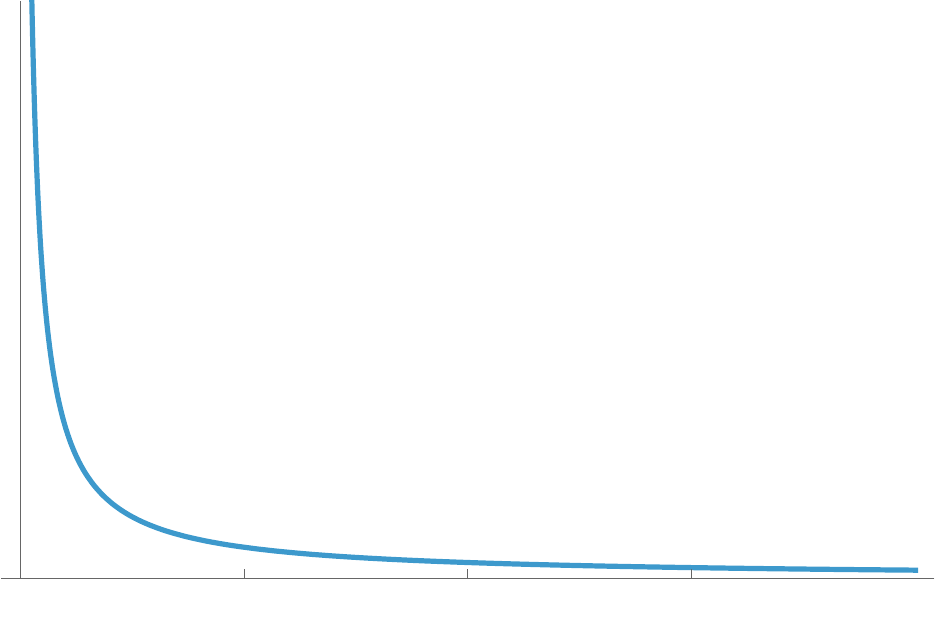}
        \put(1,135){\scriptsize $T_H$}
        \put(201,6){\scriptsize $r_+$}
    \end{overpic}
  \caption{$T_H(r_+)$ for Schwarzschild}
  \label{fig:3_Temperature_Schwarzschild}
\end{subfigure}
\begin{subfigure}{.5\textwidth}
    \centering
    \begin{overpic}[width=0.9\textwidth]{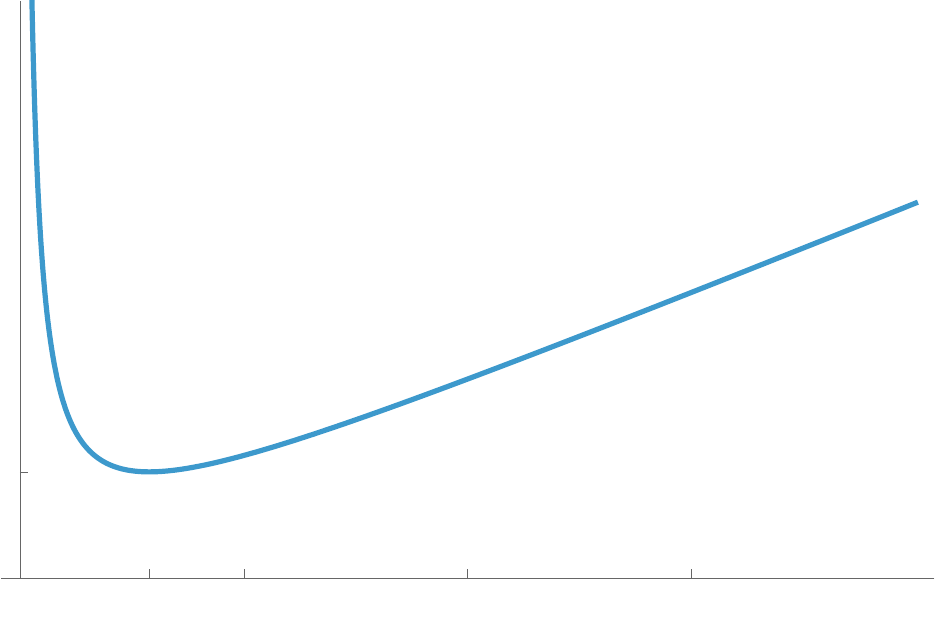}
        \put(1,135){\scriptsize $\ell T_H$}
        \put(201,6){\scriptsize $\tfrac{r_+}{\ell}$}
        \put(-10,30){\scriptsize $\tfrac{\sqrt{3}}{2\pi}$}
        \put(25,-2){\scriptsize ${1}/{\sqrt{3}}$}
    \end{overpic}
  \caption{$\ell T_H(r_+/\ell)$ for AdS--Schwarzschild}
  \label{fig:3_Temperature_Schwarzschild_AdS}
\end{subfigure}
\caption{Plots of temperature against the horizon radius for static spherical black holes in asymptotically flat and asymptotically AdS spacetime.}
\label{fig:3_Temperature_Plots}
\end{figure}

This is where the physics gets interesting! We aim to describe the canonical ensemble defined by the gravitational path integral with boundary conditions fixing the temperature, and we have found two contributions that compete at the leading order in the semiclassical limit. Since we are in the canonical ensemble, we should express the parameter of the solution $M$ in terms of $T$. In fact, we have already traded $M$ for $r_+$ via \eqref{eq:3_AdS_Schwarzschild_Mass}, so we need to invert $T=T(r_+)$ in \eqref{eq:3_AdS_Schwarzschild_Temperature}. The expression for the temperature has a competition between the two scales of the problem $(r_+,\ell)$ that is absent in Schwarzschild (which can be obtained taking $\ell\to \infty$, thus removing the cosmological constant).
The result is that there is a minimum temperature at
\[
    r_{\rm min} = \frac{\ell}{\sqrt{3}} \, , \qquad T_{\rm min} = \frac{\sqrt{3}}{2\pi \ell} \, .
\]
At fixed temperature $T > T_{\min}$, there are two black hole solutions, as we see from the plot \ref{fig:3_Temperature_Schwarzschild_AdS}. Since one branch corresponds to a smaller radius than the other branch, the two solutions are referred to as \textit{small} and \textit{large} black holes. Inverting $T(r_+)$, we find that they correspond to the two roots
\[
    r_+^{L,S} = \frac{\ell}{3} \left( 2\pi \ell T \pm \sqrt{4\pi^2 \ell^2 T^2 - 3} \right) = \frac{1}{2\pi T_{\rm min}^2} \left( T \pm \sqrt{T^2 - T_{\min}^2} \right) \, ,
\]
which are indeed real only if $T>T_{\min}$. This equation, together with the relation $M(r_+)$ in \eqref{eq:3_AdS_Schwarzschild_Mass}, allows us to write $I(\beta)$ for both branches of solutions
\begin{equation}
\label{eq:3_OnShellAction_LargeSmallBH_beta}
    I^{L,S}(\beta) = \frac{\beta_{\rm max}^2}{6\pi} \left(\left(\frac{3}{2}-\frac{\beta_{\rm max}^2}{\beta ^2}\right) \mp \frac{\left(\frac{\beta_{\rm max}^2}{\beta ^2}-1\right)^{3/2}}{\frac{\beta_{\rm max}}{\beta }}\right) \, ,
\end{equation}
where $\beta$ ranges only in $0 < \beta \leq \beta_{\rm max} \equiv 2\pi \ell/\sqrt{3}$.
We then find that
\[
    I^S(\beta) > 0 \ \Leftrightarrow \ 0 < \beta \leq \beta_{\rm max} \, , \qquad I^L(\beta) > 0  \ \Leftrightarrow \ \pi \ell < \beta \leq \beta_{\rm max} \, .
\]
The same conclusion can also be drawn less directly from the expression \eqref{eq:3_AdS_Schwarzschild_OSAction}. It's clear that the on-shell action of the black hole is negative if $r_+ > \ell \equiv r_{\rm HP}$, corresponding to the temperature
\[
    T_{\rm HP} = \frac{1}{\pi \ell} \,. 
\]
One then checks for consistency that $T_{\rm HP}>T_{\rm min}$ and that $r_{\rm HP} > r_{\rm min}$, which means that we are necessarily looking at the branch of large black holes.

\medskip

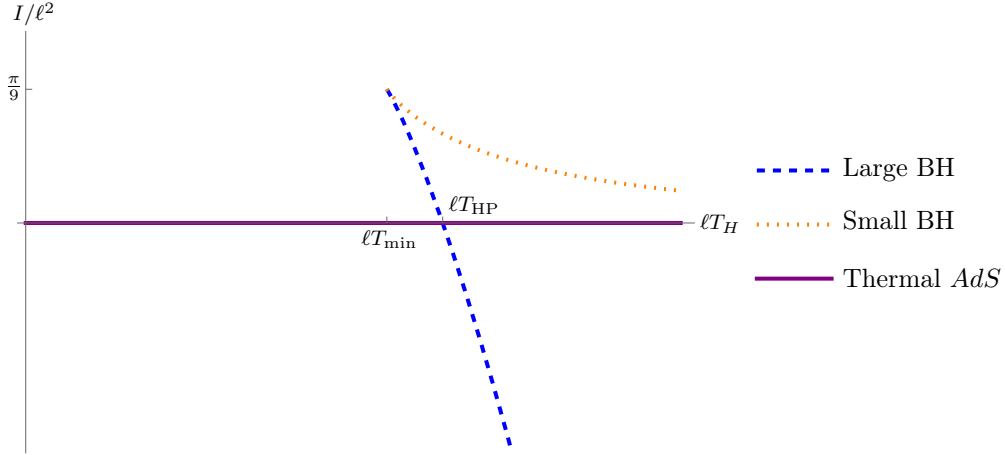
\begin{figure}
    \hspace{1.2cm}
    \begin{overpic}[width=0.7\textwidth]{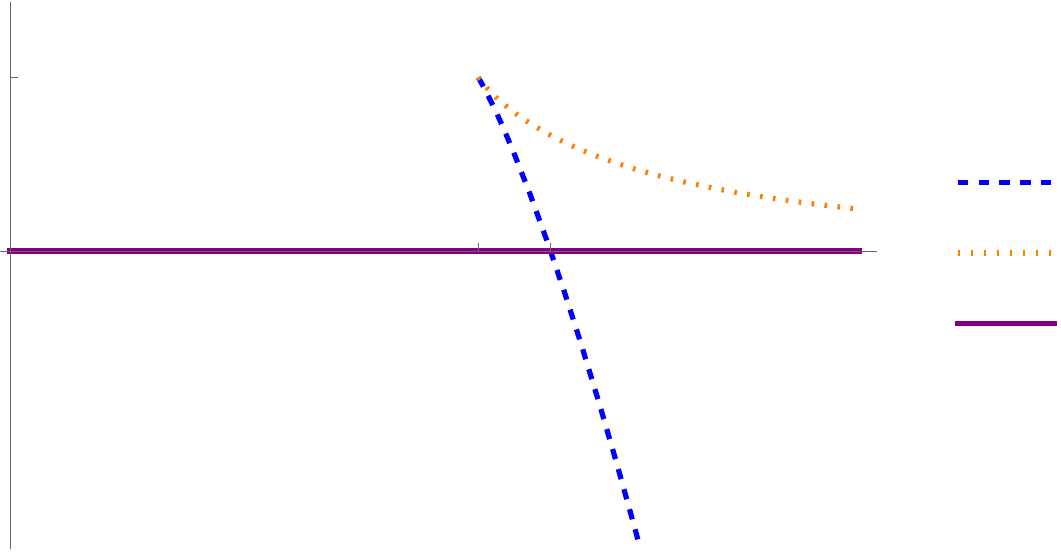}
        \put(258,86){\scriptsize $\ell T_H$}
        \put(-2,165){\scriptsize $I/\ell^2$}
        \put(-5,138){\scriptsize $\tfrac{\pi}{9}$}
        \put(130,80){\scriptsize $\ell T_{\rm min}$}
        \put(163,93){\scriptsize $\ell T_{\rm HP}$}
        \put(312,106){\footnotesize Large BH}
        \put(312,86){\footnotesize Small BH}
        \put(312,64){\footnotesize Thermal $AdS$}
    \end{overpic}
    \caption{On-shell action of the three competing solutions with conformal boundary $S^1\times S^2$ as a function of the temperature. Large and small black holes only exist if $T>T_{\rm min}$, and for $T>T_{\rm HP}$ the large black hole is thermodynamically favorable compared to thermal $AdS$.}
    \label{fig:3_HP_Transition}
\end{figure}

We are now ready to discuss the thermodynamics of the system using the quantum statistical relation \eqref{eq:3_QuantumStatisticalRelation}. We first observe that for $T<T_{\rm min}$ there is a unique solution with the appropriate boundary condition, thermal $AdS$, so the only possible equilibrium is without a black hole. As $T$ is raised above $T_{\rm min}$, there are two additional solutions that compete with thermal $AdS$ as equilibria. When $T_{\rm min} < T < T_{\rm HP}$, the free energies of both small and large black holes are positive, so it is still most favorable that the black hole evaporates leaving only thermal $AdS$. Finally, if $T>T_{\rm HP}$ the free energy of the large black hole becomes lower than that of thermal $AdS$ and the stable thermodynamic equilibrium is the large black hole state. This is represented in Figure \ref{fig:3_HP_Transition}.

Since the free energy is continuous but its first derivative is not, this change is a first order phase transition, referred to as the \textit{Hawking--Page transition} \cite{Hawking:1982dh}. A remarkable observation from the viewpoint of the gravitational path integral is that the phase transition also involves a transition between solutions with different topologies! If the length of the boundary circle is larger than $\beta_{\rm HP} = \pi \ell$, the dominant contribution to the path integral is the one ``filling the sphere'' with topology $S^1 \times \R^3$, as $\beta < \beta_{\rm HP}$, instead, the dominant contribution is that ``filling the circle'' with topology $\R^2\times S^2$.

To confirm the previous observations, we should also establish the stability of the equilibrium configurations, which is done by considering the heat capacity
\[
    C = \frac{\partial \braket{E}}{\partial T} \, ,
\]
where
\[
    \braket{E} = - \frac{\partial \ }{\partial \beta} \log Z(\beta) .
\]
We could apply the derivatives directly to the expressions for $\log Z(\beta) = - I(\beta)$ in \eqref{eq:3_OnShellAction_LargeSmallBH_beta}, but this leads to cumbersome expressions. A quicker way is to trade $\beta$ for $r_+$ using \eqref{eq:3_AdS_Schwarzschild_Temperature}, and write the action \eqref{eq:3_AdS_Schwarzschild_OSAction} in terms of $r_+$
\[
    I_{\text{AdS-Schw}} = \pi r_+^2 \frac{\ell^2 - r_+^2}{\ell^2 + 3r_+^2} \, .
\]
We then find that
\[
    \braket{E} = M \, , \qquad S = \left( - 1 + \beta \frac{\partial \ }{\partial \beta} \right) I = \pi r_+^2 \, .
\]
The first relation provides a physical interpretation for the parameter $M$ of the solution, and the second one confirms that the thermodynamical entropy agrees with the Bekenstein--Hawking formula \eqref{eq:3_AdS_Schwarzschild_Entropy}, even for asymptotically $AdS$ solutions. This is a non-trivial consistency check of the gravitational path integral approach.
Finally, we compute the heat capacity
\[
    C = 2\pi r_+^2 \frac{3 r_+^2 + \ell ^2}{3 r_+^2-\ell ^2} \, ,
\]
which is positive provided $r_+ > \ell/\sqrt{3} \equiv r_{\rm min}$. So if the black hole has $r_+ > r_{\rm min}$, that is, it's a large black hole, then its heat capacity is positive and the canonical ensemble is well-defined because it can be in equilibrium with a reservoir held at finite temperature, unlike what happens with small black holes in $AdS$ and with the Schwarzschild solution.

\subsection{Physics in AdS}
\label{subsec:3_Physics_in_AdS}

\begin{figure}
        \centering
        \begin{tikzpicture}
            \draw [densely dashed, line width=0.3mm] (-2,-3) -- (-2,3);
            \draw [line width=0.3mm] (0,-3) -- (0,3);
            \fill [fill=gray, fill opacity=.15] (-2,3) -- (-1,3) -- (-2,2) -- (0,0) -- (-2,-2) -- (-1,-3) -- (-2,-3);
            \draw [black] (-1,3) -- (-2,2) -- (0,0) -- (-2,-2) -- (-1,-3);
            \node at (0.4,0) {\small $\mathscr{I}$};
            \filldraw[black] (-1,3.6) circle (2pt) node[anchor=south]{\small $i^+$};
            \filldraw[black] (-1,-3.6) circle (2pt) node[anchor=north]{\small $i^-$};
            \draw [blue] (-2,-2) node[anchor=east,black] {\small $p$} .. controls (-0.1,0) .. (-2,2) node[anchor=east,black] {\small $q$};
            \draw [blue] (-2,-2) .. controls (-1,0) .. (-2,2);
            \node at (-2.5,0) {\small $t$};
            \node at (-2,-3.2) {\small $\theta = 0$};
            \node at (0,-3.2) {\small $\theta = \frac{\pi}{2}$};
        
        \end{tikzpicture}
    \caption{Projection of $AdS_d$ to the $(t,\theta)$ plane together with null (in black) and timelike (in blue) geodesics leaving $p$ and focusing at $q$. 
    }
     \label{fig:3_AdS_Diagram}
\end{figure}
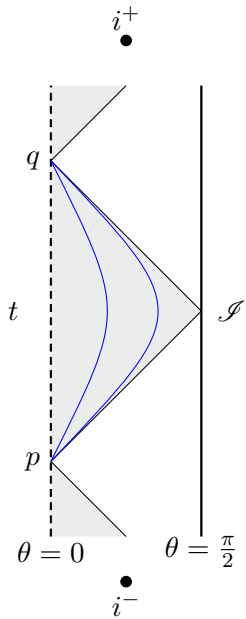

Finally, we flash some intuitive physical reasons to justify why gravity in $AdS$ behaves differently than in flat space. To do so, we look at $AdS_d$ with radius $\ell$ in Lorentzian signature. This is defined as the universal cover of the hyperboloid in $\R^{2,d-1}$
\begin{equation}
\label{eq:3_AdS_Lorentzian_Embedding_Hyperboloid}
	- X_0^2 + \sum_{i=1}^{d-2} X_i^2 - X_d^2 = - \ell^2 \, , 
\end{equation} 
with the induced metric from $\R^{2,d-1}$. We introduce the global coordinates \eqref{eq:3_AdS_Metric_GlobalCoords_1} via
\begin{equation}
\label{eq:3_AdS_Lorentzian_Embedding_Coordinates}
\begin{split}
	X_0 &= \ell \sin t \, \sqrt{1+\frac{r^2}{\ell^2}} \,, \qquad \quad X_d = \ell \cos t \, \sqrt{1+\frac{r^2}{\ell^2}} \, , \\
	X_i  &= r \, \overline{x}_i \, , \qquad i=1, \dots, d-1 \, ,
\end{split}
\end{equation}
where the $\overline{x}_i$ are functions such that $\sum_{i=1}^{d-1}\overline{x}_i^2 = 1$ and define a $S^{d-2}$. In these coordinates, the line element is
\begin{equation}
\label{eq:3_AdS_Lorentzian_Metric_GlobalCoords_1}
\begin{split}
    \rd s^2 &= - \left( 1 + \frac{r^2}{\ell^2} \right) \rd t^2 + \frac{\rd r^2}{ 1+ \frac{r^2}{\ell^2}} + r^2 \, \rd\Omega^2_{d-2} \\
	&= \left( 1 + \frac{r^2}{\ell^2} \right) \left[ - \rd t^2 + \frac{\rd r^2}{ \left( 1+ \frac{r^2}{\ell^2} \right)^2 } + \frac{r^2}{1+\frac{r^2}{\ell^2}} \rd \Omega_{d-2}^2 \right] \, .
\end{split}
\end{equation}
In the second line, we highlighted the fact that we can change coordinates using
\[
       \sqrt{1 + \frac{r^2}{\ell^2}} = \frac{1}{\cos\theta} \, , \qquad t = \ell u \, ,
\]
with $\theta \in [0,\pi/2)$, and now we have
\begin{equation}
\label{eq:3_AdS_Lorentzian_Metric_GlobalCoords_2}
\begin{split}
    \rd s^2 &= \frac{\ell^2}{\cos^2\theta} \left[ - \rd u^2 + \rd\theta^2 + \sin^2\theta \, \rd \Omega_{d-2}^2 \right] \, .
\end{split}
\end{equation}
This line element shows that $AdS_d$ is conformally related to \textit{half} of $\R_t \times S^{d-1}$, with the ``half'' arising from the fact that $\theta$ doesn't extend to $\pi$. In order to visualize this relation, we represent the strip $\R_t\times [0,\pi/2)$ in Figure \ref{fig:3_AdS_Diagram}. Importantly, the causal structure on this strip is the same as in $AdS_d$, since the two are conformally related, so null/timelike curves in one correspond to null/timelike curves in the other. In particular, null curves are straight lines at 45$^\circ$ angle, and they define lightcones at each point such that timelike (resp. spacelike) curves have tangent vectors everywhere inside (resp. outside) the lightcone. Both null and spacelike curves end on the conformal boundary at $\theta = \frac{\pi}{2}$, which therefore represents both null and spacelike infinity and is denoted $\mathscr{I}$. We refer to this boundary as ``conformal'' because it doesn't really belong to $AdS_d$, but rather it's the boundary of its closure, where the metric \eqref{eq:3_AdS_Lorentzian_Metric_GlobalCoords_2} can be extended provided it's multiplied by $\cos^2\theta/\ell^2$. This is the essence of the \textit{conformal compactification}. \\
On the other hand, timelike curves may continue to ``infinity'' and there is no conformal transformation that would take their endpoints at a finite distance on the page. Therefore, the timelike infinity $i^\pm$ are just represented as points at the two ends of the strip. Of course, it's important to remember that despite what it looks like in the figure, the ``boundary'' of $AdS_d$ is at infinite distance from any point in the interior, as measured in the physical metric \eqref{eq:3_AdS_Lorentzian_Metric_GlobalCoords_2}, because of the divergence of the defining function.

\medskip

One can also solve the geodesic equation in $AdS$, and observe that the trajectories of massive particles following timelike geodesics are periodic and those of massless particles are straight lines. Said in another way, and referring again to Figure \ref{fig:3_AdS_Diagram}, timelike geodesics from a point $p$ expand, then reconverge at the point $q$, and then repeat this behaviour, without ever reaching the (timelike) boundary, whereas null geodesics from $p$ get to $\mathscr{I}$ in a finite amount of time and draw a boundary in the future of $p$ corresponding to regions of spacetime that will not be reached by any geodesic from $p$.\footnote{This may not be true for curves that are not geodesics (that is, trajectories of accelerated observers), which can end on $\mathscr{I}$ and explore the entire future of $p$. The same behaviour also happens in flat space.} In this sense, anti-de Sitter space corresponds to having naturally created a box for gravity.\\
This also ``justifies'' why the thermal behaviour of $AdS$-Schwarzschild is very different from that of Schwarzschild itself: looking at the (analytic continuation of the) line element \eqref{eq:3_AdS_Schwarzschild_Euclidean}, we can identify the ``gravitational potential'' in the Newtonian approximation as the component $g_{tt}$ of the metric. In contrast to the asymptotically flat case, here the potential grows unbounded towards the (conformal) boundary at $r\to \infty$, thus acting as a ``well'' for massive particles. This is a defining feature of asymptotically $AdS$ spaces, not just $AdS$-Schwarzschild. Moreover, look at the temperature measured by an observer along $(\partial_t)^a$: from Tolman's law (cf. footnote \ref{footnote:2_TolmanLaw}), we find
\[
	T_{\rm obs} = \frac{T_0}{\sqrt{f(r)}} \, .
\]
In asymptotically flat spacetime, $f(r)\to 1$ as $r\to \infty$, but in asymptotically $AdS$ spaces, $f(r) \sim \frac{r^2}{\ell^2}$, so $T_{\rm obs}$ decreases as one approaches the boundary, potentially leading to a finite thermal energy without the need to put an actual box \cite{Hawking:1982dh}.

\medskip

We conclude this very brief review of gravitational properties of anti-de Sitter spacetime with a couple of hints suggesting that a holographic description of gravity in $AdS_d$ may exist in terms of a conformally-invariant field theory in $(d-1)$ dimensions (which is the essence of the $AdS$/CFT correspondence). Of course, these observations strongly motivate, but do not by themselves establish, the existence of a holographic dual description.

Compare the causal structure in Figure \ref{fig:3_AdS_Diagram} with the asymptotic structure of flat spacetime: in asymptotically flat spacetime, null infinity is composed of two \textit{null} surfaces, but in anti-de Sitter null infinity is \textit{timelike}. This has the remarkable consequence that $AdS_d$ is not globally hyperbolic (see footnote \ref{footnote:2_GloballyHyperbolic}). There are families of spacelike surfaces that cover the entire manifold, but their future and past domain of dependence do not cover the entire manifold, as there are null geodesics that do not intersect the surfaces. Specifying initial data on these surfaces is not sufficient to predict the evolution of solutions of hyperbolic equations everywhere on $AdS_d$: there will always be information coming from $\mathscr{I}$. Therefore, in order to construct a well-defined Cauchy problem, one has to specify the initial data on the \textit{timelike} boundary $\mathscr{I}$, where a field theory could actually live! Note that this is consistent with the previous observation that null geodesics reach the conformal boundary (and return) in finite time: we influence the bulk by changing the conformal boundary.

In fact, $\mathscr{I}$ itself has geometry of the Einstein static universe in one dimension lower, itself the conformal compactification of $\R^{1,d-2}$. The conformal group of $\R^{1,d-2}$, the group of spacetime symmetries of a conformal field theory in $(d-1)$ dimensions, is $SO(2,d-1)$, which is also the group of continuous isometries of $AdS_d$, as can be seen from the definition as the hyperboloid \eqref{eq:3_AdS_Lorentzian_Embedding_Hyperboloid}, which is clearly invariant under $SO(2,d-1)$. Thus, the isometries in the bulk correspond to the symmetries of the boundary field theory.

Another fun fact \cite{Witten:1998qj}. When quantizing a massive scalar field theory on $AdS_d$, with mass $m$, we find that \textit{negative} mass squared is allowed by unitarity, provided \cite{Breitenlohner:1982jf}
\[
	m^2 \ell^2 \geq - \frac{(d-1)^2}{4} \, .
\]
Moreover, one of the labels of the modes is the eigenvalue $\Delta$ of the generator $(\partial_t)^a$ of the $SO(2)$ subgroup of $SO(2,d-1)$. In fact, there are two choices: 
\[
	\Delta_\pm =  \frac{d-1}{2} \pm \sqrt{ m^2 \ell^2 + \frac{(d-1)^2}{4} } \, .
\]
The bound on $m^2$ above is equivalent to both $\Delta_\pm$ being real, but there is more: if 
\[
    m^2 \ell^2 > - \frac{(d-1)^2}{4} + 1 \, , 
\]
then the modes with $\Delta_-$ do not give a convergent energy, so they are not allowed. On the other hand, if
\begin{equation}
\label{eq:3_AlternateQuantization}
    - \frac{(d-1)^2}{4} + 1 \geq m^2 \ell^2 \geq - \frac{(d-1)^2}{4} \, ,
\end{equation}
both sets of modes are normalizable and lead to a conserved energy. The same condition can be phrased in terms of $\Delta_-$, showing that there is a lower bound
\[
	\Delta_- \geq \frac{d-3}{2} \, ,
\]
but this is precisely unitarity bound of a scalar operator of a $(d-1)$-dimensional CFT! (If we identify $\Delta$ with the scaling dimension, as we should, since the bulk isometries become boundary symmetries)

Finally, you may wonder what happens at the Hawking--Page transition: from the holographic viewpoint, the Hawking--Page transition is interpreted as a transition between a confined (thermal $AdS$) and deconfined (black hole) phase of the boundary theory, as the temperature is increased \cite{Witten:1998qj, Witten:1998zw}.

\newpage

\section{Selected topics on the gravitational path integral}
\label{sec:4_Selected_Topics}

In the previous section, we introduced the gravitational path integral as a framework to quantize gravity, and we saw that, even at the crudest approximation, keeping only the leading saddle in the semiclassical approximation in \eqref{eq:3_SemiclassicalExpansion}, yields a coherent picture of both the temperature and entropy of black holes. Moreover, we saw genuinely new physics emerging from the competition between saddles of different topology in anti-de Sitter. 
At the same time, these successes rely on a sequence of (so far hidden) assumptions, about boundary conditions, contour choices, and the treatment of gauge redundancies, that are not innocuous.

In this section, we explore some deeper aspects of the Euclidean approach: how black hole entropy can be understood topologically, why rotation forces us to consider complexified geometries, and some unresolved subtleties that ultimately limit the elevation of the gravitational path integral to a fundamental definition of quantum gravity. These questions touch on points that are still not fully understood, and are the subject of current research.

\subsection{Entropy from topology}
\label{subsec:4_Entropy_Topology}

We have seen that the Euclidean gravitational path integral is able to compute the Bekenstein--Hawking entropy, and that this is associated with black holes, in contrast to a spacetime such as thermal $AdS$. We also remarked that the \hyperlink{1_First_Law}{first law of black hole mechanics} arises from an application of Stokes' theorem giving a boundary contribution from the horizon that reduces an integral over the bifurcation surface, which can be recognized as an entropy \cite{Wald:1993nt}.

A connection between the two statements can be formalized. In all the thermal cases, we have a circle symmetry of the solution (that is, an isometry that also preserves the configurations of the matter potentially present) acting on spacetime and generated by some Killing vector $\xi^a$ with closed orbits with period $\beta$. We then compute the on-shell action of the solution, and extract the entropy by applying the differential operator
\begin{equation}
\label{eq:4_Entropy_Thermal}
	S = \left( - 1 + \beta \frac{\partial \ }{\partial \beta} \right) I \, .
\end{equation}
Note that under this operator, any contribution to the action that is linear in $\beta$ would vanish. This is crucial, as it distinguishes whether the thermal isometry acts freely or has fixed-point sets, which in turn constrains the topology of the Euclidean manifold.

Suppose first that the isometry acts freely on the spacetime $M$: there is no locus in $M$ that is fixed by the isometry, and so we can construct the quotient $B \equiv M/S^1$. The integral over $M$ splits in the integral over $S^1$ and the integral over $B$, so the on-shell action will be
\[
	I = \int_M \cL = \beta \int_{B}\cL' \, .
\]
This means that the entropy associated to a space where the symmetry acts freely is necessarily vanishing! As we saw, this is the case for the thermal $AdS$ spacetime discussed in Section \ref{subsubsec:3_ThermalAdS}, which has the topology of $S^1\times \R^3$.

\begin{figure}
    \centering
    \begin{tikzpicture}
        \draw (2,0) [partial ellipse=90:270:7cm and 2cm];
        \draw (2,0) ellipse (1cm and 2cm);
        \filldraw[black] (-5,0) circle (2pt) node[anchor=south east]{\small $r=r_+$};
        \draw [->] (1,2.3) -- (2,2.3);
        \node at (1.5,2.5) {\small $r\to \infty$};
        \draw [blue] (-4,0) [partial ellipse=117:243:0.6cm and 1cm];
        \draw [dashed] (-1.1,0) [partial ellipse=100:260:0.8cm and 1.8cm];
        \draw [->] (-1.4,0) [partial ellipse=150:210:0.8cm and 1.8cm];
        \node at (-2.5,0) {\small $\xi^a$};
        \node at (-4.3,0) {\small {\color{blue}$T_\varepsilon$}};
    \end{tikzpicture}
    \caption{The fibration induced by the symmetry $\xi^a$ degenerates at $\{r=r_+\}$, which we surround with a $\xi^a$-invariant neighbourhood with radius $\varepsilon$. Everywhere outside the neighbourhood the fibration is trivial and we can reduce the on-shell action to the integral of an exact form, which by Stokes' theorem reduces to a contribution from $T_\varepsilon$ (in blue) and a contribution from the boundary.}
    \label{fig:4_Fixed_Sets}
\end{figure}
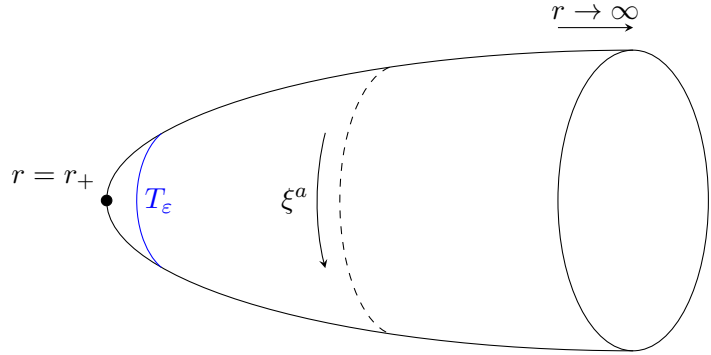

In contrast, the entropy can be non-vanishing if the symmetry has fixed sets. This is definitely the case for the black holes discussed until now: their topology is $\R^2\times S^2$, and the relevant symmetry acts as rotations of the $\R^2$ factor, fixing the $S^2$ over the origin. The fibration degenerates at the origin, so we surround it with a $\xi$-invariant tubular neighbourhood with radius $\varepsilon$, and in the limit $\varepsilon\to 0$ we recover the original integral (compare to Figure \ref{fig:4_Fixed_Sets}). The complement of the tubular neighbourhood, $M_\varepsilon$, is now a cylinder with boundary, so we can again write it as a trivial fibration over $B_\varepsilon \equiv M_\varepsilon/S^1$.

To close the argument, we use the explicit form of the Lagrangian: introducing the volume form, the Einstein--Hilbert action has the following form on-shell
\[
	\cL = - \frac{1}{16 \pi } \left( R + \frac{6}{\ell^2} \right) \vol  = \frac{1}{8 \pi } \frac{3}{\ell^2}\vol  \, .
\]
One can show, using only the Killing vector lemma and the equations of motion, that for any Killing vector $\xi^a$\footnote{Here $\hook$ denotes the contraction of a vector in a differential form: if $X^a$ is a vector, and $\omega_{b_1\cdots b_p}$ is a $p$-form, then $(X \hook \omega)_{b_1\cdots b_{p-1}} \equiv X^a \omega_{ab_1 \cdots b_{p-1}}$, $*$ is the Hodge dual, and ${}^\flat$ indicates the canonical isomorphism taking a vector to a differential form via the metric (that is, it ``lowers the index''; it's the inverse of the operation described in footnote \ref{footnote:3_MusicalIsomorphism}).}
\begin{equation}
\label{eq:4_Equivariant_Entropy_Filippo}
	\frac{3}{\ell^2} \xi \hook \vol = - \frac{1}{2} \rd * \rd \xi^\flat \, .
\end{equation}
Therefore, recalling that everywhere on $M_\varepsilon$ we can split the integration on fibre and base, and using Stokes' theorem, we can turn the integral on $M_\varepsilon$ into an integral on its boundary
\begin{equation}
\label{eq:4_Stokes_Entropy}
	I_{\rm bulk} = - \frac{1}{16 \pi } \beta \int_{B_\varepsilon} \rd * \rd \xi^\flat = - \frac{1}{16 \pi } \beta \int_{\partial B_\varepsilon} * \rd \xi^\flat \, .
\end{equation}
Here $\partial B_\varepsilon$ is the union of the (reduction of the) cutoff surface $\partial M_0/S^1$ and the boundary of the neighbourhood surrounding the origin of the ``cigar.'' Therefore, as in the discussion of the \hyperlink{1_First_Law}{first law of black hole mechanics}, we find a contribution from the asymptotic boundary (which in that context was null infinity $\mathscr{I}^+$) and a contribution from the ``horizon'': recall that the bifurcation surface in Lorentzian becomes the $S^2$ over the origin in Euclidean. In fact, this remark clarifies the connection between a non-zero entropy and the necessity of fixed points of the isometry: the Killing vector $\xi^a$ that vanishes at the origin in the Euclidean geometry is the analytic continuation of the generator of the Killing horizon in Lorentzian, where we associate the entropy to the existence of a bifurcation surface (on which the Lorentzian Killing vector necessarily vanishes).

We proceed to evaluate the two contributions to \eqref{eq:4_Stokes_Entropy}. First, we look at the fixed set. The boundary of the $\xi$-invariant neighbourhood surrounding the fixed set is a copy of $S^1\times S^2$, which we can denote by $T_\varepsilon$. The isometry generated by $\xi^a$ (by construction) doesn't act on $S^2 \cong T_\varepsilon/S^1$, and it acts on the normal space to $S^2 \cong T_\varepsilon/S^1$ as a rotation with angular velocity $\kappa$ (the surface gravity). This can be seen, for instance, looking at \eqref{eq:2_RindlerMetric_NearHorizon}: in this case, the ``thermal'' isometry is generated by 
\[
	\xi = \partial_{\tE} = \frac{2\pi}{\beta} \partial_\tau = \kappa \, \partial_\tau \, ,
\]
where $\tau$ is an angular coordinate with period $2\pi$. Therefore, in an orthonormal frame we can write
\[
	\rd \xi^\flat \rvert_{S^2} = 2 \begin{pmatrix}
		0 & 0 & 0 & 0 \\ 0 & 0 & 0 & 0 \\ 0 & 0 & 0 & \kappa \\ 0 & 0 & - \kappa & 0 
	\end{pmatrix} \, ,
\]
having chosen the normal space to be along the 3-4 directions. Substituting into \eqref{eq:4_Stokes_Entropy} yields a purely geometric contribution, independent of the details of the solution
\begin{equation}
\label{eq:4_Equivariant_Entropy_FP}
	I^{\rm FP} = \lim_{\varepsilon\to 0} I_{T_{\varepsilon}} = - \frac{1}{4} \lim_{\varepsilon\to 0}\int_{T_\varepsilon} \vol_{T_\varepsilon} = - \frac{1}{4} A_{\rm h} \, .
\end{equation}
The remaining contribution comes from the boundary of spacetime, which, as usual, we take to be a cutoff surface at $\{r = r_0\}$: the contribution from this boundary should be supplemented with the Gibbons--Hawking--York term and the counterterms, as discussed in Section \ref{subsec:3_EuclideanStatic},
\[
    I^\partial = \lim_{r_0\to \infty} \left[ - \frac{1}{16 \pi } \beta \int_{\partial M_0/S^1} * \rd \xi^\flat + I_{\rm GHY} + I_{\rm ct} \right] \, .
\]
It is non-trivial to show that, since $\xi^\flat \propto \rd \tE$, this expression corresponds to a holographic conserved charge, the mass of the spacetime, and so $I^\partial = \beta M$ \cite{Papadimitriou:2005ii}. 

Therefore, the final result for the on-shell action is
\[
	I = - \frac{1}{4} A_{\rm h} + \beta M \, .
\]
Note that we haven't used any explicit form of the solution of the equations of motion!
Now we can apply the formula \eqref{eq:4_Entropy_Thermal}. The contribution from the boundary vanishes (since it's linear in $\beta$), and we arrive at the Bekenstein--Hawking entropy formula \eqref{eq:1_Wald_Reduced_BH}. 
This result is, of course, consistent with the quantum statistical relations \eqref{eq:3_QuantumStatisticalRelation} and \eqref{eq:3_GrandCanonicalStatisticalRelation}, but the route by which it is obtained is conceptually remarkable: black hole entropy is not associated with the bulk of spacetime, nor with asymptotic data, but localizes on the locus where the thermal isometry degenerates. In Lorentzian signature, this locus is the bifurcation surface of a Killing horizon; in Euclidean signature, it appears as the sphere over the tip of the cigar. 

Two final comments. First, this reformulation of Wald’s argument (see \cite{Wald:1993nt, Iyer:1995kg, Papadimitriou:2005ii}) highlights a deep connection between symmetry, fixed points, and entropy. It also provides, especially via \eqref{eq:4_Equivariant_Entropy_Filippo} and \eqref{eq:4_Equivariant_Entropy_FP}, a first glimpse of the role played by equivariant localization techniques in gravitational theories, a theme that reappears, in a systematic and unavoidable way, in supersymmetric settings \cite{BenettiGenolini:2023kxp}. Second, the systems we discussed so far enjoy a $U(1)$ symmetry, which we have associated to a thermal circle and thus to an entropy. However, as mentioned in Section \ref{subsec:2_FourLaws}, entropy should ultimately be studied in the context of information theory. If one takes this viewpoint, it is possible to generalize the notion of thermal entropy to systems without a continuous $U(1)$ symmetry but admitting a $\mathbb{Z}_n$ symmetry \cite{Lewkowycz:2013nqa}. Quite remarkably, this ``generalized'' entropy is still captured by the area of a codimension-2 surface where a circle shrinks to zero size (not necessarily smoothly)!

\subsection{Rotation, charge and complex metrics}
\label{subsec:4_Rotation}

So far, we have only considered static Lorentzian spacetime, with vanishing mixed terms $g_{tx^i}$, so that, after the analytic continuation $t=-\ii\tE$, the resulting metric was positive-definite (commonly referred to as ``Euclidean'' or, more properly, Riemannian). We now introduce rotation and charge, and this will lead to some modifications.

\medskip

In Lorentzian signature, the presence of angular momentum is described in terms of a spacelike Killing vector $m^a$ generating a $U(1)$ action on the space. If we impose that the metric near the boundary is that on $S^1_\beta \times_f S^2$ (where the subscript $f$ signals the possibility of a cross term), then the $U(1)$ generated by $m^a$ could be the azimuthal rotations of the $S^2$ factor. We can combine $m^a$ with the timelike Killing vector $k^a$ to define a Killing vector $\xi^a = k^a + \Omega \, m^a$. The coefficient $\Omega$ is interpreted as angular velocity: we construct adapted coordinates $k=\partial_t$ and $m=\partial_\phi$, then along an orbit of $\xi^a$, up to a choice of integration constant, we find $\phi = \Omega t$, and we identify $\Omega$ as an angular velocity. To start seeing some complex quantities, notice that after Wick rotation, $\xi^a$ becomes a complex vector field, though it is pure imaginary if $\Omega$ is pure imaginary as well.

To discuss charge, we should include an electromagnetic field: our Lorentzian action will be the Einstein--Hilbert action \eqref{eq:3_SEH}, supplemented by the Maxwell action
\[
    S = \frac{1}{16\pi} \int \left( R + \frac{6}{\ell^2} - F_{ab}F^{ab} \right) \sqrt{-g} \, \rd^4 x \, ,
\]
and we should impose gauge-invariant boundary conditions for the electromagnetic field. On the boundary, with topology $S^1 \times S^2$ after Wick rotation and Euclidean identification, the gauge field is specified by imposing the flux of the curvature through $S^2$ (corresponding to the magnetic charge, which we set to zero), and the holonomy of the gauge potential around $S^1$: $\exp \left( \ii \int_{S^1}A \right) \equiv \exp \left( \ii \beta \Phi_E \right)$, and we will identify $\Phi_E = \ii \Phi$, where $\Phi$ is the electrostatic potential. Again, notice that $\Phi_E\in\R$ if and only if $A_{\tE}$ is real as well, which means that the gauge field in Lorentzian signature is complex. Conversely, one could start from a real Lorentzian gauge field with $A_t \in \R$, but then the Wick-rotated gauge field would be complex!

\medskip

We now show more formally that introducing rotation means that the periodic identification of the fields in the KMS condition \eqref{eq:2_KMS_v1} is modified \cite{Gibbons:1976pt}. To do so, we consider the grand canonical Gibbs formula with two operators generating $U(1)$ symmetries (which generalizes \eqref{eq:2_GibbsFormula}): the charge and angular momentum with corresponding potentials $\Phi$ and $\Omega$
\begin{equation}
\label{eq:4_GibbsFormula_GrandCanonical}
    \braket{\cO}_{\beta, \Phi, \Omega} \equiv Z^{-1} \Tr \left( \cO \e^{ - \beta ( H - \Phi Q - \Omega J ) } \right) \, ,
\end{equation}
and consider the Wightman functions of a complex scalar field $\varphi(t, \phi, \mb{x})$ (where $\mb{x}$ are $r$, $\theta$) on which $Q$ and $J$ act as
\[
\begin{split}
    \e^{\ii \alpha Q} \varphi(t, \phi, \mb{x}) \e^{ - \ii \alpha Q} &= \e^{\ii \alpha q} \varphi(t, \phi, \mb{x}) \,, \\ 
    \e^{ \ii \alpha J} \varphi(t, \phi, \mb{x}) \e^{ - \ii \alpha J} &= \varphi(t, \phi - \alpha, \mb{x}) \, .
\end{split}
\]
Note that the second one is a spacetime $U(1)$ symmetry, in contrast to the first one.
Then, using the Heisenberg evolution of the fields and the cyclicity of the trace, we find\footnote{Compare to \eqref{eq:2_Correlators_v1} and \eqref{eq:2_WightmanFunctions_Scalars} for the notation.}
\begin{align}
    G_+^{\beta, \Phi, \Omega}(z, \phi, \mb{x}, \mb{y}) & \equiv Z^{-1} \Tr \left[ \varphi(z,\phi, \mb{x}) \varphi^\dagger(0,0,\mb{y}) \e^{ - \beta( H - \Phi Q - \Omega J)} \right] \nonumber \\
    &= \e^{\beta \Phi q} \, Z^{-1} \Tr \left[ \varphi^\dagger(0,0,\mb{y}) \varphi(z + \ii \beta ,\phi + \ii \beta \Omega, \mb{x}) \e^{- \beta (H - \Phi Q - \Omega J)} \right] \nonumber \\
    \label{eq:4_KMS_Twisted}
    &= \e^{\beta \Phi q} \, G_-^{\beta, \Phi, \Omega}(z + \ii \beta, \phi + \ii \beta\Omega, \mb{x}, \mb{y}) \, ,
\end{align}
which is a generalization of \eqref{eq:2_KMS_v1}. In Section \ref{subsec:2_Near_Horizon}, we argued that the Euclidean counterpart of the ``periodicity'' of the propagators given by the KMS condition is the periodic identification of the coordinate $\tE$ (imaginary part of $z$) along the thermal circle, $\tE \sim \tE + \beta$. This leads to a gravitational path integral with boundary conditions on $S^1_\beta \times \Sigma$, describing the partition function of the system in the canonical ensemble (see discussion around \eqref{eq:3_FreeEnergy_Def}). Here, we have a generalization. By analogy, the KMS condition \eqref{eq:4_KMS_Twisted} means that we should be imposing a \textit{twisted} identification of the coordinates on the boundary $\partial M = S^1_\beta \times S^2$: if the metric is given by the product of the two round metrics, then
\begin{equation}
\label{eq:4_KN_Identifications_Twisted}
    (\tE, \phi) \sim (\tE, \phi + 2\pi) \sim (\tE + \beta, \phi + \ii \beta\Omega) \, .
\end{equation}
Compared to the case of the static black hole, we specify additional chemical potentials at the boundary: $\Omega$, corresponding to a $U(1)$ isometry of the solution, and $\Phi$, due to the presence of a $U(1)$ gauge field. We are thus in a grand canonical ensemble and the gravitational path integral computes the characteristic state function $\beta F(\beta, \Omega,\Phi) = - \log Z(\beta, \Omega, \Phi)$.

\medskip

Concretely, the paradigmatic spacetime describing a rotating electrically charged object in $AdS$ is the $AdS$-Kerr--Newman solution, which has the line element
\begin{align}
\label{eq:4_AdSKN_LineElementBLCoordinates}
\begin{split}
	\rd s^2 &= - \frac{\Delta_\theta  }{\Xi ^2 W} \left( \Delta_\theta  \Delta_r - a^2 \sin ^2\theta \left( 1 + \frac{r^2}{\ell^2}\right)^2 \right) \rd t^2 - 2 \sin^2\theta \, a \Delta_\theta \frac{2Mr - Q^2}{ W \Xi^2} \, \rd t \rd\phi \\
	& \ \ \ + \frac{\Delta_\theta  \left(a^2+r^2\right)^2-a^2 \Delta_r \sin ^2\theta }{\Xi ^2 W} \sin^2\theta \, \rd\phi^2 + W \left( \frac{\rd r^2}{\Delta_r} + \frac{\rd \theta^2}{\Delta_\theta}  \right)
\end{split}
\end{align}
and the gauge field
\begin{equation}
    A = \frac{ Q r }{W\Xi}\Big( \Delta_\theta \, \rd t - a \sin^2\theta \, \rd \phi \Big) + c \, \rd t \, .
\end{equation}
Here $c$ is a constant, and
\[
\begin{split}
& \Delta_r = (r^2 + a^2) \left( 1 + \frac{r^2}{\ell^2} \right) - 2 M r + Q^2 \, , \\
& \Delta_\theta = 1 -  \frac{a^2}{\ell^2} \, \cos^2 \theta \, , \qquad W = r^2 + a^2 \cos^2 \theta \, , \qquad \Xi = 1 - \frac{ a^2}{\ell^2} \, .
\end{split}
\]
The metric and gauge field are real if the parameters $(M,Q,a)$ are real, which means that upon Wick rotation $t=-\ii \tE$, this metric will not be Riemannian but complex, since the terms $g_{\tE \phi}$ will be pure imaginary, and the gauge field as well. It is sometimes called a ``quasi-Euclidean'' metric.\footnote{More generally, this is a class of metrics that when put in the ADM form have pure imaginary shift vectors $N^i$ (see \eqref{eq:4_ADM_Decomposition}).} How can we analyze this case? What is the regularity that one should impose on a complex solution?

\medskip

One way out of this impasse is to analytically continue $a$, $Q$ and $c$: we define $a = \ii a_E$, $Q=\ii Q_E$, $c=\ii c_E$, with real $a_E$, $Q_E$ and $c_E$, thus obtaining a Riemannian metric and a real gauge field \cite{Gibbons:1979xm}. At this point, we can use the framework of the Euclidean gravitational path integral. We pick as $\partial M$ a surface of constant $r=r_0$, and consider the induced metric and gauge field. In fact, in order to match the boundary conditions, we need to do one further coordinate change: as $r_0\to \infty$, we find
\begin{align}
	\rd s^2 &\sim \ell^2 \frac{\rd r^2}{r^2} + \frac{r^2}{\ell^2} \left[ \frac{ 1+a_E^2\cos^2\theta/\ell^2 }{1+a_E^2/\ell^2} \rd \tE^2 + \ell^2 \frac{\rd\theta^2}{1+a_E^2\cos^2\theta/\ell^2} + \ell^2 \frac{\sin^2\theta}{1+a_E^2/\ell^2}\rd\phi^2 \right] \, ,  \nonumber \\
\label{eq:4_KN_hE_Infinity_1}
	A &\sim c_E \, \rd \tE \, .
\end{align}
To show that this space is actually asymptotically $AdS$, we need to exchange $(r,\theta)$ for $(y,\Theta)$
\[
	r^2 = y^2 ( 1 + a_E^2 \sin^2\Theta/\ell^2 ) \, , \qquad \cos^2\theta = \frac{\cos^2\Theta}{1 + a_E^2 \sin^2\Theta/\ell^2} \, ,
\]
and then \eqref{eq:4_KN_hE_Infinity_1} becomes
\begin{equation}
\label{eq:4_KN_hE_Infinity_2}
\begin{split}
	\rd s^2 \sim \ell^2 \frac{\rd y^2}{ y^2} + \frac{y^2}{\ell^2} \left( \rd \tE^2 + \ell^2  (\rd \Theta^2 + \sin^2\Theta \, \rd\phi^2) \right) \, .
\end{split}
\end{equation}
Finally, to match the boundary conditions, we should impose the identification \eqref{eq:4_KN_Identifications_Twisted}, and in order to keep $\phi$ real, we introduce $\Omega_E \equiv \ii \Omega\in\R$. This shows that, upon setting $c_E = \Phi_E$, the analytic continuation of the $AdS$-Kerr--Newman solution is a semiclassical saddle of the gravitational path integral with the required boundary conditions. According to the argument above, we expect to be able to use this framework to describe the grand canonical ensemble of a thermal gravitational system with charge and angular momentum. To do so, we should first make sure that the classical solution is smooth. After the analysis, we would analytically continue the parameters back to their Lorentzian values.

Imposing smoothness of the Riemannian metric means that the topology is locally the product of a cigar and a 2-sphere, and allows us to identify the thermodynamical potentials appearing in the identifications \eqref{eq:4_KN_Identifications_Twisted} as
\begin{equation}
\label{eq:4_KN_beta_Omega_Phi}
    \beta = 4\pi \frac{r_+^2 - a_E^2}{\Delta'(r_+)} \, , \qquad \Omega_E = a_E \frac{1 + r_+^2/\ell^2}{r_+^2 - a_E^2} \, , \qquad c_E = - \frac{Q_E r_+}{r_+^2 - a_E^2} = \Phi_E \, .
\end{equation}
The Killing vector
\begin{equation}
\label{eq:4_KN_EuclideanKV}
    \xi_E = \frac{\partial \ }{\partial \tE} + \Omega_E \frac{\partial \ }{\partial \phi}
\end{equation}
generates a $U(1)$ isometry of the spacetime: its orbits are circles in the cigar factor that shrink as we go to $r= r_+$, since $\abs{\xi_E}^2\rvert_{r=r_+} = 0$, and thus $\xi_E$ vanishes there: the fixed point set is the $S^2$ in the transverse space. This corresponds to the fact that in the Lorentzian metric \eqref{eq:4_AdSKN_LineElementBLCoordinates}, the locus $\{r= r_+ \}$ is an event horizon for the spacetime, and it is generated by the Killing vector
\[
    \xi = \frac{\partial \ }{\partial t} + a \frac{1 + r_+^2/\ell^2}{r_+^2 + a^2} \frac{\partial \ }{\partial \phi} = \ii \xi_E \rvert_{\tE = \ii t, a_E = - \ii a, Q_E = - \ii Q}\, . 
\]
One can then check that we can indeed provide a ``physical'' (i.e. Lorentzian) interpretation for the quantities in \eqref{eq:4_KN_beta_Omega_Phi}: $\beta = 2\pi/\kappa$, where $\kappa$ is the surface gravity, consistently with Hawking's temperature formula \eqref{eq:2_HawkingTemperature}; $\Omega_E = \ii \Omega$, where $\Omega$ is the angular velocity of the horizon introduced at the beginning of the section; and $\Phi_E = \ii \Phi $, where $\Phi$ is the electrostatic potential, defined as $\Phi \equiv \xi \hook A \rvert_{r=r_+} - \xi \hook A \rvert_{r\to \infty}$. Moreover, we can compute the entropy using the Bekenstein--Hawking definition in terms of the area of the bifurcation surface in the horizon (which is a non-round 2-sphere), obtaining 
\begin{equation}
\label{eq:4_KN_BH_entropy}
	S_{\rm BH} = \pi \frac{r_+^2 - a_E^2}{1+a_E^2/\ell^2} \, .
\end{equation}

\medskip

Since we are working in the grand canonical ensemble, we should express the free energy, or, equivalently, the on-shell action, in terms of the potentials $(\beta, \Phi_E, \Omega_E)$. Therefore, we should ultimately view the formulae \eqref{eq:4_KN_beta_Omega_Phi} as fixing the parameters of the solution $(M,Q_E,a_E)$ in terms of the boundary data $(\beta, \Phi_E, \Omega_E)$. Inverting \eqref{eq:4_KN_beta_Omega_Phi} is in fact impossible to do analytically. However, we can compute the on-shell action using holographic renormalization (using the counterterm \eqref{eq:3_Counterterms}), finding
\[
    I = \frac{\beta}{2 \Xi}\left[ M - \frac{r_+}{\ell^2} (r_+^2 - a_E^2) + \frac{r_+ Q_E^2}{r_+^2 - a_E^2} \right]  \, .
\]
This satisfies the quantum statistical relation \eqref{eq:3_QuantumStatisticalRelation} in the form
\begin{equation}
\label{eq:4_KN_QuantumStatisticalRelation}
    I(\beta, \Phi, \Omega) = - S_{\rm BH} + \beta ( \braket{E} - \Phi_E \braket{Q_E} - \Omega_E \braket{J_E} ) \, ,
\end{equation}
where
\[
    \braket{E} = \frac{M}{\Xi^2} \, , \qquad \braket{Q_E} = \frac{Q_E}{\Xi} \, , \qquad \braket{J_E} = - \frac{a_E M}{\Xi^2} \, ,
\]
and $S_{\rm BH}$ is the Bekenstein--Hawking entropy \eqref{eq:4_KN_BH_entropy}. 
We can also check that the variables $(\braket{E}, \braket{Q_E}, \braket{J})$ are indeed canonically conjugate to the thermodynamic potentials $(\beta, \Phi, \Omega)$. We conclude that we can now interpret \eqref{eq:4_KN_QuantumStatisticalRelation} as a Legendre transform from the grand canonical to the microcanonical ensemble, and read off the thermodynamic entropy, which (again) coincides with the Bekenstein--Hawking formula \eqref{eq:4_KN_BH_entropy}, confirming once again that the Euclidean gravitational path integral approach gives thermal properties that are consistent with the KMS definitions in Lorentzian spacetime.

\medskip

Despite the remarkable success of the computation, we should recall that we bypassed the subtleties of the complex metric by performing analytic continuation of the parameters, with the result that the interpretation of the ensemble is now quite obfuscated \cite{Witten:2021nzp}. The problem we started with was the description of the grand canonical ensemble \eqref{eq:4_GibbsFormula_GrandCanonical} in the \textit{real} Lorentzian background \eqref{eq:4_AdSKN_LineElementBLCoordinates}, that is, the matrix element of the (un-normalized) density matrix
\[
    \rho = \e^{ - \beta (H - \Phi Q - \Omega J) } \, .
\]
The computations done with the (real) Riemannian metric lead to a consistent picture, but note that the potentials appearing in the density matrix above are now pure imaginary
\[
\begin{split}
    \beta &= 2 \pi \frac{ r_+^2 - a_E^2}{ r_+ - M + ( 2r_+^2 - a_E^2) r_+/\ell^2 } \, , \\ 
    \Phi &= - \ii \Phi_E = \ii \frac{Q_E r_+ }{r_+^2 - a_E^2} \, , \\ 
    \Omega &= - \ii \Omega_E = - \ii a_E \frac{1 + r_+^2/\ell^2}{r_+^2 - a_E^2} \, .
\end{split}
\]
Therefore, the Riemannian metric actually corresponds to a saddle of a putative grand canonical ensemble based on the (un-normalized) density matrix
\begin{equation}
\label{eq:4_DensityMatrix_ComplexEnsemble}
    \rho = \e^{ - \beta (H - \ii \Phi_E Q - \ii \Omega_E J) } \, .
\end{equation}
Not only is the physical meaning of this operator not clear, but even its mathematical properties are not obvious, since evaluated on the states it would correspond to complex weights in the Gibbs formula \eqref{eq:4_GibbsFormula_GrandCanonical}, so it would not be clearly convergent.

An alternative approach is to include in the path integral the complex ``quasi-Euclidean'' metric obtained performing the Wick rotation $t = - \ii \tE$ without analytic continuation of the constants $a$ and $Q$. This is not implausible: after all, in the saddle-point approximation of an integral along the real line, the leading contributions may come from complex saddles reached by deforming the integration contour into the complex plane. However, once we open the Pandora's box of complex metrics in the path integral, we need to define the rules for their inclusion. That is, we need to be able to find a criterion to define the correct integration contour for the gravitational path integral (though we shall see in Section \ref{subsubsec:4_ConformalFactor} that there are reasons to doubt that it may exist).

One such criterion was put forward in \cite{Kontsevich:2021dmb, Witten:2021nzp}. It would take us too far to discuss the criterion in detail, but it is remarkable that when applied to the Wick rotation of the $AdS$-Kerr--Newman black hole \eqref{eq:4_AdSKN_LineElementBLCoordinates} it states that the solution is an ``allowed'' saddle of the gravitational path integral provided $\abs{\Omega}<1$. This is equivalent to the condition that makes the thermal partition function of a conformal field theory on $\partial M \cong S^1\times S^2$ converge! The $AdS$/CFT correspondence guarantees that this latter partition function should be equal to the gravitational path integral with the appropriate holographic boundary conditions, so in this case the criterion of \cite{Kontsevich:2021dmb, Witten:2021nzp} is consistent with the ``microscopic'' description of the gravitational path integral provided by the CFT on the conformal boundary.

\medskip

We conclude with a remark. Though the physical meaning of boundary conditions leading to the density matrix \eqref{eq:4_DensityMatrix_ComplexEnsemble}, with purely imaginary angular velocity and electric potential, is not transparent, partition functions of this type nevertheless arise naturally in the context of supersymmetry-protected observables.
For instance, in order to preserve supersymmetry after the Wick rotation of the $AdS$-Kerr--Newman black hole \eqref{eq:4_AdSKN_LineElementBLCoordinates}, we find that we should impose the following constraint among the chemical potentials
\begin{equation}
\label{eq:4_SUSY_Constraint}
    \beta ( 1 - 2 \Phi + \Omega) = 2\pi \ii ( 1 + 2n ) \qquad n \in \Z \, ,
\end{equation}
and the density matrix \eqref{eq:4_DensityMatrix_ComplexEnsemble} becomes
\[
    \rho_{\rm SUSY} = (-1)^{2J} \e^{ - \beta ( H - J - Q) + \beta (\Phi-1) ( Q + 2J) } \, , 
\]
which after trace over the Hilbert space leads to the so-called superconformal index
\[
    \cI = \Tr (-1)^{2J} \e^{ - \beta (H-J-Q) + \beta(\Phi-1) (Q + 2J)} \, .
\]
This observable has remarkable properties in a supersymmetric theory, which follow from the fact that $H-J-Q$ is the anti-commutator of two supercharges, that is, $\{\cQ, \tilde{\cQ}\} = H-J-Q \equiv \Delta$, and $(Q-2J)$ is a combination of bosonic generators that commutes with both $\cQ$, $\tilde{\cQ}$. This guarantees that states with $\Delta > 0$ show exact boson-fermion degeneracy, appearing in boson-fermion pairs, and therefore do not contribute to $\cI$, because of the $(-1)^{2J}$ factor \cite{Witten:1982df}. The only contributions to $\cI$ come from \textit{supersymmetric} states that are annihilated by $\cQ$ and $\tilde{\cQ}$. Supersymmetric indices are crucial observables for the counting of microstates of supersymmetric black holes, but the supersymmetry constraint \eqref{eq:4_SUSY_Constraint} \textit{cannot} be solved with real $(\beta, \Phi, \Omega)$. Therefore, consistency with supersymmetry compels us to include in the gravitational path integral complex metrics with \textit{complex} angular velocity or electrostatic potential \cite{Cabo-Bizet:2018ehj, Iliesiu:2021are}, and to check whether the criterion of \cite{Kontsevich:2021dmb, Witten:2021nzp} is satisfied by these metrics \cite{MasterThesis, BenettiGenolini:2025jwe, BenettiGenolini:WIP}.

\subsection{Some subtleties}
\label{subsec:4_Subtleties}

We delved into the ``allowability'' of complex metrics in the path integral arising in the context of rotating solutions, but there are more subtleties in the Euclidean quantum gravity approach, which we have until now deliberately avoided and remain the subject of active research. We will mention here a few without attempting to propose any solution.

\subsubsection{Inclusion of multiple topologies}

In the discussion of quantum field theory on a Lorentzian curved spacetime, we usually restrict ourselves to a globally hyperbolic one. As mentioned in footnote \ref{footnote:2_GloballyHyperbolic}, a globally hyperbolic spacetime is foliated by surfaces of constant ``time'', and the topology of spacetime is $\R_t \times \Sigma$.
After Wick rotation, this leads us naturally to consider Euclidean solutions with (asymptotic) boundary $S^1_{\tE} \times \Sigma$, which encode thermal properties of the Lorentzian system. More generally, the $S^1_{\tE}$ factor heuristically represents the existence in Lorentzian of a trace over a Hilbert space defined on the spatial slice $\Sigma$. 

However, there are plenty of \textit{gravitational instantons}: Riemannian regular solutions with finite Einstein--Hilbert action (with GHY term and regulator) and (asymptotic) boundary whose topology is different from $S^1\times \Sigma$.\footnote{Their name comes from the analogy with \textit{instantons} in quantum field theory (and quantum mechanics), where the name denotes solutions to the Euclidean equations of motion with finite action. Schematically, these represent tunneling effects between different vacua of the theory.} For instance, we could have smooth solutions which interpolate between different topologies (think about a pair of pants) that would be singular in Lorentzian, or even smooth solutions with boundaries that do not at all have a well-defined Lorentzian continuation (for instance, they could have topology $M \cong \R^4$ and fill a $\partial M \cong S^3$, or even more exotic three-manifolds).\footnote{The existence of a time-orientable Lorentzian structure on a manifold requires certain topology, whereas this is not true of Riemannian metrics, which exist on any smooth manifold.} What should we make of these solutions?

\subsubsection{Conformal factor problem}
\label{subsubsec:4_ConformalFactor}

One of the reasons we put forward as an argument for the use of the Euclidean action in the path integral for quantum gravity was that in Euclidean signature the path integral of quantum field theory converges. Whilst this does hold for the theories of fields of spin $0$ and $1$ and for anti-commuting fields of spin $\frac{1}{2}$, whose Euclidean actions define positive semi-definite quadratic forms, it is not true for gravity, because the Euclidean action 
\[
    S_E[g_E] = - \frac{1}{16\pi} \int_M R \, \sqrt{g_E} \, \rd^4 x - \frac{1}{8\pi} \int_{\partial M} K \, \sqrt{h_E} \, \rd^3 x 
\]
does not have a definite sign \cite{Gibbons:1978ac}. Indeed, under a Weyl rescaling of the metric $g_E$ to $\Tilde{g}_E = \Omega^2 g_E$ (which is \textit{not} a diffeomorphism but a change in the space of metrics), we have
\[
    \Tilde{R} = \Omega^{-2} \left( R - 6 \Omega^{-1} \nabla^2 \Omega \right) \, , \qquad \Tilde{K} = \Omega^{-1} \left( K + 3\Omega^{-1} n^a \nabla_a \Omega \right) \, .
\]
Therefore (integrating by parts and using the divergence theorem)
\[
    S_E[\Tilde{g}_E] = - \frac{1}{16\pi} \int_M \left( \Omega^{2}  R + 6 \abs{\rd \Omega}^2 \right) \, \sqrt{g_E} \, \rd^4 x - \frac{1}{8\pi} \int_{\partial M} \Omega^{2} K \, \sqrt{h_E} \, \rd^3 x \, ,
\]
and we can make this arbitrarily negative by choosing $\abs{\rd \Omega}^2$ sufficiently large, so the Euclidean gravity action is unbounded below. This problem does not arise at the leading order in the semiclassical expansion \eqref{eq:3_SemiclassicalExpansion}, as one is considering only a solution of the equations of motion, but it does become crucial when including the one-loop contribution. 

Recent work emphasizes that the conformal factor problem of the Euclidean gravitational action is not resolved by imposing constraints or restricting the class of metrics \cite{Horowitz:2025zpx}. Even off-shell, the Einstein--Hilbert action remains unbounded below under conformal deformations, and no canonical prescription for the integration contour in complexified metric space is currently known (as remarked in Section \ref{subsec:4_Rotation}). This suggests that Euclidean saddle points should be treated as useful semiclassical tools rather than as defining a fundamental, well-defined path integral formulation of quantum gravity.

\subsubsection{Gauge group}
\label{subsubsec:3_GaugeGroup}

Another problem that arises only when going beyond the leading order in \eqref{eq:3_SemiclassicalExpansion} is that of gauge redundancy. In defining the gravitational path integral, one (formally) integrates over all field configurations, including those that are physically equivalent due to redundancies in the description. In ordinary gauge theories, this overcounting is handled by systematic procedures such as the BRST or BV formalism. Before applying similar tools to gravity, however, one must first clarify in what sense general relativity is a gauge theory, and which transformations should genuinely be regarded as gauge redundancies. Loosely speaking, general relativity is indeed a gauge theory of the diffeomorphism group, but the definition of its gauge group is subtler than in Yang--Mills theory:\footnote{For the \textit{cognoscenti}: at the end of the day, a difference with, say, Yang--Mills theory of the Lie group $G$, which is based on a $G$-principal bundle $P\to M$ is due to the fact that in general relativity the relevant principal bundle is the frame bundle itself.} not all diffeomorphisms act trivially on the physical phase space or preserve boundary data, and so, only a subset should be viewed as redundancies to be factored out in the path integral. In particular, diffeomorphisms acting as redundancies necessarily belong to the infinite-dimensional component of the diffeomorphism group connected to the identity, Diff$_0(M)$.\footnote{The Lie group of diffeomorphisms is not necessarily connected: for instance, if $M$ is a Lie group $G$, then $G$ acts on itself via left-translation, so Diff$(G)$ contains $G$, and thus may have multiple connected components.} Components that are not connected to the identity are considered to act non-trivially on the theory, that is, they are not redundancies to be ``gauged away.'' In fact, even the action of diffeomorphisms in the component connected to the identity may be non-trivial on the boundary conditions. 

Let $X^a$ be a vector representing the infinitesimal action of a diffeomorphism in Diff$_0(M)$, and we let $X^a$ act on the Lagrangian density $\cL_{\rm EH} = (R - 2\Lambda) \sqrt{g_E}$ via Lie derivative. Recalling that $\cL_{\rm EH}$ is a tensor density of weight 1, we have
\[
\begin{split}
    \cL_X \left[(R - 2\Lambda) \sqrt{g_E}\right] &= X^a \partial_a (R - 2\Lambda) \, \sqrt{g_E} + \nabla_a X^a (R - 2\Lambda) \, \sqrt{g_E} \\
    &= \nabla_a \left[ X^a (R - 2\Lambda) \right] \sqrt{g_E} \, ,
\end{split}    
\]
so we can use the divergence theorem to conclude that the variation of the Einstein--Hilbert action is a boundary term
\[
    \delta_X \int_M (R- 2\Lambda) \, \sqrt{g_E} \, \rd^4 x = \int_{\partial M} n_a X^a (R - 2\Lambda) \, \sqrt{h_E} \, \rd^3 x \, .
\]
If the vector field does not have compact support, or more generally, fails to vanish sufficiently fast near the boundary, the action is not invariant under the diffeomorphism it generates, and so it is \textit{not} a redundancy in the description. The diffeomorphisms that act in a non-trivial way on the physical phase space by changing boundary data or conserved charges, whether because they are not connected to the identity or because the generating vector field does not vanish at the boundary, are sometimes called \textit{large} diffeomorphisms (in contrast to \textit{small} diffeomorphisms).

\medskip

The same distinction between gauge redundancies and physical transformations also appears in the Hamiltonian formulation of general relativity. In a globally hyperbolic spacetime, as already mentioned, we have a global time function $t:M\to \R$ such that $\{ t = 0 \}$ is a Cauchy surface, and from this surface we can construct additional coordinates $x^i$ such that the metric has the form
\begin{equation}
\label{eq:4_ADM_Decomposition}
	\rd s^2 = - N^2 \rd t^2 + h_{ij} ( \rd x^i + N^i \rd t)( \rd x^j + N^j \rd t) \, .
\end{equation}
This is referred to as ``$3+1$'' or ADM decomposition (Arnowitt--Deser--Misner), and shows that we can rewrite the action in terms of the \textit{lapse function} $N(t,\mb{x})$, the \textit{shift vector} $N^i(t,\mb{x})$ and the induced metric on the Cauchy surface at constant $t$, $h_{ij}(t,\mb{x})$. They play the role of canonical variables for the initial value problem. One then computes the canonically conjugate momenta, and it turns out that the only non-vanishing canonical momentum is that conjugate to $h_{ij}$, which is $\Pi^{ij} \sqrt{h}$, where $\Pi^{ij}$ is the Brown--York tensor that already appeared in \eqref{eq:3_Piab} (basically with a computation analogous to that leading to \eqref{eq:3_VariationSEH_SGHY}). The resulting Hamiltonian is composed by a bulk term and a boundary term proportional to $\Pi_{ij}$, but the bulk term, which is proportional to the constraints, \textit{vanishes} on-shell, and the only contribution comes from the boundary.

This should not come to you as a surprise. Defining energy in general relativity is not easy, because of the equivalence principle and the properties of small diffeomorphisms: in a neighbourhood of a point, you can always define normal coordinates such that any quantity defined with the metric and its first derivatives would vanish, thus ``gauging away'' gravity. However, as we just found out, large diffeomorphisms are not just gauge redundancies, and correspondingly the Hamiltonian does receive contributions from the boundary that can be evaluated. In fact, one can show that the variation of the action with respect to the boundary metric, that is $\Pi_{ij}$ in \eqref{eq:3_Piab}, behaves as a stress-energy tensor (finally explaining its name \textit{Brown--York stress-energy tensor}).
Again, this is expected: our (admittedly brief) study of black holes suggested that gravitational physics in a region should be described in terms of its boundary (the holographic principle mentioned in Section \ref{subsec:2_Where_to_now}), which perfectly resonates with what we just described. This is borne out in the context of the $AdS$/CFT correspondence: the dynamics of gravity in $AdS$ is actually described by a theory ``living'' on its boundary for which the Brown--York tensor is the actual stress-energy tensor.


















\newpage

\bibliographystyle{ytamsalpha}
\bibliography{gravity.bib}

\end{document}